\documentclass[acmsmall,screen,nonacm]{acmart}

\AtBeginDocument{%
  \providecommand\BibTeX{{%
    \normalfont B\kern-0.5em{\scshape i\kern-0.25em b}\kern-0.8em\TeX}}}

\setcopyright{cc}
\setcctype{by-sa}
\acmDOI{10.1145/3763106}
\acmYear{2025}
\acmJournal{PACMPL}
\acmVolume{9}
\acmNumber{OOPSLA2}
\acmArticle{328}
\acmMonth{10}
\received{2025-03-25}
\received[accepted]{2025-08-12}

\usepackage{xspace}
\usepackage{pifont}
\usepackage{wrapfig}
\usepackage{fancyvrb}
\usepackage{algorithmicx}
\usepackage{algpseudocode}
\usepackage{algorithm}
\usepackage{amsfonts}
\usepackage{amsthm}
\usepackage{semantic}
\usepackage{listings}
\usepackage{array} %
\usepackage{stmaryrd}
\usepackage{syntax}
\usepackage{proof}
\usepackage{color}
\usepackage{mathtools}
\usepackage{multirow}
\usepackage{bbding}
\usepackage{soul} %
\usepackage{fontawesome}
\usepackage{gensymb}
\usepackage{subcaption}
\usepackage{enumerate} %
\usepackage{graphbox} %

\def\sectionautorefname{Section}
\def\subsectionautorefname{Section}
\def\subsubsectionautorefname{Section}
\theoremstyle{definition}

\makeatletter
\AtBeginDocument{%
    \catcode`_=12
    \begingroup\lccode`~=`_
    \lowercase{\endgroup\let~}\sb
    \mathcode`_="8000
    \immediate\write\@auxout{\catcode`_=12 }%
    \immediate\write\@auxout{\catcode`^=12 }%
}
\makeatother
\begin{document}
\title{Formalizing Linear Motion G-Code for Invariant Checking and Differential Testing of Fabrication Tools}

\author{Yumeng He}
\orcid{0009-0002-1171-4191}
\email{u1528477@umail.utah.edu}
\affiliation{%
  \institution{University of Utah}
  \city{Salt Lake City}
  \state{UT}
  \country{USA}
}

\author{Chandrakana Nandi}
\orcid{0000-0001-8633-8413}
\email{cnandi@cs.washington.edu}
\affiliation{%
  \institution{Certora Inc. and University of Washington}
  \city{Seattle}
  \state{WA}
  \country{USA}
}

\author{Sreepathi Pai}
\orcid{0000-0002-3691-7238}
\email{sree@cs.rochester.edu}
\affiliation{%
  \institution{University of Rochester}
  \city{Rochester}
  \state{NY}
  \country{USA}
}

\def\sectionautorefname{Section}
\def\subsectionautorefname{Section}
\def\subsubsectionautorefname{Section}

\newcommand{\todo}[1]{{\color{red}#1}}
\newcommand{\changesadd}[1]{{\color{blue}#1}}
\newcommand{\changesremove}[1]{{\color{red} \sout{#1}}}

\newcommand{\ymhcomment}[1]{{\color{blue}[#1]}}
\newcommand{\sree}[1]{{\color{olive}[#1]}}

\newcommand{\gcode}{G\hbox{-}code\xspace}
\newcommand{\tool}{\textsc{GlitchFinder}\xspace}
\newcommand{\glitchfinder}{\textsc{GlitchFinder}\xspace}
\newcommand{\glitchrunner}{\textsc{GlitchRunner}\xspace}

\newcommand{\gone}{\lstinline[language=gcode]!G1!\xspace}
\newcommand{\gzero}{\lstinline[language=gcode]!G0!\xspace}
\newcommand{\F}{\texttt{F}\xspace}
\newcommand{\E}{\texttt{E}\xspace}
\newcommand{\X}{\texttt{X}\xspace}
\newcommand{\Y}{\texttt{Y}\xspace}
\newcommand{\Z}{\texttt{Z}\xspace}

\newcommand{\prog}{\textsf{gcode$_{d}^h$}\xspace}
\newcommand{\denote}[1]{\llbracket~ #1 ~\rrbracket}

\newcommand{\mypara}[1]{\textbf{\textit{#1.}}}
\newcommand{\none}{\texttt{None}\xspace}
\newcommand{\infinity}{\texttt{Infinity}\xspace}

\newcommand{\invariant}{$\mathcal{I}$\xspace}

\lstdefinelanguage{gcode}{
  morekeywords={},
  literate=
    {G0}{{{\color{blue}\bfseries G0}}}2
    {G1}{{{\color{blue}\bfseries G1}}}2
    {X}{{{\color{blue}\bfseries X}}}1
    {Y}{{{\color{blue}\bfseries Y}}}1
    {Z}{{{\color{blue}\bfseries Z}}}1
    {E}{{{\color{blue}\bfseries E}}}1
    {F}{{{\color{blue}\bfseries F}}}1,
  sensitive=true,
}
\lstset{
  language=gcode,
  basicstyle=\small
}

\renewcommand{\changesadd}[1]{#1}
\renewcommand{\changesremove}[1]{}


\begin{abstract}
    The computational fabrication pipeline for 3D printing is much like a compiler ---
   users design models in Computer Aided Design (CAD) tools
   that are lowered to polygon meshes to be ultimately
   compiled to machine code by 3D slicers.
For traditional compilers and programming languages,
  techniques for checking program invariants are well-established.
Similarly, methods like differential testing are frequently used
 to uncover bugs in compilers themselves, which makes them more reliable.

The fabrication pipeline would benefit from similar techniques but
  traditional approaches do not directly apply to the representations
  used in this domain.
Unlike traditional programs, 3D models exist both as geometric objects (a CAD model or a polygon mesh) as well as machine code that ultimately runs on the hardware.
The machine code, like in traditional compiling, is affected by many factors like the model, the slicer being used, and numerous user-configurable parameters that control the slicing process.

In this work, we propose a new algorithm
  for lifting \gcode
  (a common language used in many fabrication pipelines)
  by denoting a \gcode program to a set of cuboids, and
  then defining an approximate point cloud representation
  for efficiently operating on these cuboids.
Our algorithm opens up new opportunities:
  we show three use cases that demonstrate
  how it enables
  (1)~error localization in CAD models through invariant checking,
  (2)~quantitative comparisons between slicers, and
  (3)~evaluating the efficacy of mesh repair tools.
We present a prototype implementation of our algorithm in a tool, \tool, and
  evaluate it on 58 real-world CAD models.
Our results show that \tool is particularly
  effective in identifying slicing issues due to small features,
  can highlight differences in how popular slicers (Cura and PrusaSlicer) slice the same model, and can identify
  cases where mesh repair tools (MeshLab and Meshmixer)
  introduce new errors during repair.

\end{abstract}


\maketitle

\section{Introduction}
\label{sec:intro}

Programming language and compiler researchers have noted
  similarities between traditional compiler
  toolchains and
  computational fabrication pipelines.
Akin to source code, designers prepare a 3D model in computer-aided design (CAD) programs, export it to a polygon mesh which functions like a conventional intermediate representation (IR), that is then ``compiled'' to instructions for a computer-aided manufacturing (CAM) tool which runs these instructions to fabricate the physical model.
Over the years, there has been work on
  correct-by-construction computer-aided design (CAD) compilers,
  tools that reverse engineer polygonal meshes to CAD models, program synthesizers for
  CAD, and domain specific languages (DSLs) targeting
  various fabrication related tasks~\cite{verso, snapl, inverse, reincarnate, taxon, imprimer, surynek2025objectpackingschedulingsequential, sherman}.

In this work, we use techniques from program semantics and differential testing to automate the detection of errors in models as well as in tools used in the 3D printing fabrication pipeline.

\textbf{The 3D printing fabrication pipeline.}
First, a user designs their model using geometric constructs in a CAD software.
These can be textual programmatic systems like OpenSCAD~\cite{openscad}, or interactive 3D geometric manipulation software like
SolidWorks~\cite{solidworks}, Onshape~\cite{onshape},
and Rhino~\cite{rhino}.
The CAD software then generates a low-level polygon mesh
  by compiling the user's high-level design
  into a set of triangles that
  represent the surface of the model. Next, a \textit{slicer} (e.g., 
  PrusaSlicer~\cite{prusaslicer}, Cura~\cite{cura},
  Simplify3D~\cite{simpl}, Slic3r~\cite{slic}) slices the mesh, usually to
  generate 2D horizontal layers from which it
  then emits \gcode~\cite{gcode, gcodestandard}
  for a 3D printer to execute.
  
\gcode is a programming language for numerical control,
developed in the 1960s
and defined by the RS-274 standard~\cite{gcodestandard}.
It is the de facto standard for
   computer aided manufacturing on machines like
   3D printers, laser cutters, and CNC (Computer Numerical Control) mills.
\gcode is analogous to low-level machine instructions but
  omits constructs like loops and conditionals.
The 3D printers we focus on are fused-deposition modeling (FDM) printers,
  which heat and melt polymer filament and lay it down
  on a ``print-bed'' through an extruder to create a 3D object.
The process of laying down the plastic is specified by commands embedded in the \gcode that move the extruder in 3D space while extruding plastic.  
The 3D printer's firmware~\cite{marlinfirmwaremarlin2024, klippercommunityklipper2024}
  interprets these commands.
Other \gcode commands heat
  the print bed and the extruder,
  select and switch between possibly multiple extruders,
  perform calibration, etc.~\cite{gcode2025}.
  
\begin{figure}
\begin{minipage}{0.3\linewidth}
    \includegraphics[scale=0.12]{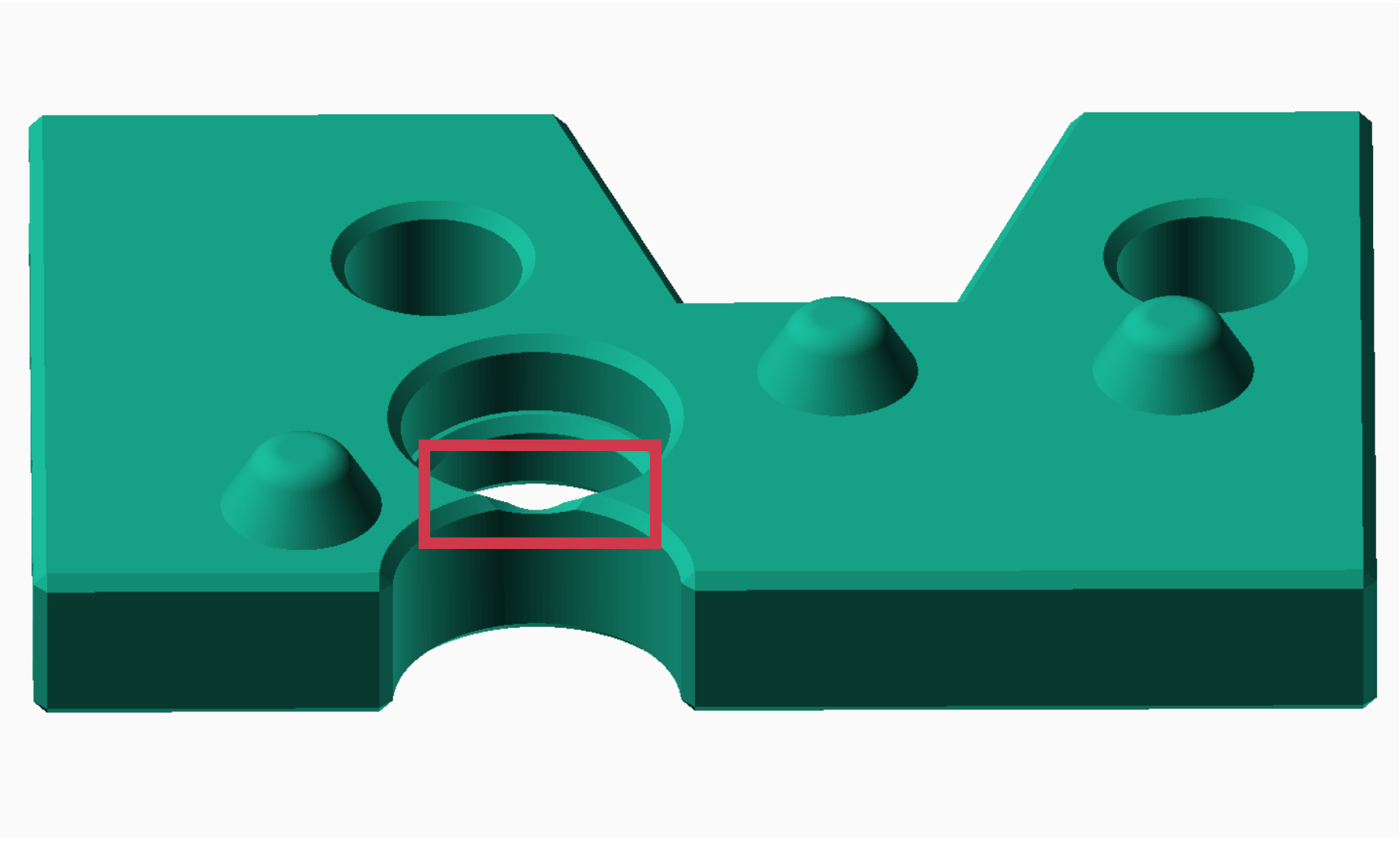}
\end{minipage}
\hfill
\begin{minipage}{0.3\linewidth}
    \centering
    \includegraphics[scale=0.12]{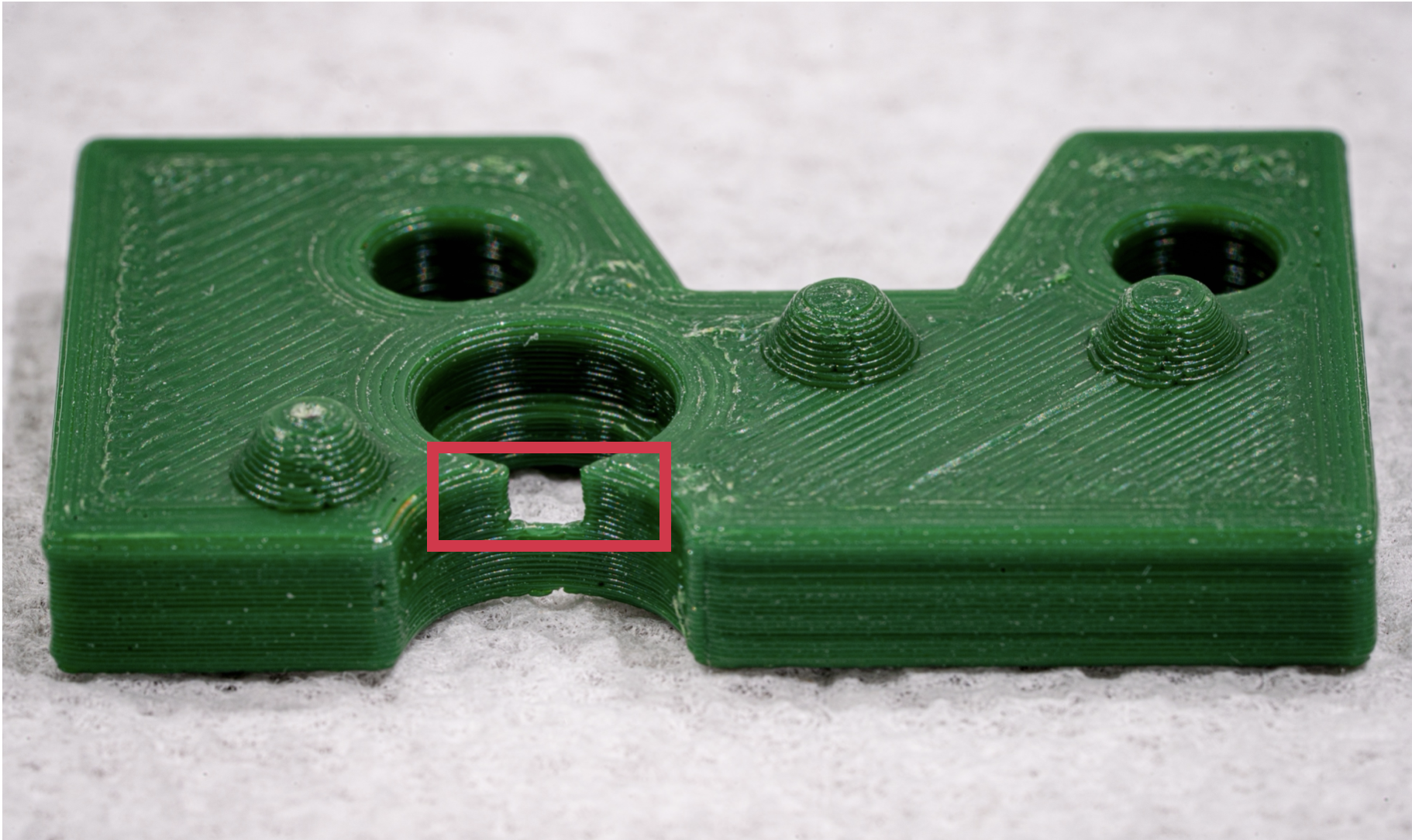}
\end{minipage}
\hfill
\begin{minipage}{0.3\linewidth}
    \centering
    \includegraphics[scale=0.08]{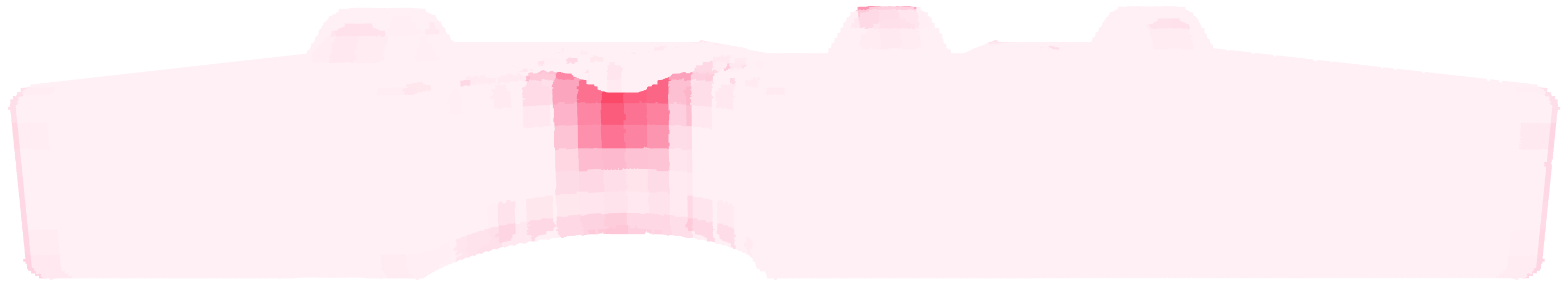}
\end{minipage}

    \caption{3D printed mechanical part (center) diverges from the original design (left) due to thin walls. The slightly
    curved gap in the model (highlighted by the red square) when printed, appears as a rectangular gap. The heatmap generated by our work (right) reveals this deviation before the model is printed.}
    \label{fig:nice-ercf-photo}    

\end{figure}

\textbf{Challenges with 3D printing.}
In practice, 3D printing requires repeated attempts before a successful print~\cite{chilana1}.
A part may fail to print correctly due to various reasons.
The design itself may have issues like feature sizes that are too small for a given hardware.
The mesh may have problems like
  flipped normals due to which the
  ``inside'' and ``outside'' of the model may be swapped.
And finally, the slicer may fail to generate \gcode
  due to bugs in the slicer itself or
  in some cases the generated \gcode may not accurately represent the original model.
To overcome these challenges,
  users may have to change/fix the design, fix mesh errors
  (e.g., holes,
  duplicate vertices and faces, invalid 2-manifold)
  using (often multiple!) mesh repair tools like MeshLab~\cite{meshlab} and Meshmixer~\cite{meshmixer}
  before slicing, adjust various settings that are used to configure slicers like
  filament extrusion rate,
  temperature, change filament, and so on, or use different slicers to overcome slicing errors and differences due to heuristics used in slicing algorithms.
Since 3D printers are slow, a single 3D print can take
  from many hours to days, making this iterative process time and resource intensive.

Debugging \gcode is like debugging low-level assembly.
Originally intended for small, low-cost
  microcontrollers optimizing for factors like frame resonance,
  acceleration, and toolhead inertia,
  standard \gcode lacks control flow and uses fully unrolled loops.
This results in extremely long, straight-line programs.
Commands specify absolute or relative positions via floating-point
  constants derived from high-level 3D models,
  making it hard to trace code back to the original geometry.

\textbf{The Problem: Lack of \gcode analysis tools.}
Being able to (1) check invariants of the
  \gcode before starting a print, and (2) compare the output of various tools
  that are used for slicing and repairing meshes
  via differential testing
  can help make the \gcode more reliable and
  the tools more robust, ultimately leading to a
  more efficient fabrication process.
In traditional compilers and programming languages,
  there is decades of research on program analysis and differential testing;
  similar techniques would benefit
  computational fabrication.
  
However, traditional approaches do not directly
  apply to this domain because representations
  like CAD designs, polygon meshes, and \gcode are different
  from typical programs.
CAD designs and meshes are for example representations
  of an object's geometry,
  whereas the \gcode that ultimately
  runs on CNC machines is expected to produce the
  \textit{same} geometry (modulo physical constraints) while also being affected
  by factors like 
  the slicer being used,
  the printer the \gcode is being generated for, and
  the settings used to slice the model~\cite{nadyachi2025}.
Two \gcode programs that correspond to the same model can also
   be different based on the slicing algorithm used,
   making it even harder to compare \gcode programs from
   two different slicers.
  
The choice of representation being analyzed or used for
  comparing slicers and mesh repair tools
  is an important one --- a model
  may represent a valid geometry at the CAD or mesh level but fail to slice
  or print correctly
  (\autoref{fig:nice-ercf-photo} shows a real-world example).
  Thus using a mesh representation for %
  differential testing of slicers and mesh repair tools may
  reveal little about whether the
  output of these tools will ultimately
  lead to successful printing and slicing.

\begin{figure}
    \centering
    \includegraphics[width=\linewidth]{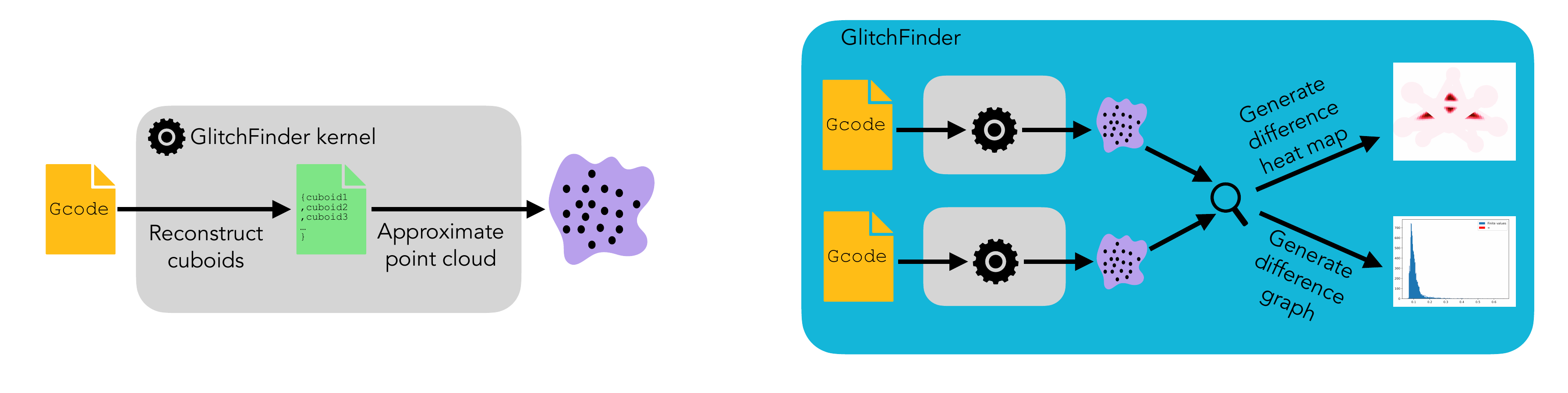}
    \caption{(Left) \tool's kernel first reconstructs a set of cuboids from a \gcode program to denote it. From the reconstructed cuboids,
    \tool then generates a point cloud. (Right) This kernel can then be used for comparing two \gcode programs to generate a difference heatmap
    and a difference distribution graph.
    As later sections will show, both model invariant checking and differential testing can be reduced to a comparison of two \gcode programs.}
    \label{fig:glitchflow}
\end{figure}

\textbf{Our solution and insights.}
We address these challenges by proposing
  a new technique for
  analyzing \gcode programs.
Our work is guided by the following three key insights.

Our \textit{first} insight is that analyzing
  \gcode captures the combined effect of
  the CAD design, the slicer used, and
  slicer settings.
This is analogous to analyzing low-level bytecode
  as opposed to source code.
We show how
  this ``analyze-what-executes'' approach
  helps check key invariants and
  localize potentially problematic regions of a model that
  may not print successfully,
  even if the model at the CAD design or mesh level
  represents a valid geometry.

Our \textit{second} insight is a \gcode program
  can be lifted by constructing
  a set of cuboids that \textit{denotes} the program.
  Being able to represent a \gcode program in this manner
  gives us a way to analyze \gcode programs and compare them
  by comparing the set of cuboids they represent.

Our \textit{third} insight is that by comparing two \gcode
  programs we can differentially test
  mesh repair tools and slicers.
We show that this is an effective approach
  for comparing the
  behavior of different slicers, and
  for evaluating
  the efficacy of popular mesh repair tools.
  
As shown in the heatmap in \autoref{fig:nice-ercf-photo} (right),
  our technique successfully localizes the part of the
  model in \autoref{fig:nice-ercf-photo} (left) that 
  failed to print correctly (dark red in \autoref{fig:nice-ercf-photo}, center).
Crucially, it finds the problem by statically
  analyzing the \gcode,
  \textit{before} the model is printed.

We built a prototype tool, \tool, to implement our approach.
\tool reconstructs
  a set of cuboids that \textit{denotes} a \gcode program (\autoref{fig:glitchflow}, left).
To efficiently compare two \gcode programs,
  \tool uses a sampling-based approach to approximate
  a point cloud from the cuboid set
  (\autoref{fig:glitchflow}, left).
\tool uses a second algorithm to efficiently
  compare two point clouds to generate a difference heatmap
  for visualizing the differences between two programs
  (\autoref{fig:glitchflow}, right).
We use \tool to (1)~localize errors on 56 complex, real-world,
  3D models, (2)~compare two popular slicers
  to identify different behaviors on the same models,
  and, (3)~compare two popular mesh repair tools to
  evaluate their efficacy.
This paper makes the following contributions:
\begin{itemize}
    \item A new method
    for analyzing \gcode programs by
    reconstructing a cuboid set which is then 
     approximated as a point cloud
    and an algorithm for comparing two point clouds.
    \item Using point cloud comparison for invariant checking
     to localize problematic parts of a 3D model that cannot be identified from the CAD design or mesh representation.
    \item Using point cloud comparison to differentially test slicers and mesh repair tools.
    \item A prototype implementation in a tool dubbed \tool, and its evaluation over
    real-world models, popular
    slicers, and mesh repair tools.
\end{itemize} 

The rest of the paper is structured as follows:
\autoref{sec:background} provides background
  on slicing and \gcode,
\autoref{sec:line-rec} presents
  big step operational semantics for the linear motion subset of \gcode
  this work targets and our cuboid reconstruction algorithm,
\autoref{sec:comp} presents a sampling-based point cloud
  generation approach from the reconstructed cuboids and
  a new algorithm for
  comparing two \gcode programs,
\autoref{sec:visual} discusses showing the output
  of comparing two \gcode programs,
\autoref{sec:impl} has implementation details and limitations,
\autoref{sec:invcheck} shows how
  \gcode analysis helps check invariants of 3D models,
\autoref{sec:difftest} shows how \gcode comparison enables differential testing
  of slicers and mesh repair tools,
\autoref{sec:related} presents related work, and
\autoref{sec:conclusions} concludes.

\section{Background on \gcode}
\label{sec:background}

\begin{wrapfigure}{r}{0.35\textwidth}
\vspace{-30pt}
\lstset{language=gcode}
\begin{lstlisting}[basicstyle=\tiny\ttfamily]
...
G1 X151.801 Y158.165
G1 F1800 E-0.75
G1 F600 Z2.3
G0 F18000 X150.411 Y157.446 Z2.3
G0 X159.094 Y155.912
G1 F600 Z2.1
G1 F720 E0.75
G1 F2400 X160.02 Y156.838 E0.147
G1 X159.914 Y156.974 E0.01936
...
\end{lstlisting}
\vspace{-10pt}
\end{wrapfigure}

\begin{figure}
\includegraphics[width=\linewidth]{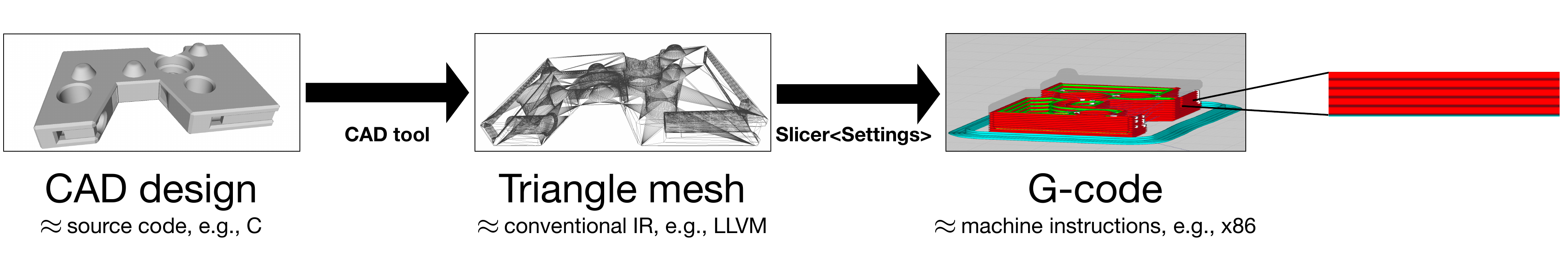}
\caption{Left to right: high-level CAD design, triangle mesh, a preview of the generated \gcode for the mechanical part shown in
  \autoref{fig:nice-ercf-photo}, and a
  zoomed in view of the cuboids
  represented by each \gcode instruction which forms the basis of our semantics. Below each representation, we provide an analogy from a traditional compiler pipeline.
The extruder of a 3D printer deposits filament along
  the horizontal lines in each layer of the \gcode
  to manufacture the object.}
\label{fig:rec-rx15}
\label{fig:bg}
\end{figure}

\noindent \autoref{fig:bg} shows three different representations of the
mechanical part in \autoref{fig:nice-ercf-photo} alongside analogies to a traditional compiler pipeline.
The high-level CAD design
  must first be converted into a 
  triangle mesh representation from which
  a slicer (e.g., Cura~\cite{cura}) generates
  \gcode, a preview of which is shown to the right
  in \autoref{fig:bg} (including a close up view of the extruded lines in the preview).
A slicer must first be configured with a variety of
  settings before it can generate \gcode~\cite{nadyachi2025} with the most important one
  being the model of the 3D printer
  for which the \gcode is to be generated.
This, in turn, dictates additional settings like
  the diameter, $d$,
  of the nozzle on the printer's extruder,
  from which material is deposited.
This work focuses on \gcode representing linear motion, 
  generated through the commonly used approach of
  \textit{uniform, planar slicing}~\cite{3dp-overview, DOLENC1994119}.
For a model of height $H$ and for a uniform layer height $h$,
  this method
  generates $N$ slices where $N = \lceil H / h \rceil$.
The layer height $h$ is another parameter that is set to a fixed value in the slicer
  before slicing a model.
These slices are parallel to a horizontal ``print bed'' as shown in \autoref{fig:bg}~(right) ---
  the extruder of a 3D printer
  deposits filament along the 2D lines in each layer.
  
The snippet above shows examples of
  \gcode instructions from the
  preview shown in \autoref{fig:bg}~(right).\footnote{The
  full \gcode program contains other non-motion commands for
  setting temperature, selecting from among multiple extruders, and controlling peripherals such as lights.}
A single straight line may be broken down into multiple instructions depending on the specific
  slicing algorithm used in a slicer. 
\lstinline[language=gcode]!G0! represents movements that do not extrude filament,
  whereas \lstinline[language=gcode]!G1!  represents extruding movements.
In the instruction,
\lstinline[language=gcode]!G0  F18000 X150.411 Y157.446 Z2.3!,
  \lstinline[language=gcode]!F! represents the ``feed rate'' which indicates how fast the extruder must move (18000mm/min),
  and \lstinline[language=gcode]!X!, \lstinline[language=gcode]!Y!, and \lstinline[language=gcode]!Z!
  indicate the absolute (or relative\footnote{\gcode commands \texttt{G90} and \texttt{G91} switch between absolute and relative positioning respectively})
  position of the extruder after this instruction is executed.
For example, in the absolute setting,
  after executing this instruction the extruder would end up at the 3D coordinate:
  $(150.411,~ 157.446,~ 2.3)$.

When an argument is omitted, its last value is used.
Therefore, in the next instruction in the snippet,
  \lstinline[language=gcode]!G0  X159.094 Y155.912!,
  the extruder moves to the coordinate $(159.094,~ 155.912,~ 2.3)$ at the same speed
  as the previous instruction.
Here, the \lstinline[language=gcode]!Z! coordinate also remains the same.
The difference between any two consecutive distinct values of
  \lstinline[language=gcode]!Z! should always be $h$, the
  layer height chosen for this model (in this case, 0.2mm).
The \lstinline[language=gcode]!E! in the instruction
  \lstinline[language=gcode]!G1  F2400 X160.02 Y156.838 E0.147!
  states that 0.147mm of filament is extruded during this movement. 
Commands like
\lstinline[language=gcode]!G0! and \lstinline[language=gcode]!G1! are similar to
assembly instructions and intended to run on CNC machines.
\gcode programs are \textit{stateful}: both instructions update the
  position of the toolhead (e.g., extruder of a 3D printer) and
  \lstinline[language=gcode]!G1! additionally
  also adds filament either
  directly on the print bed or on top of previous layers.
This observation guides how we define the state of a \gcode
program in \autoref{sec:line-rec1}.
  
\section{Formalizing Linear Motion \gcode Program}
\label{sec:line-rec}

Our key idea is that by analyzing \gcode programs,
  (1)~we can check invariants of models and localize problematic parts, and
  (2)~compare two \gcode programs
  for differentially testing slicers and mesh repair tools.
To analyze a \gcode program comprised only of linear motion,
  we lift the program to a set of cuboids and
  approximate it as a point cloud.
To that end, we first provide formal semantics for \gcode.

\subsection{Semantics of \prog that Guides Cuboid Reconstruction}
\label{sec:line-rec1}
We define \prog to be
  a \gcode program that is parametrized by
  $d$, the extruder nozzle diameter on the 3D printer for which
  some slicer generated the \gcode, and
  $h$ the fixed layer height used when configuring the slicer.
\autoref{fig:syntax} shows the formal syntax
  of \prog, specifically the
  subset of \gcode this paper targets:
  linear motion \gcode.
In \autoref{fig:syntax}, \textsf{pos} represents the
  position arguments of \lstinline[language=gcode]!G0!
  and \lstinline[language=gcode]!G1! and \textsf{attribute}
  is used for arguments \lstinline[language=gcode]!E! and \lstinline[language=gcode]!F!.
A \prog can be \textsf{empty} or a list of commands
  (we model linear motion commands \lstinline[language=gcode]!G0! and \lstinline[language=gcode]!G1!).

\renewcommand{\syntleft}{}
\renewcommand{\syntright}{}

\begin{figure}
\sf
\begin{minipage}{0.8\linewidth}
\setlength{\grammarparsep}{5pt plus 1pt minus 1pt}
\setlength{\grammarindent}{5em}
\begin{grammar}

<pos> ::= (\ensuremath{\mathbb{R}},~ \ensuremath{\mathbb{R}},~ \ensuremath{\mathbb{R}}) \quad \quad
len ::= {\ensuremath{\mathbb{R}}} \quad \quad
rate ::= {\ensuremath{\mathbb{R}}} \quad \quad
attribute ::= "E" len | "F" rate 

<cmd> ::= G0 (pos, attribute*) | G1 (pos, attribute*) \quad \quad

<\prog> ::= empty | cmd :: \prog

\end{grammar}
\end{minipage}
\caption{Syntax of the subset of \gcode this paper targets: linear motion generated by 3D printing slicers.
We define \prog to be a \gcode program whose layer height set to $h$ and
  which is generated for a 3D printer whose extruder nozzle has diameter $d$.}
\label{fig:syntax}
\end{figure}

These commands (\textsf{cmd}) are instructions for an extruder
  on a 3D printer to deposit filament
  in a 2D horizontal layer such that
  the resulting material (e.g., plastic) deposition constructs the 3D model.
The extruder has a heating element that melts the material which
  comes out from a nozzle which typically has a round hole.
If the nozzle was to extrude filament in free space,
  the filament would therefore form a \textit{cylindrical} shape along a line segment (similar to toothpaste out of a tube in air).
However, typically the nozzle extrudes on a
  horizontal print bed (subsequent layers are extruded on top of the previous layer).
The extruded cylinder will thus be ``squished'' between the nozzle and the print bed (or previous layer)
  such that the top and bottom faces of the extruded material are parallel to each other. 
\autoref{fig:bg} (right) shows each extruded line up close for visual understanding.
Thus, the horizontal cross-section of an extruded line of material is a flat rectangle with round corners.

\begin{figure}
\begin{subfigure}{0.35\textwidth}
    \includegraphics[width=\linewidth]{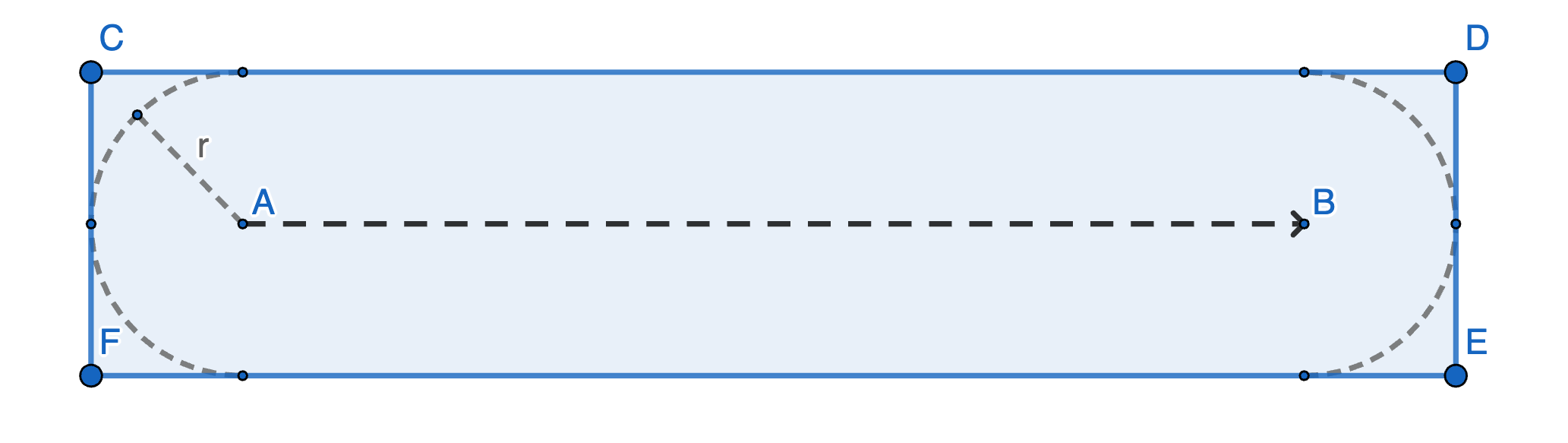}
\end{subfigure}
\hspace{40pt}
\begin{subfigure}{0.35\textwidth}
\centering
    \includegraphics[width=0.8\linewidth]{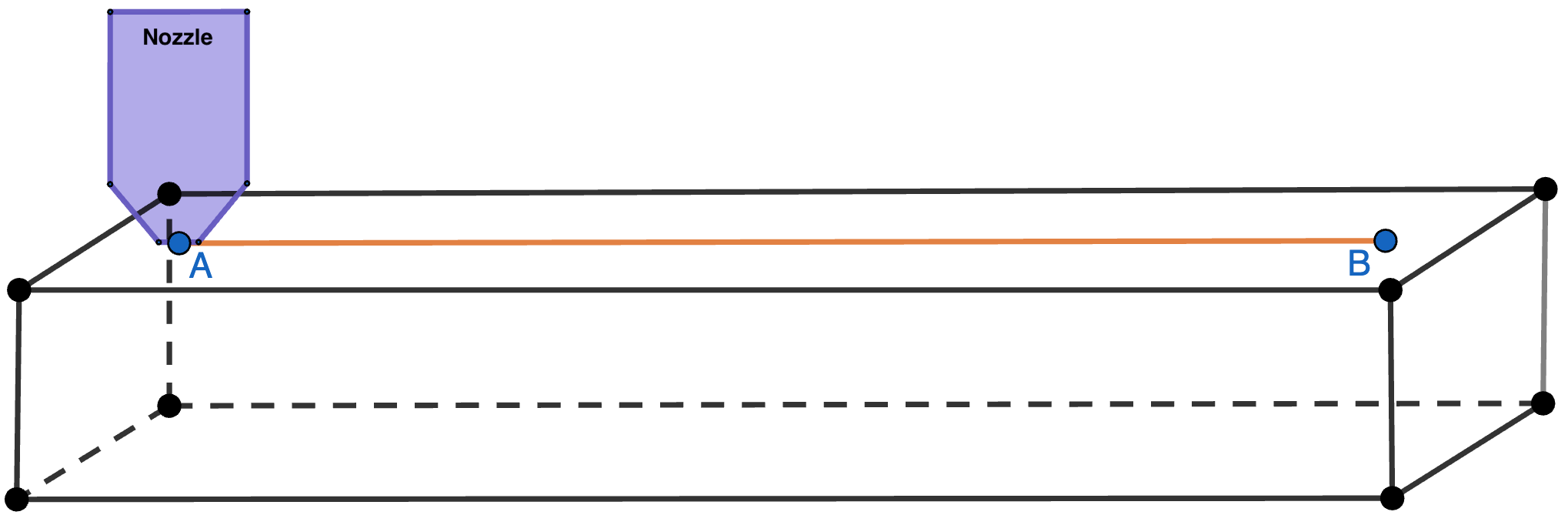}
\end{subfigure}
\caption{(Left) Top view of the cuboid shown to the right.
    $r = d/2$ is the radius of the nozzle on an extruder
    that is moving from point $A$ to $B$.
    The outer rectangle $CDEF$ approximates
    a rectangle of length $AB$ together with two
    semicircles protruding from either end.
    Intuitively, imagine these rounded ends are
    made by a circle of radius $r$
    centered at $A$ that is dragged to the point $B$.
    The dashed arrow depicts the trajectory of the nozzle's center --- it moves from left to right.
    (Right) Reconstructed cuboid
    that denotes the line AB extruded by the purple nozzle whose diameter is $d$.}
\label{fig:cubAB}
  \label{fig:coord}
\end{figure}

\begin{figure}
\sf
\[\begin{array}{lll}
x,~ y,~ z \in \mathbb{R} & \mathsf{vertex} ::= (x,~ y,~ z) & \mathsf{origin} ::= (0, 0, 0) \\ \\

\mathsf{face} ::=  (\mathsf{vertex},~ \mathsf{vertex}, \mathsf{vertex},~ \mathsf{vertex})  &  \mathsf{cuboid} ::= (\mathsf{face}$\ensuremath{_t}$,~ \mathsf{face}$\ensuremath{_b}$) &  \\ \\
\mathsf{state} ~\sigma ::= (p{_e},~ \{~ \mathsf{c} ~|~ \mathsf{c} ~is~ ~a~ \mathsf{cuboid} ~\}) & \quad p_e ::= \mathsf{vertex} &

\end{array}\]

\newcommand\semrulesep{7em}
\begin{equation*}
\begin{array}{cc}
    \inference{\sigma;~ s ~\Downarrow_{d,h}~ \sigma' \quad \quad \quad \sigma';~ ss ~\Downarrow^p_{d,h}~ \sigma''}{\sigma;~ s ~::~ ss ~\Downarrow^p_{d,h}~  \sigma''}[\prog]
     & 
     \inference{}{(p_e,~ cs);~ \gzero(p,~ \_) ~\Downarrow_{d,h}~ (p,~ cs) }[G0]
\end{array}
\end{equation*}

\begin{equation*}
\inference{cs' = cs ~\cup~ \{(f_t,~ f_b)\}}{(p_e,~ cs);~ \gone(p,~ \_)  ~\Downarrow_{d,h}~ (p,~ cs') }[G1]  \quad \text{where}
\end{equation*}

\[\begin{array}{rcl}
    f_b &=&
       (p_e  ~+_v~  (-k\cdot\cos(-\theta+\tfrac{\pi}{4}),~  k\cdot\sin(-\theta+\tfrac{\pi}{4}),~ {p_e}_z), \\
       && \;\;p \;\;+_v~  ( k\cdot\cos(\theta+\tfrac{\pi}{4}),~   k\cdot\sin(\theta+\tfrac{\pi}{4}),~ {p_e}_z),  \\
       && \;\;p \;\;+_v~  ( k\cdot\cos(-\theta+\tfrac{\pi}{4}),~ -k\cdot\sin(-\theta+\tfrac{\pi}{4}),~ {p_e}_z), \\
       && \;\;p_e  ~+_v~ (-k\cdot\cos(\theta+\tfrac{\pi}{4}),~  -k\cdot\sin(\theta+\tfrac{\pi}{4}),~ {p_e}_z)),
       \quad k = \sqrt{2} (d/2), \quad \theta = \tan^{-1}(\mathit{slope}(\overrightarrow{p_ep})) \\
      \\
    f_t &=& \mathsf{map}(\lambda v.~ (v ~+_v~ (0,~ 0,~ h)),~ f_b) 
    
\end{array}\]

\caption{Big step operational semantics of \prog.
From \autoref{fig:coord} (Left), $A$ is $p_e$, $B$ is $p$,
$CDEF$ is $f_t$, and $r = d/2$.
We define $+_v$ as the addition operator over two 3D vertices.
As is standard, \textsf{map} applies its first argument, a function, to all vertices in $f_b$.
The $z$ component of $p_e$ is shown by ${p_e}_z$.}
\label{fig:semantics}
\end{figure}

\autoref{fig:coord}~(left) demonstrates this visually --- 
  $AB$ represents the top view of an extruded line
  between two consecutive \gcode instructions.
The rounded ends can be imagined as being constructed by
  dragging a circle of radius $r$ centered at $A$ to the point $B$,
  where $r = d / 2$ is the radius of the nozzle.
We ignore the rounded corners and
  approximate the line segments with the outer cuboids (top view shown by $CDEF$ in Figure~\ref{fig:coord}, left)
   whose height is set to be the uniform layer height $h$ in the slicer
   when generating the G-code for the model.
\autoref{fig:cubAB}~(right) shows the 3D view of the reconstructed cuboid whose top \textit{face} is $CDEF$.

\autoref{fig:semantics} shows the big step operational semantics of \prog.
We define the state $\sigma$ of a \prog program to be a tuple whose first element,
  $p_e$, is the current position of the extruder nozzle in 3D space. 
We assume that initially it is at \textsf{origin} (see top of \autoref{fig:semantics}).
The second element in the tuple is a set of cuboids that
  denotes the 3D object being constructed by \prog.
We represent a \textsf{cuboid} with its top face ($f_t$) and bottom face ($f_b$)
  where a \textsf{face} is a 4-tuple made from \textsf{vertex} as shown.
  
We define two judgments: $\Downarrow_{d, h}$ and $\Downarrow_{d, h}^p$.
The first, $\Downarrow_{d, h}$, evaluates a 
  \textsf{cmd} (as shown in \autoref{fig:syntax}) to produce a new state.
The attributes do not affect the semantics,
  as shown by $\_$ in the inference rules for \lstinline[language=gcode]!G0! and \lstinline[language=gcode]!G1!.
In the case of \lstinline[language=gcode]!G0!,
  no new \textsf{cuboid} is added in the new state: it remains $cs$ which was the set of cuboids in the previous state.
In the case of \lstinline[language=gcode]!G1!,
  a new \textsf{cuboid} is added to $cs$ shown by $cs ~\cup~ \{(f_t, f_b)\}$.
The vertices for $f_b$ are computed as shown in \autoref{fig:semantics}
  using the rules of trigonometry.
The variable $k$ in \autoref{fig:semantics} represents the distance from a
  line's endpoint to its nearest cuboid vertex --- in \autoref{fig:coord}~(left),
  $k$ is the length of $AC$, $AF$, $BD$, and $BE$.
  
We use $+_v$ to indicate addition of two 3D vertices: $v_1 +_v  v_2 = ({v_1}_x + {v_2}_x, {v_1}_y +  {v_2}_y, {v_1}_z + {v_2}_z)$.
Here, $\theta$ is the gradient (computed using the helper function \textit{slope} in \autoref{fig:semantics})
  of the line joining $p_e$ and $p$, shown by $\overrightarrow{p_ep}$, i.e., 
    $\theta$ is the angle between the $x$-axis
    and $\overrightarrow{p_ep}$.  

Since the top and bottom face are parallel,
the coordinates of the vertices of $f_t$
only differ from those of $f_b$
  in the $z$ coordinate (as we denote each instruction as a cuboid),
  which is offset by the layer height, $h$.
For both \lstinline[language=gcode]!G0! and \lstinline[language=gcode]!G1!, 
  $p_e$ is updated to $p$ which is the position argument.
This indicates that after this command is executed,
  the new position of the nozzle is $p$.

The judgment $\Downarrow_{d, h}^p$ evaluates a \prog program
  as shown in \autoref{fig:semantics} by folding
  over all the commands in the program.
The rule states that at state $\sigma$,
  if the command $s$ evaluates to produce the $\sigma'$
  following the rules of $\Downarrow_{d, h}$, and
  at state $\sigma'$, the rest of program, $ss$ evaluates to $\sigma''$,
  then the entire program \prog evaluates to generate the final state $\sigma''$.

This explains how, given a \prog program,
  \tool reconstructs
  a set of cuboids that denotes the 3D model that would
  be printed upon executing \prog on a 3D printer (\autoref{fig:glitchflow}, left).

\subsection{Approximating \prog as a Point Cloud to Facilitate Comparison}
\label{subsec:cloud}
\begin{algorithm}
\caption{Point Cloud Comparison Algorithm}\label{alg:comp}
\sf \footnotesize
\begin{algorithmic}
\Procedure{point_cloud_segmentation} {pc1, pc2, ubox}
\State bbox $\gets$ getBbox(pc1, pc2)
\State n $\gets$ countUnitBox (ubox, bbox)
\State boxed\_point\_sets $\gets$ [ [ [], [] ] $\times$ n]
\ForAll {p $\in$ P | P $\in$ \{pc1, pc2\}}
    \State offset $\gets$ findBoxIndex (p, ubox, bbox)
    \State boxed\_point\_sets [offset][i].append (p)
    \Comment{i $\in$ \{0, 1\} depends on if p $\in$ pc1 or pc2}
\EndFor
\State \textbf{return} (boxed\_point\_sets, n)
\EndProcedure
\end{algorithmic}
\begin{algorithmic}
\Procedure{compare_point_clouds}{boxed\_point\_sets, n}
    \State hd\_list $\gets$ []
    \ForAll{i $\in$ range(n)}
        \State P_1 $\gets$ boxed\_point\_sets [i][0]
        \State P_2 $\gets$ boxed\_point\_sets [i][1]
        \State P_1' $\gets$ [] \Comment{P_1's neighborhood}
        \State P_2' $\gets$ [] \Comment{P_2's neighborhood}
        \ForAll{j $\in$ {neighboringBox} (i)}
            \State P_1'.extend(boxed\_point\_sets[j][0])
            \State P_2'.extend(boxed\_point\_sets[j][1])
        \EndFor
        \State hd\_list.append(max (oneWayHD(P1, P2'), oneWayHD(P2, P1')))
    \EndFor
    \State \textbf{return} hd\_list
\EndProcedure
\end{algorithmic}
\label{alg:compalg}
\end{algorithm}

Once a 3D model has been reconstructed from a \prog program as described in \autoref{sec:line-rec1},
 it can be used for many downstream applications. This work demonstrates two:
  (1)~invariant checking of 3D models, and
  (2)~differential testing of mesh repair tools and slicers.
Our key observation
  is that \textit{both} tasks can be
  performed by \textit{comparing} two \prog programs. 
This comparison process is non-trivial and naive approaches   
  suffer from
  poor scalability,
  numerical errors,
  imprecise localization of differences, and
  hard-to-interpret results. 
This section and the ones that follow 
  tackle each of these problems in turn.

A naive approach for comparing two \prog programs, \textsf{p1} and \textsf{p2}, 
   once a 3D model is reconstructed from \prog as a set of cuboids,
   is to use set-theoretic constructive solid geometry operations
  like \textit{difference}~\cite{openscad, reincarnate, tsoutsou-17} that are available in CAD tools: 
  \textsf{difference (union (c11, c12, ..., c1n), union (c21, c22, ..., c2m))}, where
  \textsf{union (c11, c12, ..., c1n)} is the
  set of cuboids for \textsf{p1}, and 
  \textsf{union (c21, c22, ..., c2m)} is the set of cuboids for \textsf{p2}.
If the result is $\emptyset$, that means the two programs
   \textit{denote} the same 3D solid.
A nonempty result indicates differences.
However, early experimentation showed that
  for large models this fails to scale.
  
Therefore, in \tool, we present a more practical
  algorithm for comparing two \prog programs.
Our algorithm discretizes the cuboids into a set of points
  yielding a \textit{point cloud}
  for the reconstructed model (\autoref{fig:glitchflow}, left).

To generate a point cloud to approximate
  the reconstructed cuboid set,
  \tool samples the cuboids.
A straightforward strategy is to use uniform sampling where
  the same number of points are sampled from each cuboid.
However, this would oversample short cuboids and undersample long cuboids.
We therefore define
$$\Omega_g : \mathsf{cuboid} \rightarrow \mathsf{points}$$
  as a function that \textit{proportionally samples}
  different numbers of points from differently sized cuboids,
  depending on its dimensions.
We use the parameter $g$ to indicate the \textit{sampling gap},
  which represents the distance between any pair of neighboring points.
$\Omega_g$ then samples
$$\left(\lceil (l+d) / g\rceil + 1\right)\cdot\left(\lceil d / g\rceil + 1\right)\cdot
\left(\lceil h / g\rceil + 1\right)$$
points for a cuboid with length $l + d$, where $d$ is the nozzle diameter,
  $l$ is the length of the extruded line, and
  $h$ is the layer height. 
From \autoref{fig:coord}, the dimensions of the reconstructed cuboid are
  length: $(l + d)$, breadth: $d$, and height: $h$.
This method produces point clouds that are usually smaller
  than what uniform sampling from the cuboids would produce,
  speeding up point cloud comparison.
To summarize, if $G$ is a \prog program and $G ~\Downarrow_{d, h}^p~ (\_,~ C)$,
  then $\mathsf{flatmap}(\Omega_g,~ C)$
  is the point cloud \tool generates for $G$.
Next, we show how \tool compares two point clouds using an \textit{augmented} Hausdorff distance metric.

\section{Comparing Point Clouds}
\label{sec:comp}
\label{sec:hd}

To compare two point clouds %
  obtained as described in Section~\ref{subsec:cloud},
  we treat each point cloud as a \textit{set} of points and
  introduce an augmented version of Hausdorff distance~\cite{hausdorff1914grundzuge}.
  
The Hausdorff distance is widely used for evaluating global similarity by calculating the maximum of nearest-neighbor distances between two point clouds~\cite{hd1, hd2, hd3}.  
However, because it yields a single metric for the entire comparison,  we cannot  localize faults to regions of the point cloud.
To address this, 
  \tool segments each point cloud into multiple \textit{unit boxes} and
  performs a finer-grained analysis by comparing the subset of points within the corresponding unit boxes.
This segmentation, done naively, causes wild swings in the Hausdorff metric as points on the boundaries of partitions get ascribed to different unit boxes due to floating-point errors.
To counter this, we introduce a novel \textit{augmented} Hausdorff metric that 
  compensates for points displaced into adjacent unit boxes due to numerical errors.

\subsection{Overall Approach}
\label{subsubsec:overall}

Our approach consists of two main steps:
(1) point cloud segmentation followed by
(2) point cloud comparison by Hausdorff distance computation within each segment.
To perform point cloud segmentation, 
  \tool first creates a single bounding box that encompasses the larger of the  
  two point clouds to be compared (\textsf{getBbox}) if the extents of the point clouds are different. 
The bounding box is then uniformly divided into multiple smaller unit boxes.
Users can specify the dimensions of a unit box,
  which typically depend on the scale of the smallest feature differences users aim to capture.
We then place each point in each point cloud into a unit box depending on its spatial position.
This procedure is shown in \textsc{\footnotesize{\textsf{POINT\_CLOUD\_SEGMENTATION}}} in ~\autoref{alg:comp}. 
The inputs to the procedure are the point clouds \textsf{pc1} and \textsf{pc2},
  and dimensions of the unit box, \textsf{ubox}.
The function \textsf{countUnitBox} computes the total number of
  unit boxes based on the dimensions of the bounding box
  (\textsf{bbox}) and \textsf{ubox}.
Then, the function \textsf{findBoxIndex} finds the closest unit box
  in which to place a point.

\begin{figure}
    \includegraphics[width=0.7\textwidth]{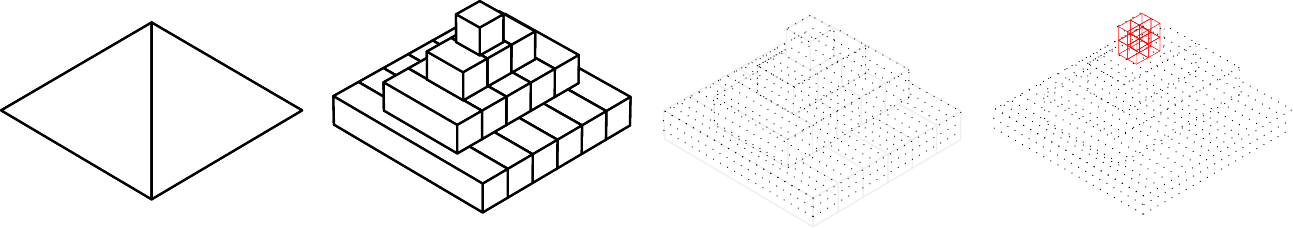}
    \caption{\label{fig:cuboid-cloud-ubox} Artifacts at different stages of processing: the input model whose \gcode we are lifting, the reconstructed cuboid set, the approximated point cloud, and a subset of unit boxes overlaid on the point cloud.}
\end{figure}

Now we can compare two point clouds by computing
  the distance between corresponding subsets of points in each unit box.
Figure~\ref{fig:cuboid-cloud-ubox} visualizes the inputs and outputs of the different stages of reconstruction and comparison. 
  
In our context, 
  the distance can be a numerical value, \texttt{none}, or $\infty$,
  depending on the number of points in the unit box, $u$.
When both point clouds contain points in $u$, 
  the distance is a numerical value. 
If $u$ is empty for both point clouds
 (i.e., neither point cloud has points in $u$),
  we define the the distance to be \texttt{none}.
The $\infty$ case occurs when only one point cloud has points in $u$, 
  while the other does not. 
However, we observe that many such cases 
  arise as side effects of discretization and quantization~\cite{curvislicer, pravnovich}
  or floating-point errors. 

To avoid labeling these cases as having $\infty$ distance
  and to better highlight the primary differences in model features, 
  we adopt the following approach: 
  for the point cloud that has no points in $u$, 
  we construct a \textit{neighborhood} of points centered around $u$. 
This neighborhood includes points from all adjacent boxes, i.e., 
  boxes that either share a face, vertex, or edge with $u$.
Including $u$, there are 27 such boxes.
We then compute the Hausdorff distance 
  from this neighborhood 
  to the subset of points in $u$
  from the other point cloud.
This is explained in detail in the following section (\autoref{subsubsec:fperror}).

\subsection{Handling Floating-point Error} 
\label{subsubsec:fperror}

\begin{figure}
\begin{subfigure}[t]{0.4\linewidth}
    \includegraphics[width=\linewidth]{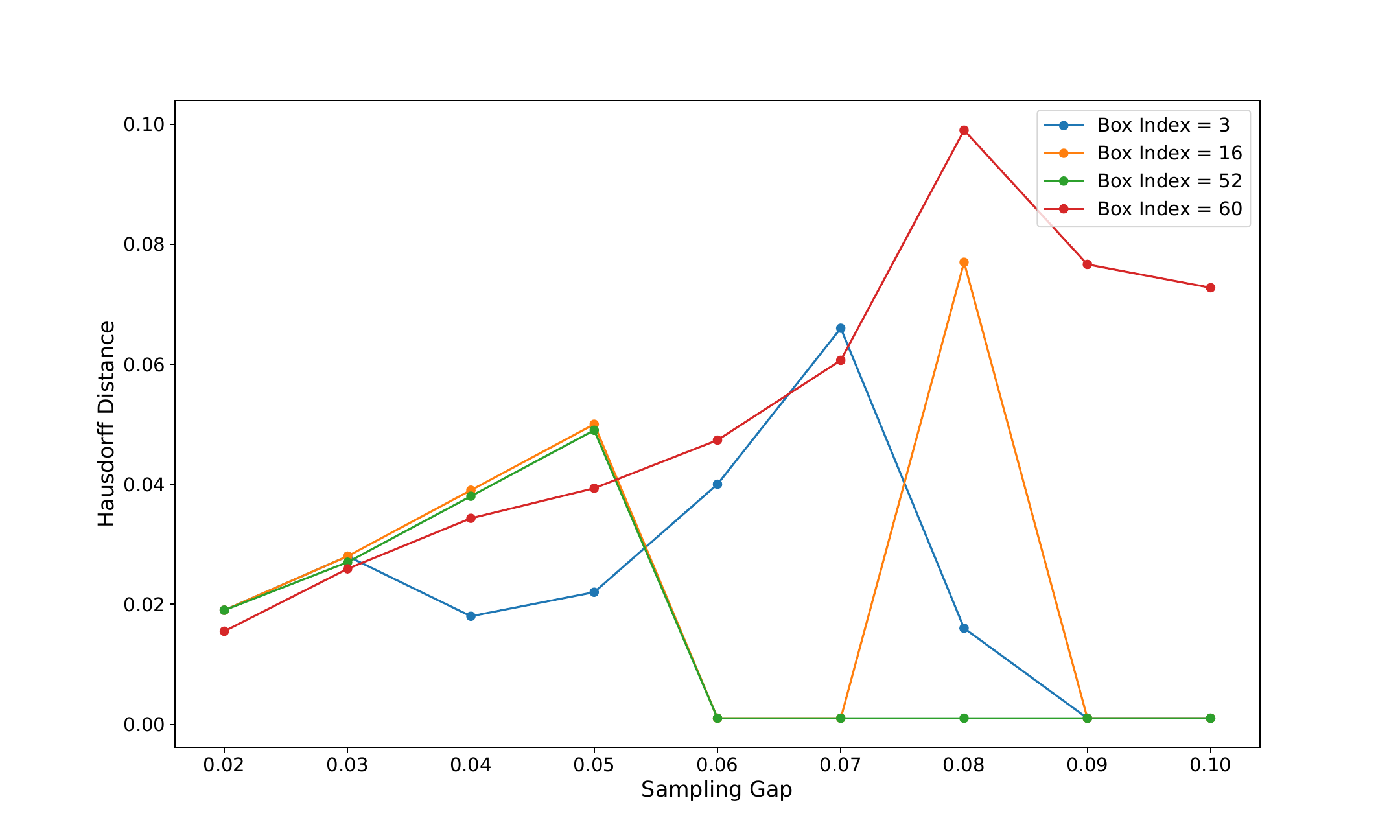} %
    \caption{\label{fig:non-monotonicity} Hausdorff distance for points in four random unit boxes under varying sampling gaps.
    As the sampling gap changes, the Hausdorff distance within all boxes fluctuates, showing a 10$\times$ change in value.}
\end{subfigure}%
\hfil
\begin{subfigure}[t]{0.4\linewidth}
    \includegraphics[width=\linewidth]{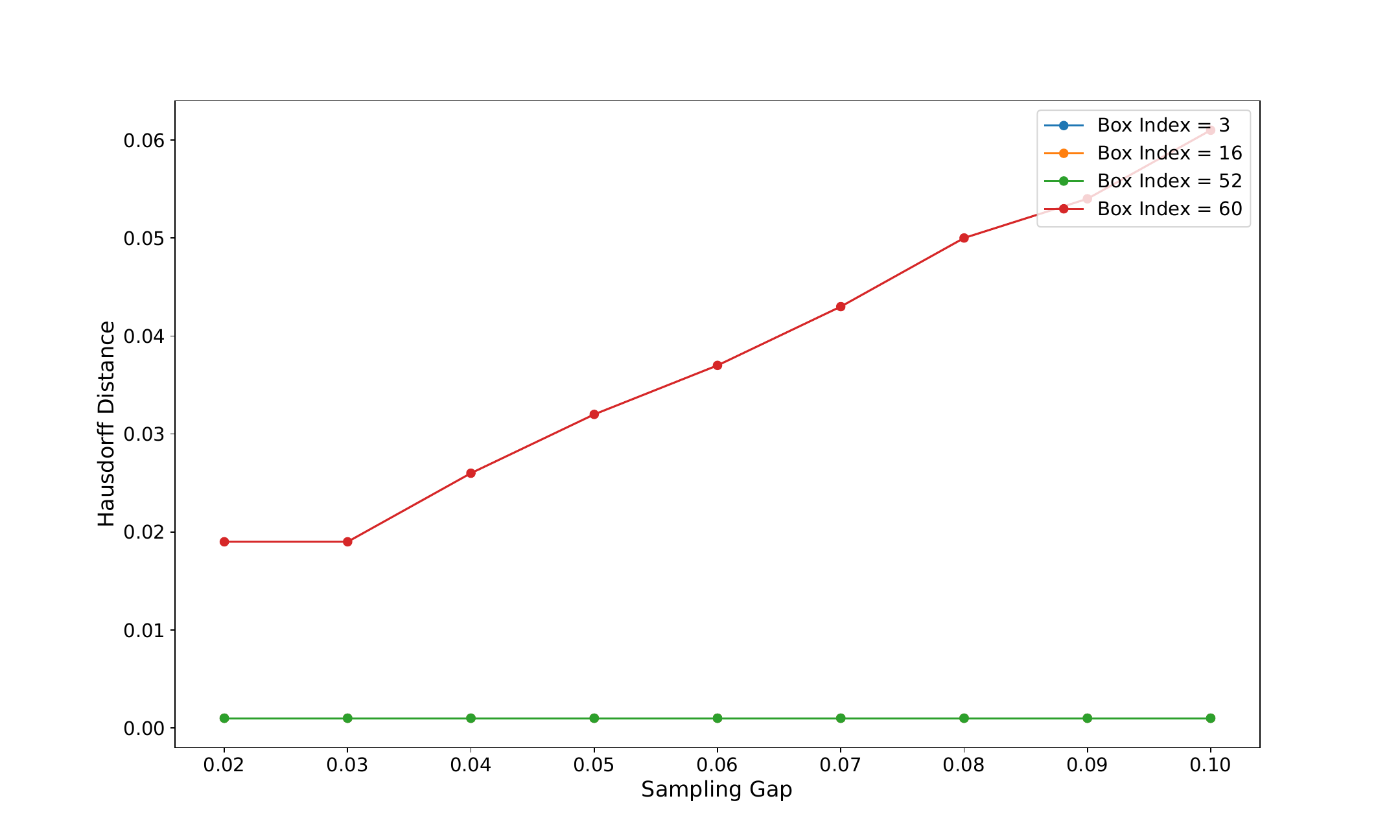} %
    \caption{\label{fig:monotonicity} Hausdorff distance within the same four boxes in ~\ref{fig:non-monotonicity},
    obtained by using Equation~\ref{eq:new-hd}.
    The Hausdorff distance now exhibits a monotonic, non-increasing trend as the sampling gap decreases.
    }
\end{subfigure}
\caption{Hausdorff distance
($y$-axis) vs sampling gap
($x$-axis) with and without floating point error handling.}
\end{figure}

When comparing two very similar point clouds,
  one would expect the Hausdorff distances to be relatively small
  if there is no significant difference in model features. 
Additionally, the distance should decrease as the sampling gap becomes smaller, 
  since the sampled point cloud more closely resembles the original model.
However, this expected behavior is not observed 
  when the standard Hausdorff distance metric is applied to points within each unit box.
To understand this phenomenon better,
  we compared two point clouds,
  both reconstructed from the \prog program
  of a simple rectangular prism model.
One was the point cloud of the original model, and the other was the point cloud of a rotated version of the model which was rotated back after reconstruction.
Ideally, these should be the same point cloud.
  
Figure~\ref{fig:non-monotonicity} shows the Hausdorff distance between points in the two models within four
  randomly selected, nonempty unit boxes.
At the smallest sampling gap (0.02),
  the Hausdorff distance is 10$\times$ higher than the distance at the largest sampling gap (0.10). %
We discovered that 
  these non-monotonic fluctuations are caused by floating-point errors 
  during point cloud segmentation. 
When \textsf{findBoxIndex} determines the unit box a point belongs to,
  imprecise floating-point arithmetic can result in a point from box \( A \) 
  being incorrectly assigned to a different box \( B \) in the other model. 
Since Hausdorff distance~\cite{hausdorff, hausdorff1914grundzuge}
  takes the maximum distance from any point in one point set to the other point set, 
  it is highly sensitive to 
  inaccuracies arising from missing points in one of the sets.
Therefore, 
  we develop the following technique to mitigate the effect of
  floating-point errors in \textsc{\footnotesize{\textsf{POINT\_CLOUD\_SEGMENTATION}}}.

We present a new augmented Hausdorff distance formula
  as follows:
  the distance between two point sets $X$ and $Y$ in a unit box $B$ is defined as
\begin{equation}
\label{eq:new-hd}
    d'_H(X,~ Y) ~=~ \max\{\sup_{x\in X}~ d(x,~ Y'),~ \sup_{y\in Y}~ d(X',~ y)\}
\end{equation}
where $d$ denotes the standard Hausdorff distance formula between a point and a set of points.
$X'$ and $Y'$ are the respective \textit{neighborhoods} of $X$ and $Y$ (defined in \autoref{subsubsec:overall}),
  where a neighborhood of a point set within a unit box $B$
  contains points from all neighboring boxes of $B$,
  including points from $B$ itself.
This definition of $d'_H(X,~ Y)$ ensures that, 
  even if a point $x$ in $X$
  cannot find its closest point $y$ in $Y$
  due to misclassification of $y$, 
  $x$ can still match with $y$ 
  in $Y$'s neighborhood $Y'$.
This procedure is shown in \textsc{\footnotesize{\textsf{COMPARE\_POINT\_CLOUDS}}} in ~\autoref{alg:comp}.
It takes the segmented point clouds \textsf{boxed\_point\_sets} and number of unit boxes \textsf{n} as input
  and returns a list of augmented Hausdorff distances, \textsf{hd_list}.
In the algorithm, \textsf{oneWayHD(X,Y)} is equal to ${\sup_{{x\in X}}}~ d(x,~ Y)$. 
The Hausdorff distance computed with Equation~\ref{eq:new-hd} 
  now exhibits the expected monotonic behavior (\autoref{fig:monotonicity}).

\section{Visualizing True Differences}
\label{sec:visual}

Reconstructing point clouds from two \prog programs enables
  invariant checking and differential testing of fabrication tools by allowing \prog comparison.
So far, we have described how \tool performs this reconstruction and comparison.
We now discuss how the results are presented to the user.
We visualize distances per unit box as a heatmap.
However, mapping magnitudes linearly to colors is ineffective because, alongside the \textit{true} differences caused by faulty models or slicer/mesh tool variations,
there are also \textit{unwanted} errors from discretization and quantization~\cite{cura, pravnovich}
  that are either inherent to slicing,
  and can be introduced during point cloud reconstruction,
  as well as floating-point errors inevitable in finite-precision arithmetic.
Ideally, \tool should highlight only true differences while filtering out these inherent artifacts.

\subsection{Spatial Averaging to Reduce Unwanted Errors}
\label{subsec:spatialavg}
On examining the distribution of distances computed using ~\autoref{alg:comp}, 
  we observed that 
  discretization and quantization errors are typically confined to specific regions of the model,
  such as boundaries and areas connecting the inner wall and infill, 
  while floating-point errors appear relatively randomly.
Therefore, 
  \tool averages distances spatially 
  by computing the mean distance within each unit box and its immediate neighborhood.
Specifically, 
  the distance for unit box $B$ is now calculated as
  $\left[\sum_{i=1}^{n} d_i\right]/n$,
  where $n$ is the number of neighboring boxes of $B$ 
  that contain a numerical distance value. 
We preserve all distances that are \texttt{none} or $\infty$ without modification.

While this averages out the effect of random unwanted errors in the heatmap, 
  it can also reduce the distances representing true differences. 
To address this, 
  using box size 
  smaller than the size of the feature difference that user wants to detect 
  is preferable. 
As demonstrated in our
  evaluation
  (\autoref{tab:model-param1}, \autoref{tab:model-param2} in \autoref{sec:appexdixparams}), 
  with sufficiently small unit box sizes, 
  the heatmaps consistently highlight areas with true differences.

\subsection{Choosing Colors for Visualization}

Heatmap colors are drawn from a continuous spectrum. Points in each nonempty unit box share one color.
Nonempty boxes that have $\infty$ distance are
  assigned the darkest possible color
  in the spectrum.
Otherwise, 
  users set a \textit{threshold percentile} parameter 
  to determine the cut-off value for
  distinguishing distances representing true differences 
  from those representing unwanted differences.
For example,
  a 90$^{th}$ threshold percentile means that 
  all distances smaller than the 90$^{th}$ percentile of the distances
  are considered difference-free 
  and are assigned the lightest possible color.
This is used to 
  highlight true differences 
  by filtering out other unwanted errors.
We devised this strategy
  after examining the distribution of true errors in detail,
  as described in Section~\ref{subsubsec:dist}. 
For distances between the defined threshold and $\infty$, 
  we normalize their difference relative to the threshold over the entire distance range 
  and assign colors based on the normalized distances.

\subsection{Analyzing Trends in Distance Distributions}
\label{subsubsec:dist}
\begin{figure}
\begin{subfigure}{0.25\linewidth}
\centering
\includegraphics[scale=0.18]{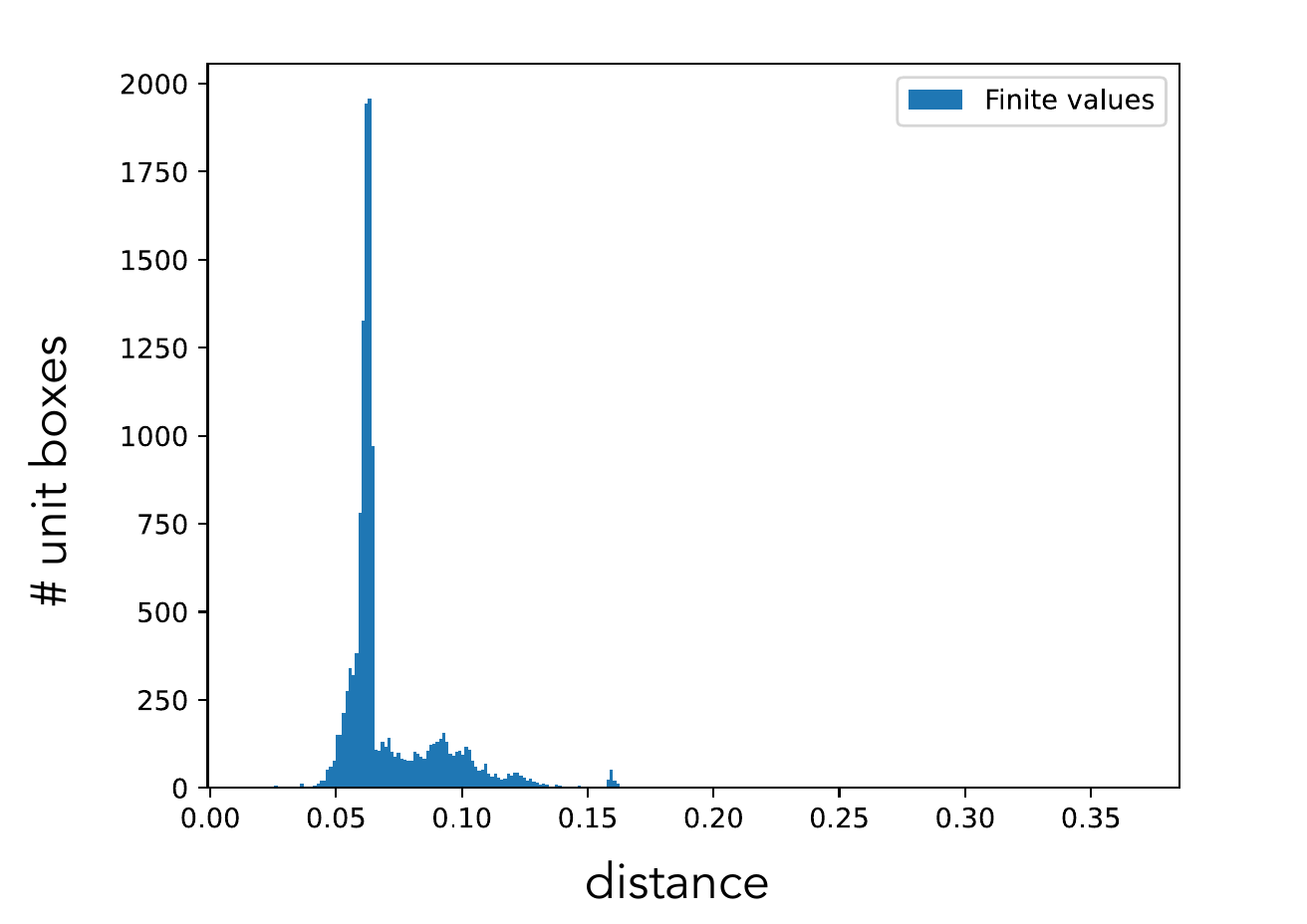}
\caption{Orig Distribution}
\end{subfigure}
\begin{subfigure}{0.25\linewidth}
\centering
\includegraphics[scale=0.05]{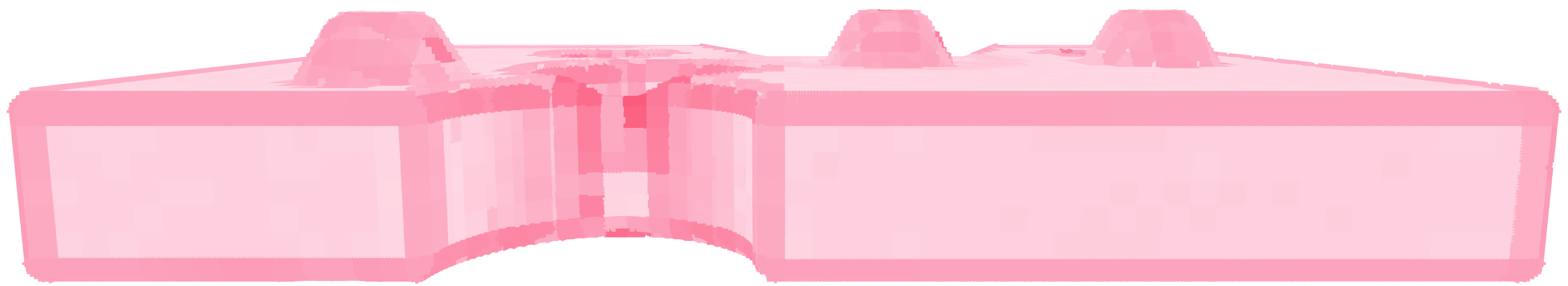}
\vspace{1em}
\caption{Heatmap}
\end{subfigure}
\begin{subfigure}{0.25\linewidth}
\centering
\includegraphics[scale=0.05]{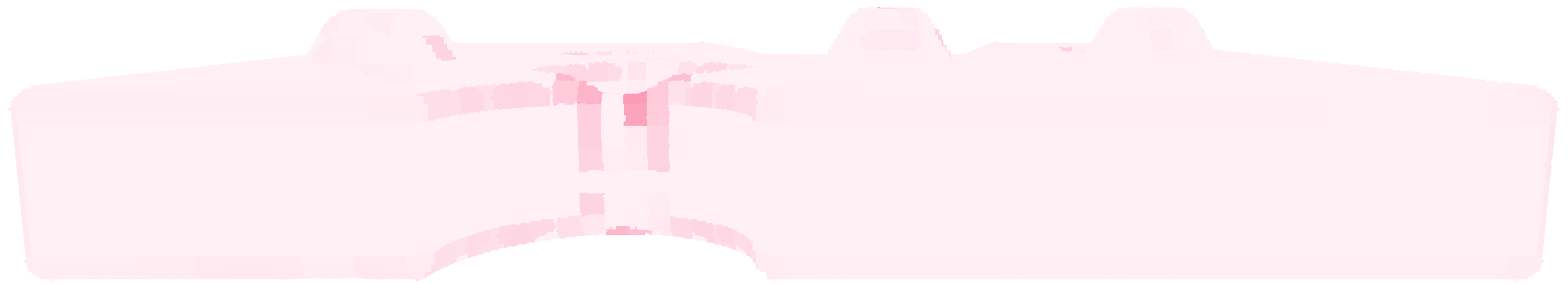}
\vspace{1em}
\caption{Heatmap+threshold}
\end{subfigure}   
\begin{subfigure}{0.25\linewidth}
\centering
\includegraphics[scale=0.18]{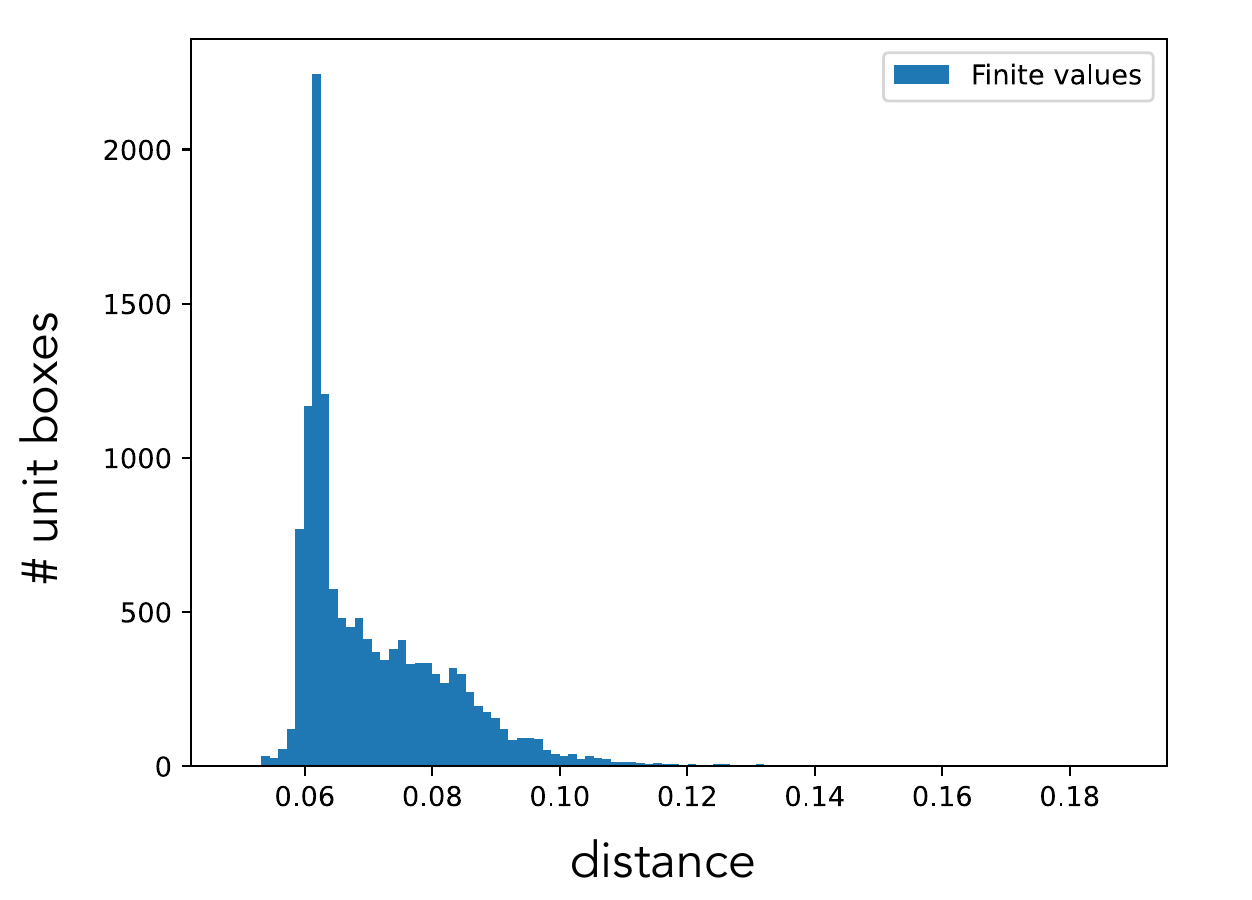}
\caption{Avg Distribution}
\end{subfigure}
\begin{subfigure}{0.25\linewidth}
\centering
\includegraphics[scale=0.05]{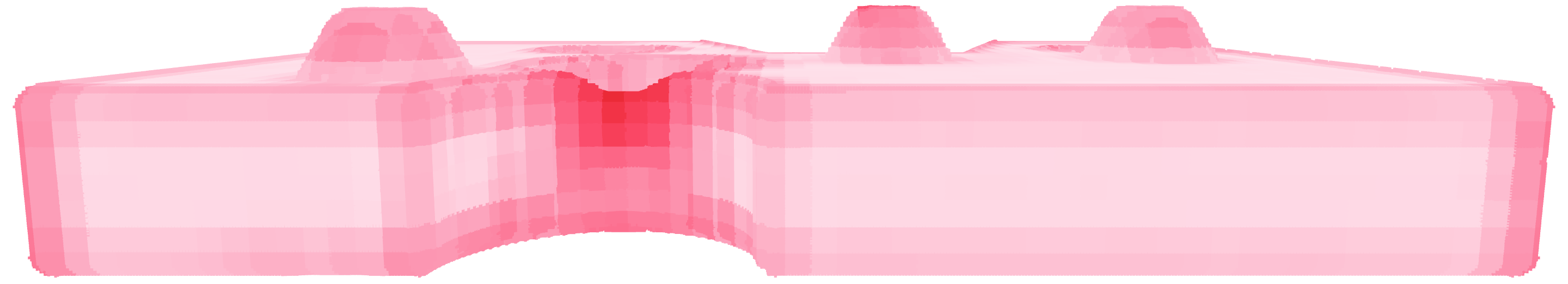}
\vspace{1em}
\caption{Avg Heatmap}
\end{subfigure}
\begin{subfigure}{0.25\linewidth}
\centering
\includegraphics[scale=0.05]{heatmap-generation/ERCF/latest-ercf-avg-thresh.png}
\vspace{1em}
\caption{Avg Heatmap+threshold \label{fig:visualization:avgthreshold}}
\end{subfigure}  
\caption{Heatmap and distance distribution visualization: the bottom row uses spatial averaging.}
\label{fig:visualization}
\end{figure}

Across benchmarks,
  both raw (\autoref{alg:comp}) and spatially averaged distance distributions (\autoref{subsec:spatialavg}) show that
  most unwanted errors, 
  caused by discretization, quantization, or floating-point arithmetic, 
  tend to have small magnitudes.
True differences, 
  on the other hand, 
  are relatively infrequent compared to unwanted errors 
  and typically exhibit much larger magnitudes.
A distribution graph of distances 
  between two models with notable feature differences 
  often exhibits a right-skewed shape, 
  where the maximum distance is significantly larger than the minimum.
A high percentile threshold therefore effectively separates true differences from noise.
In contrast, a distribution graph between two models \textit{without} significant feature differences 
  often tends to exhibit a nearly normal or slightly
  left-skewed distribution shape. 
The distribution may become more right-skewed as the unit box size decreases though still within a narrow range
  relative to the sampling gap.
\tool visualizes these differences via both heatmaps and distribution graphs to help distinguish true faults.
\autoref{fig:visualization} visualizes the difference heatmap for the model in \autoref{fig:nice-ercf-photo}.
The final result (bottom right) demonstrates how averaging and using a threshold ($90^{th}$ percentile)
  allows immediate identification of the fault in the model.

\section{Implementation}
\label{sec:impl}
\label{sec:limitations}

We implemented \tool in Python in approximately 1,200 LOC and it is publicly available~\cite{glitch}.
The \tool``kernel'' takes as input a \gcode file (with \texttt{.gcode} extension)
  and following the semantics in \autoref{sec:line-rec},
  reconstructs a set of cuboids.
From that, the kernel generates an approximate point cloud as explained in \autoref{sec:comp}.
The kernel can be used to compare two \gcode programs and output a heatmap image and a
  distance distribution graph. %
This can be used for various downstream applications;
  this work demonstrates two (\autoref{sec:invcheck}, \autoref{sec:difftest}).

\subsection{Complexity}
\label{subsec:scalability}
We assess \tool’s scalability by first 
  deriving the asymptotic cost of each of its components 
  and 
  showing that every stage grows linearly with the input geometry.
\autoref{subsec:runtimes} provides running times from our evaluation. 

\tool parses the input \gcode files; 
  for each linear motion command (with extrusion),
  it computes vertices of the corresponding cuboid,
  making this phase run in time linear in the number of commands.
The proportional sampler is also linear in the number points generated.
If the two input models are misaligned,
  we rotate the second point cloud;
  this operation touches each of its points only once
  and therefore runs in time proportional to the number of points in that cloud.
In point cloud segmentation (\autoref{alg:comp}),
  the only non-constant work is the loop that assigns
  each point to its corresponding unit box, so the running time is linear
  in the total number of points across the two clouds. 
For point cloud comparison we iterate over the unit boxes;
  in each box we compute an augmented Hausdorff distance
  between the two point subsets it contains using
  SciPy’s\footnote{https://scipy.org/} \texttt{scipy.spatial.KDTree}:
  building a tree for a subset of $m$ points and querying it once per point costs $\mathcal{O}(m\log{}m)$,
  yielding an overall $\mathcal{O}(N\log{}N)$ across all points $N$.
The pipeline finishes with error distribution and heatmap generation, 
  each running in time linear in the number of unit boxes.

In general, point cloud comparison is the most time-consuming phase, 
  with runtime dominated by SciPy’s \texttt{KDTree} operations. 
Faster nearest-neighbor structures or optimized libraries could reduce this time. 
The rest of the runtime is largely I/O.
Additional implementation-level optimizations are also possible:
  for example in the invariant checking application (\autoref{sec:invcheck}), 
  several comparisons share the same original G-code; 
  parsing and sampling that file once and reusing the resulting point cloud across all comparisons would eliminate redundant work.

\subsection{Limitations}
\tool's cuboid reconstruction algorithm uses a rectangular cuboid as an approximation for each deposited plastic segment,
  which may introduce inaccuracies in modeling the \gcode. 
In particular, this prevents \tool from modeling \gcode programs that use 
non-planar or non-uniform slicing. %
\gcode commands that encode non-linear movement (e.g., circular motion) are also not supported;
   we only model \lstinline[language=gcode]!G0! and \lstinline[language=gcode]!G1!.
We produce a logical approximation of the
  actual 3D model and do not consider the influence of
  temperature variations or filament properties.
\tool does not model friction, vibration, or mechanical properties,
  and focuses only on the
  ``core'' program
  that is responsible for constructing the model
  ignoring non-motion
  instructions that are present in all \gcode programs.
This means that \tool targets
  errors that are directly attributable to digital artifacts:
  3D designs, meshes, slicers, or a combination thereof.
There may however be other kinds of errors that can be attributed to the hardware, like loose printer belts, insufficient (or excessive) cooling, etc. that \tool cannot identify.

\section{Invariant Checking for 3D Models}
\label{sec:invcheck}
\label{subsec:claim1}

This section evaluates our first claim that by
  analyzing \prog program using \tool,
  we can statically check invariants of models to detect
  regions that are likely to print incorrectly,
  even if they are valid 3D solids.
First, we define a specific invariant of \prog programs 
  checking which has helped us identify problems in many benchmarks.
  
\textbf{Motivation.} In traditional compilers, the idea of
  equivalence modulo inputs~\cite{vule} has been used for
  testing compilers by perturbing
  programs to obtain semantically equivalent, yet syntactically different inputs.
This idea can be instantiated for the 3D printing domain as well. For example,
  rotating a model should not affect the shape represented by the generated
  \gcode; it should only change its orientation in 3D space.
In fact, it is common knowledge~\cite{warrior, truss, holygrail}
  among users interacting with
  3D printing tools that
  comparing \gcode programs 
  sliced in different orientations 
  can reveal problems in 3D models and
  even expose bugs in slicers.\footnote{\url{https://github.com/Ultimaker/Cura/issues/16726\#issuecomment-1719546101}}
Inspired by these observations,
  we define the following invariant for \gcode programs.

\textbf{Rotation Invariant, \invariant.}
We define a 3D triangle $\mathsf{mesh}$ as $(\mathbb{R}^3, \mathbb{R}^3, \mathbb{R}^3)$.
Let $R$  be a function that rotates a \textsf{mesh} by a 3D vector and
  slices it (by invoking a slicer) to produce \prog.
Let $r_c$ be a function that rotates a \textsf{cuboid} set by a 3D vector.
Let $r_p$ be a function that rotates a point cloud by a 3D vector.
\[
\begin{array}{lll}
     R &: (\mathbb{R},~ \mathbb{R},~ \mathbb{R}) ~\rightarrow~ \mathsf{mesh} &\rightarrow~ \prog  \\
     r_c &: (\mathbb{R},~ \mathbb{R},~ \mathbb{R}) ~\rightarrow~ \mathsf{cuboids} &\rightarrow~ \mathsf{cuboids} \\
     r_p &: (\mathbb{R},~ \mathbb{R},~ \mathbb{R}) ~\rightarrow~ \mathsf{points} &\rightarrow~ \mathsf{points}
\end{array}
\]

Let $M$ be a $\mathsf{mesh}$ that produces a \prog program $G$; let $G ~\Downarrow_{d,h}^p~ (\_,~ C)$ and let $P ~=~ \mathsf{flatmap}(\Omega_g,~ C)$ be the point cloud approximated from $C$.
Let $G_v$ be the \prog obtained by
  slicing the rotated mesh, $R(v, M)$; let $G_{v} ~\Downarrow_{d,h}^p~ (\_,~ C_v)$,
and let $P_v ~=~ \mathsf{flatmap}(\Omega_g,~ C_v)$ be the point cloud approximated from $C_v$.
The rotation invariant states:
$$r_c (v, C) ~=~  C_v$$
i.e., rotation and denotation must commute.
However, as mentioned in \autoref{subsec:cloud},
  checking this does not scale to complex models.
We therefore define and check the following approximate version of
  this invariant over the point clouds:\footnote{We emphasize that this does \textit{not} mean the 3D model must be
  physically \textit{printed} in the rotated orientation; indeed many models
  are designed such that they can only successfully be printed
  in a particular orientation (e.g., due to overhang).}
$$r_p (v, P) ~\overset{\epsilon}{\approx}~  P_v$$

Observe that checking \invariant
  has now reduced to comparing two \prog programs by
  comparing their point clouds which \tool enables.

Other affine transformations like translation and scaling
  also define valid invariants, i.e., they commute with denotation.
For example, translating a 3D mesh ($M$) along the $x$-axis, slicing it, computing the denoted cuboid set,
  and then translating the cuboids back should yield the original cuboid set corresponding to $M$.
The same holds for scaling and rotation.
However, we hypothesize that \invariant is more effective for revealing model bugs and aiding differential testing of slicers and mesh repair tools.
Unlike translation and scaling which involve simple arithmetic,
  rotation requires trigonometric functions %
 which empirically appear to expose more errors.

The rest of this section evaluates our claim that
  by checking \invariant,
  \tool can localize problematic parts in large,
  real-world 3D models for which
  we analyze the difference heatmap and distance distribution graph (\autoref{fig:glitchflow}, right) produced by \tool.

\subsection{Benchmarks}
We used \tool to check \invariant for {56} real-world models, 
  including {50} problematic models that failed to produce
  correct \gcode programs as reported by users, 
  and {6} error-free models.
Among the {50} problematic models, 
  one is shown in \autoref{fig:nice-ercf-photo}, 
  a mechanical part from an open-source project~\cite{ercf}.
To select the remaining models, 
  we manually examined GitHub issues and
  user forums from two popular slicers, 
  UltiMaker Cura~\cite{cura}\footnote{We used the ``\textsf{Slicing Error}'' filter on Github.} and Prusa~\cite{prusaslicer}.
We selected {50} issues / posts that 
  included a model that we could use for our evaluation.
We obtained the {6} error-free models by inspecting
  the Voron-0 3D printer parts repository~\cite{voron} because
  parts for building 3D printers are more likely to be error-free due to widespread use and extensive testing.

\subsection{Experimental Setup}
Our \tool results were obtained on CloudLab~\cite{Duplyakin+:ATC19} \texttt{xl170} machines which are
  10-core Intel E5-2640v4 running at 2.4 GHz with 64GB RAM and 6G SATA SSDs.
To generate \gcode, we used the
  popular and actively maintained
  Cura slicer~\cite{cura} running on Linux.
We configure the slicer with the following settings:
  0.4mm nozzle diameter, 
  0.2mm layer height
  100\% infill density, 
  and no support structures or build plate adhesion 
  (all other settings remained had default values).\footnote{
These settings are not intended for actual printing;
  they are only used for statically comparing two \gcode programs using \tool.}

The axes and angles of rotation are parameters to \tool and therefore user configurable.
By inspecting problematic models,
  we observed that rotations
  about the major axes
  often lead to different behaviors
  in problematic models.
Since the slices are parallel to the $XY$ plane (\autoref{sec:background}),
  we found that most of the interesting errors are exposed when
  the rotation changes the orientation of the object
  such that its base is no longer ``on'' the $XY$ plane.
As a result, rotation about the $z$-axis
  does not expose as many errors.
Therefore, we chose to rotate each model by two different vectors:
  by $\left(90, 0, 0\right)$ and 
  by $\left(0, 90, 0\right)$
  (i.e., rotation about $x$ and $y$ axes).
We also found that rotating by other angles like $45$ degrees tends to
  increase unwanted errors (like quantization errors)
  that do not correspond to true differences.

For streamlining our evaluation, 
  we wrote an automation script (called \glitchrunner)
  to which we pass 
  the 3D triangle mesh from the original model,
  the rotation angles,
  the sampling gap ($g$),
  the unit box dimensions
  (see \autoref{alg:compalg} for point cloud segmentation), and
  a threshold percentile
  (see \autoref{sec:visual} for heatmap and distribution graph).
These arguments are all user configurable.
\autoref{sec:appexdixparams} lists the parameters
  for all our benchmarks.
\glitchrunner automates the rotation,
  the slicing (by invoking the Cura slicer's \texttt{CuraEngine} directly), and ultimately
  invokes \tool.
For some models,
  \glitchrunner first scales them to fit within Cura's build plate
  for before slicing.
\tool's kernel (\autoref{fig:glitchflow}, left) 
  reconstructs the cuboid set and approximates the point cloud for each model,
  performs \textsc{\footnotesize{\textsf{POINT\_CLOUD\_SEGMENTATION}}},
  and ultimately runs \textsc{\footnotesize{\textsf{COMPARE\_POINT\_CLOUDS}}}
  (\autoref{alg:comp}).
Since each model is rotated both about the $x$ and $y$ axes,
  \tool produces two heatmaps and distribution graphs for each benchmark.
\glitchrunner takes these and generates a combined heatmap and distribution graph by
  computing the average distance
  for each unit box.\footnote{This requires that all \textsf{hd_list} are generated 
  using the same dimensions of \textsf{ubox} and bounding box. 
During the averaging, 
  distances with a \texttt{none} value are ignored, 
  and whenever any distance within a unit box is $\infty$, 
  the average distance for that box is set to $\infty$.}

\subsection{Results}
In all 50 problematic benchmarks, 
  \tool successfully identified the problematic areas. 
Based on the original discussion about the model posted by users,
  we classify these models into three categories based
  on the kind of problem they exhibit: 
  models with small feature sizes (12),
  non-watertight models (33), and
  models with flipped normals (5), 
  and discuss our findings.

\subsubsection{Models with Small Features}
\label{subsubsec:small-feature}

\begin{figure}
    \centering
    \includegraphics[scale=0.2]{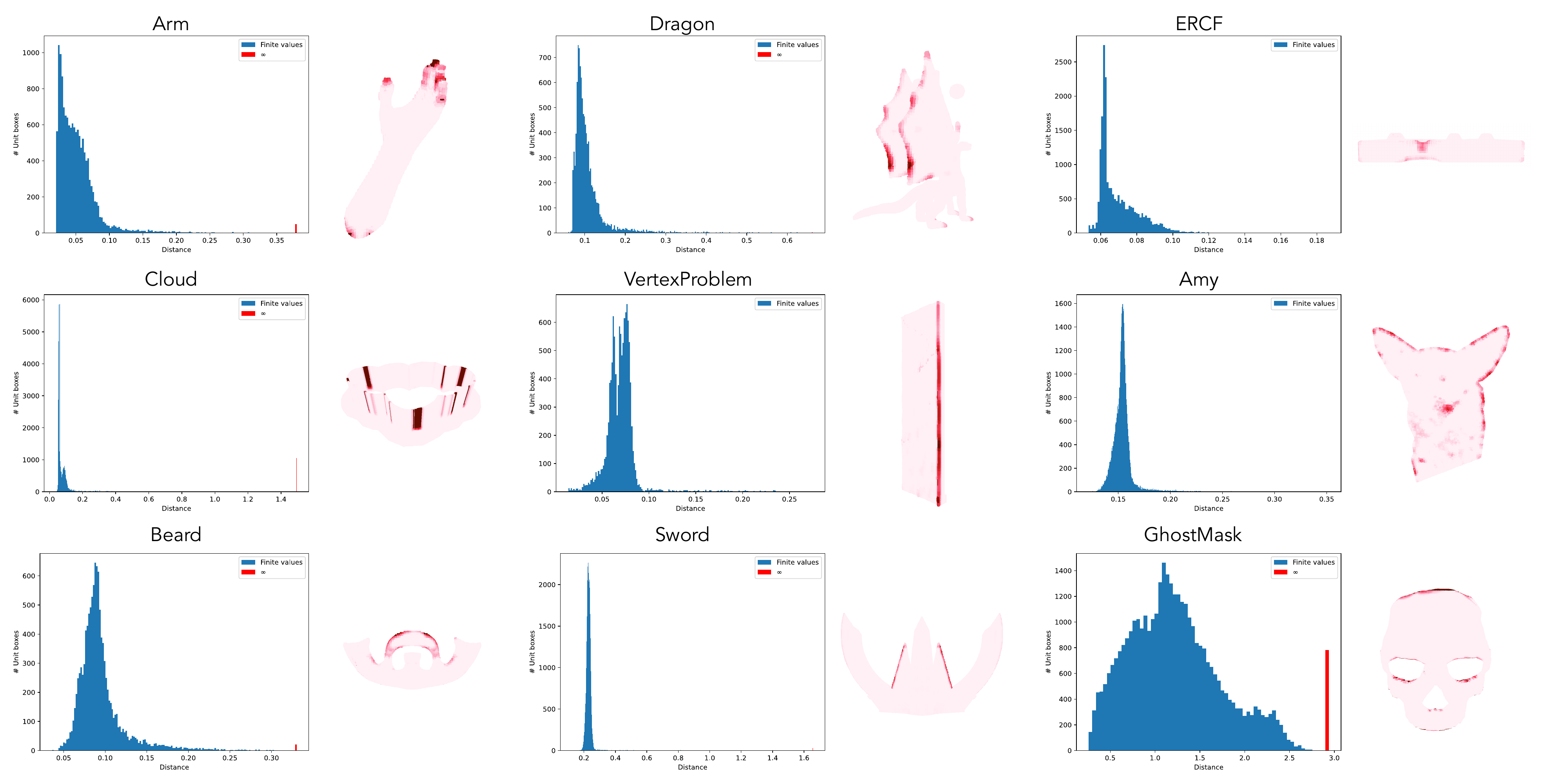}
    \includegraphics[scale=0.2]{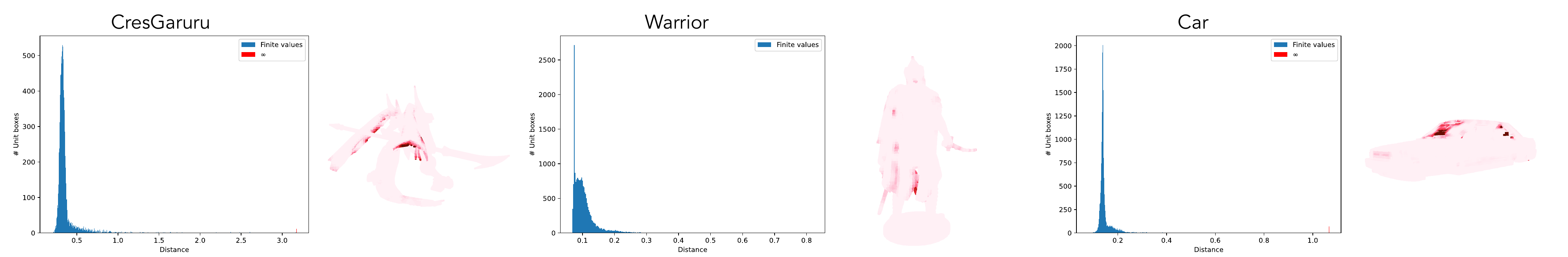}
    \caption{Output of \tool on 12 benchmarks with small feature sizes. We omit the axis labels for space --- $x$-axis is distance and $y$-axis is \#unit boxes (same as elsewhere). Dark red regions highlight small features due to which they failed to be sliced correctly. Lines to the right in the distribution graphs indicate that there were unit boxes that had large distances between the two point clouds. Red lines in the distribution graphs indicate that some unit boxes had $\infty$ distance, 
    meaning that those boxes contained no points in one of the two point clouds.
    Both of these indicate true differences.
    The bar for $\infty$ is chosen such that it is
    slightly farther on the $x$-axis from the
    maximum numeric value distance.}
    \label{fig:eval-feature}
\end{figure}

\autoref{fig:eval-feature} presents the overall heatmaps and distribution graphs 
  for a collection of 12 models with small geometries. 
\tool is particularly effective in these scenarios, 
  as the differences in \gcode
  resulting from small features 
  are often too subtle 
  and difficult to detect manually.
Cura failed to slice the
  \texttt{Amy} and \texttt{Warrior} models 
  entirely in their original orientations. 
To address this, 
  we applied a 90-degree rotation about the $z$-axis
  and treated this rotated orientation
  as the original for these models.
  
We highlight four models 
  (\texttt{Arm}, \texttt{VertexProblem}, \texttt{Dragon}, \texttt{Amy}):
  these exhibit varying degrees of mesh flaws, 
  which are detectable by
  existing mesh tools
  according to the discussions in the
  Github issue where we found the models. 
However, in all these cases,
  the only flaw in the generated \gcode
  was that it was missing the parts
  that contained small or intricate features.
This suggests that
  the slicer was able to correct for the bad geometry
   but failed to handle the small feature size. 
For example, as reported in the Github issue for the \texttt{Arm} model, 
  the user observed that even after repairing the mesh, 
  there was no significant difference in the
  generated \gcode, rendering the use of 
   mesh repair tools on these models ineffective.

This highlights a broader challenge in 3D printing: 
  many imperfections in 3D printing arise
  due to fine features,
  yet, existing mesh tools 
  do not typically check for this issue.
  The reason is that
  the smallness of a feature is \textit{relative}
  to the nozzle size and slicing parameters and therefore
  its effects only manifest \textit{after}
  slicing (generating \gcode). 
\textit{\tool can detect this class of problems precisely because it
  targets \gcode},
  which captures the interplay between the 3D model,
  the choice of slicing software, and
  the settings used to configure the slicer.
\tool is capable of identifying differences arising from this combination of factors, 
  all of which play a crucial role 
  in the actual 3D printing process.
This is analogous to static analysis tools that
 verify bytecode instead of source code,
 thereby not needing to trust the compiler.

\subsubsection{Non-Watertight Models}
\label{subsubsec:notwatertight}

\begin{figure}
    \centering
    \includegraphics[scale=0.17]{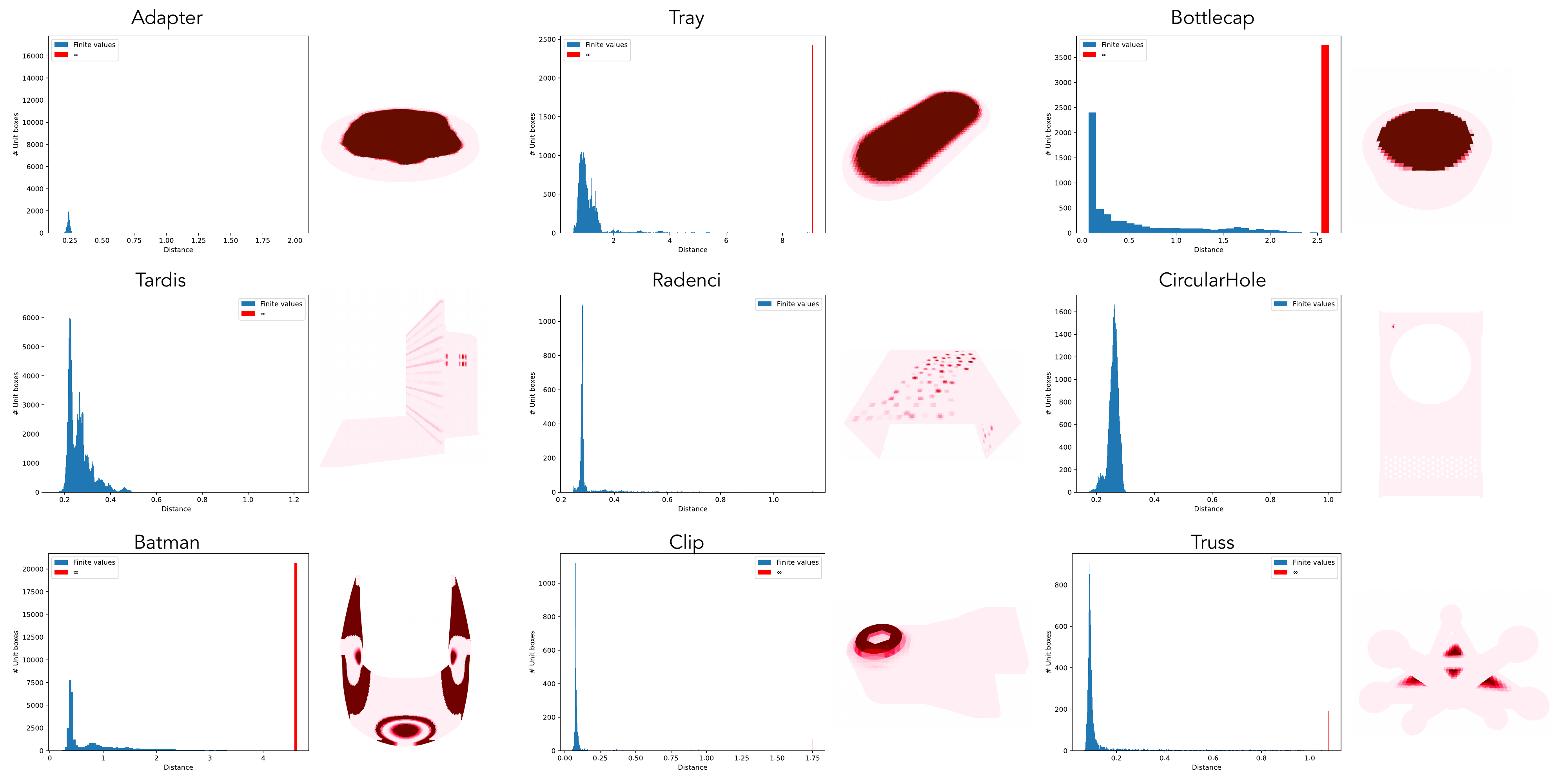}
    \includegraphics[scale=0.17]{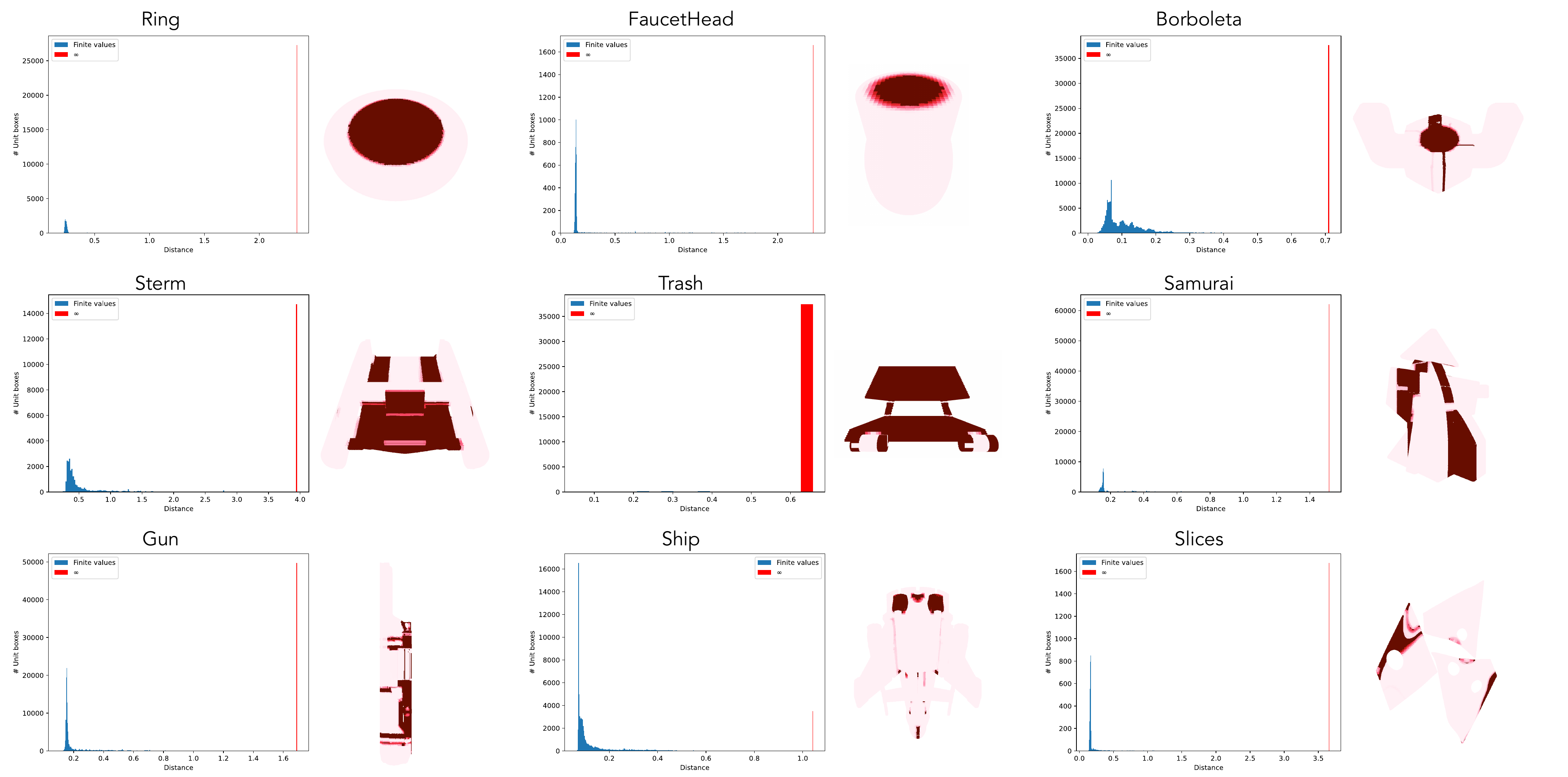}
    \includegraphics[scale=0.17]{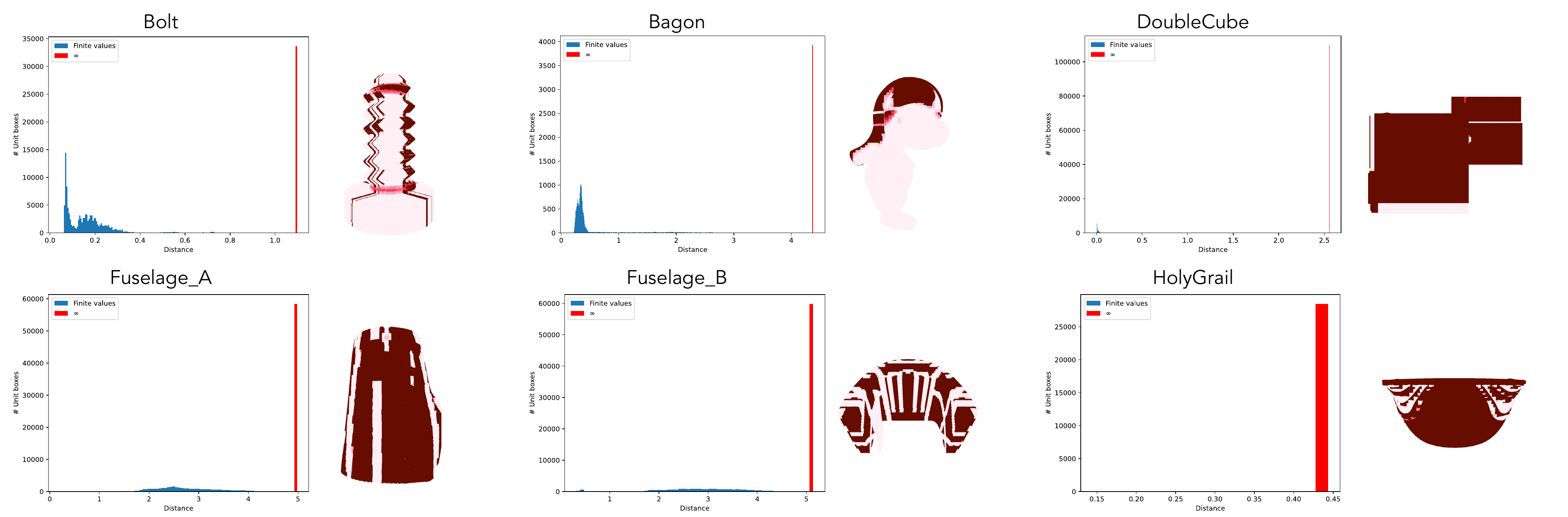}
    \caption{Output of \tool on the 24 of 33 non-watertight benchmarks.
    The dark red regions highlight parts that caused slicing to fail. 
     We omit the axis labels for space --- $x$-axis is distance and $y$-axis is \#unit boxes.
     Lines to the right in the distribution graphs and red lines have the same meaning as \autoref{fig:eval-feature}.}
    \label{fig:eval-water1} 
\end{figure}

\begin{figure}
    \centering
    \includegraphics[scale=0.17]{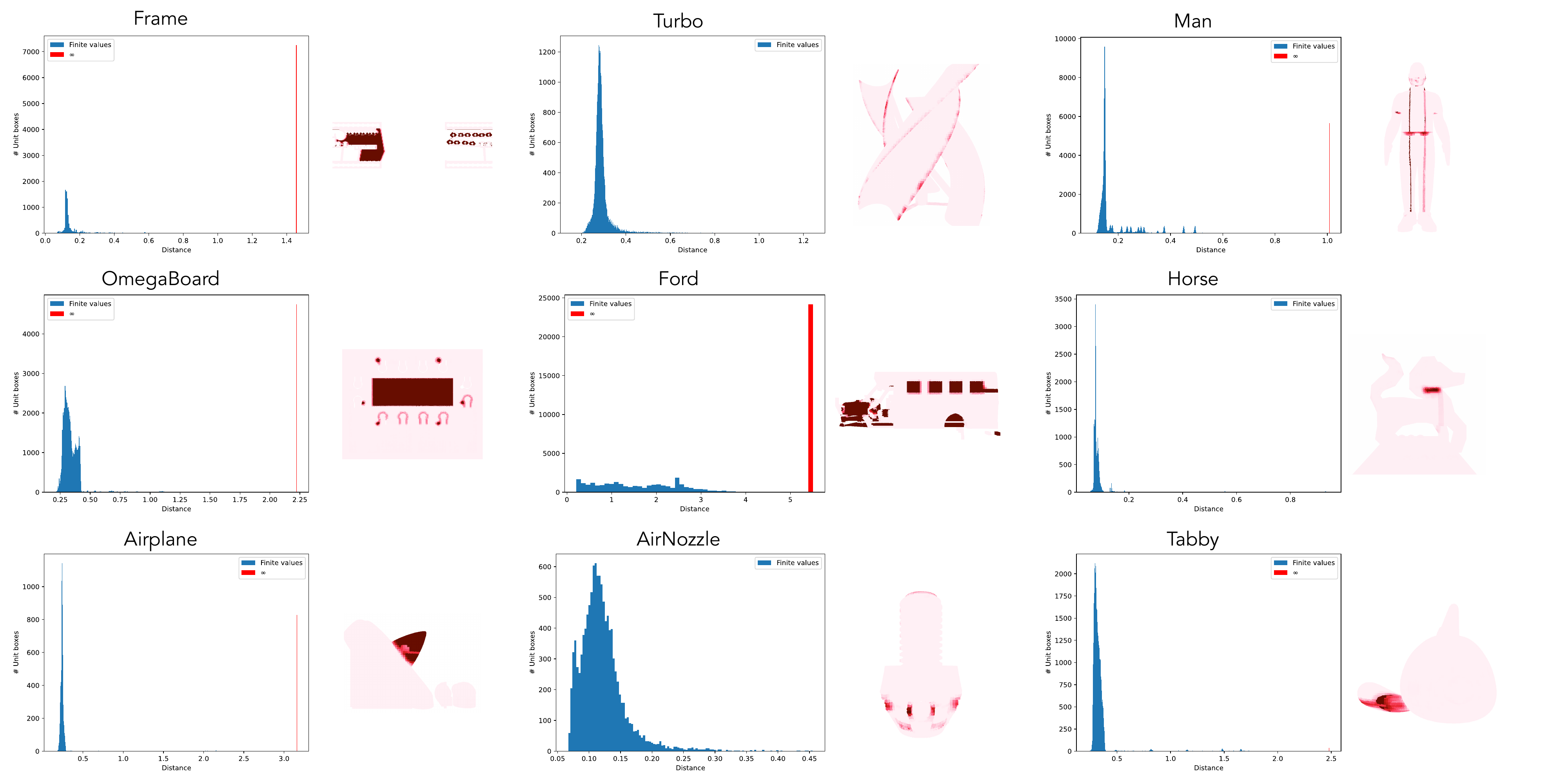}
    \caption{Continuation of \autoref{fig:eval-water1}:
     Output of \tool on the remaining 9 of 33 non-watertight benchmarks.
    The dark red regions highlight parts that caused slicing to fail. 
     We omit the axis labels for space --- $x$-axis is distance and $y$-axis is \#unit boxes.
     Lines to the right in the distribution graphs and red lines have the same meaning as \autoref{fig:eval-feature}.}
    \label{fig:eval-water2} 
\end{figure}

\autoref{fig:eval-water1} and \autoref{fig:eval-water2}
  show the overall heatmaps
  and distribution graphs for 33 models which have
   geometric defects preventing watertight closure.
Of the 33 models, 7 
  (\texttt{Airplane}, \texttt{Frame}, \texttt{Gun},
  \texttt{Man}, \texttt{Tabby}, \texttt{Tray}, and \texttt{Truss})
  were scaled by users to expose issues; we replicated those scalings.
Non-water-tightness accounts for a
  significant fraction of slicing failures, often preventing
 slicers from reliably generating \gcode. 
While some mesh repair tools~\cite{netfabb, meshlab, meshmixer}
  can correct for non-water-tightness, 
  they do not all check for the same errors, 
  and there is no guarantee that the repaired models
  will always result in a correct slicing (\autoref{subsec:mesh-tool-comp}). 
\glitchrunner bypasses these limitations by
  enabling checking \invariant
  rather than locating specific flaws: 
  it identifies regions violating \invariant,
  regardless of the underlying nature of the mesh defect.
User reports also indicate that 
  some models in this category 
  (e.g., \texttt{Tardis} and \texttt{Ford}) 
  could be successfully repaired by specific mesh repair tools,
  yielding correct slicing results. 
However, other models like \texttt{Fuselage\_A} and \texttt{Man}
  are still incorrectly sliced even after mesh repair.

\subsubsection{Models with Flipped Normals}
\label{subsubsec:flipped-normal}
\begin{figure}
    \centering
    \includegraphics[scale=0.17]{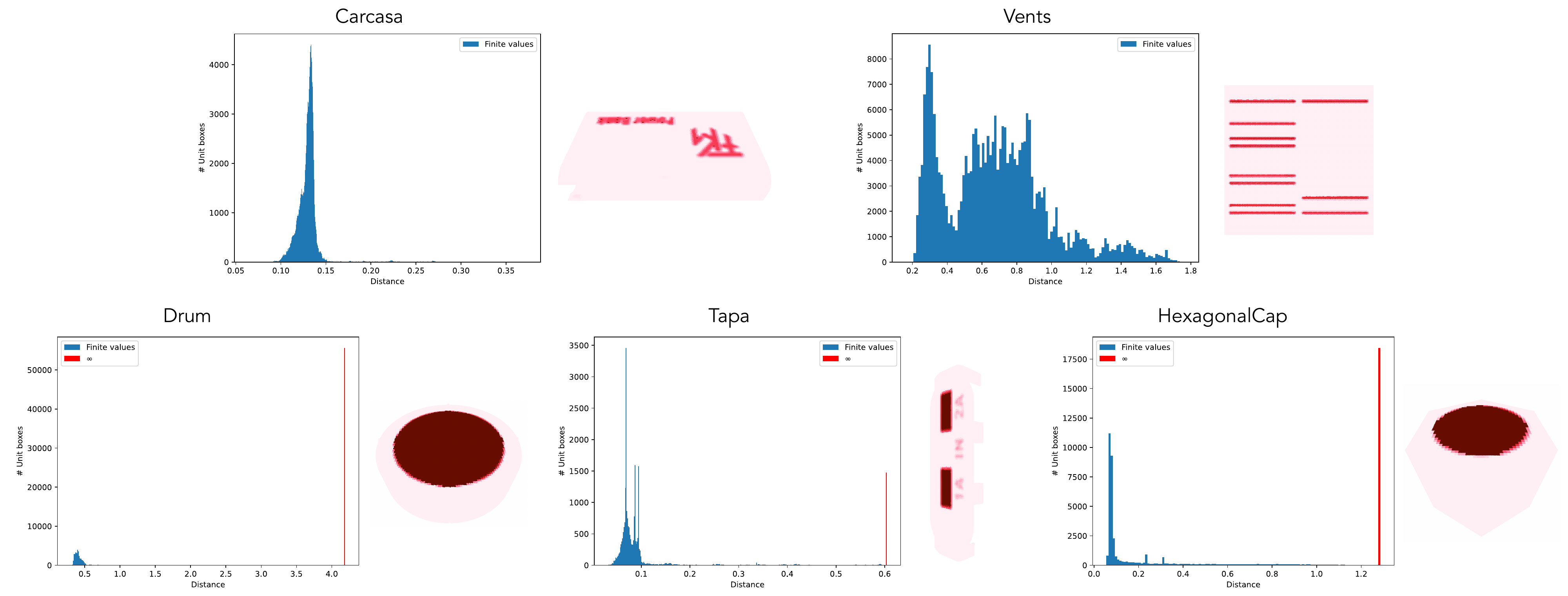}
    \caption{Output of \tool on the 5 benchmarks with flipped normals.
    The dark red regions highlight parts that caused slicing to fail. Lines to the right in the distribution graphs and red lines have the same meaning as \autoref{fig:eval-feature}.}
    \label{fig:eval-normal} 
\end{figure}

\autoref{fig:eval-normal} shows the heatmaps and distribution graphs for 5 out of the 50 models for
  which the only issue was ``flipped normals.''
A normal is a vector perpendicular to the faces of a triangle mesh and typically points outward from the surface,
  indicating the ``outside'' of the model.
In case of a flipped normal, this perpendicular vector 
  points ``inwards'' which can confuse slicers into filling a region which is supposed to be a  hole.
Applying a rotation that orients the hole
  in a direction that is not orthogonal to the
  slicing direction
  can frequently resolve the issue. 
Therefore, checking \invariant is well-suited 
  for detecting models with such problems. 
In Cura, 
  flipped normal issues are not immediately evident upon loading the model 
  unlike non-watertight models 
  which are more visibly flagged by an error message. 
Instead, users must manually identify these problems by inspecting
  the preview of the \gcode (that most slicers support) to
  detect ``overhangs'' that appear in unusual locations,
  such as on the top surface of the model.
This is particularly tedious 
  for large or complex models but something that checking \invariant with \tool can easily detect.

\subsubsection{Addressing False Positives with Distribution Graphs}
By design, \tool highlights regions with
  relatively significant deviations which could
  indicate true differences
  but also unwanted errors (\autoref{subsubsec:overall}, \autoref{sec:visual}).
  These unwanted errors can be viewed as being
  analogous to false positives in traditional
  program analyses.
The heatmaps on their own cannot distinguish between these
  two cases. To address this,
  \tool also generates the distance
  distribution graph.

\begin{figure}
    \centering
    \includegraphics[scale=0.23]{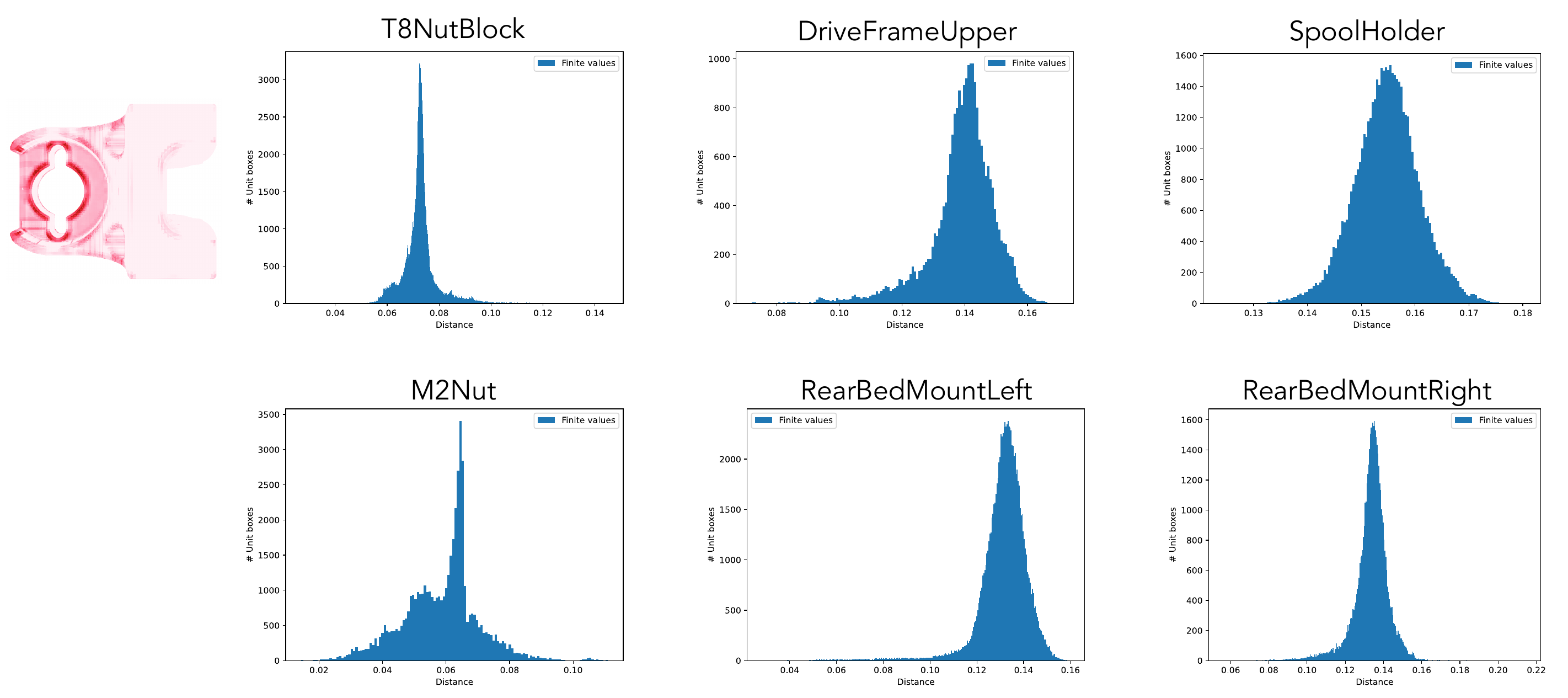}
    \caption{Output of \tool on 6 error-free models: the slightly left-skewed / approximate normal distance graph suggests that the differences are due to unwanted errors.}
    \label{fig:eval-errorfree} 
\end{figure}

We ran \tool on {6} well-tested, error-free models
  obtained from a repository that contains
  parts of a 3D printer~\cite{rearLeft, rearRight, m2Nut, t8NutBlock, driveFrameUpper, spoolHolder} (such mechanical parts are reasonably error-free)
  to demonstrate
  that heatmaps alone are not sufficient for
  identifying true differences and
  the distribution graphs are required for effectively distinguishing
  between true differences and unwanted errors (\autoref{fig:eval-errorfree}).
All distributions appear approximately normal or slightly
  left-skewed, 
  which is consistent with our earlier observations (\autoref{subsubsec:dist}) that 
  true differences tend to have an outliers characteristic.
For one example (\texttt{T8NutBlock}), 
  we show both the heatmap and distribution graph ---
  the heatmap shows that
  the areas with the darkest colors correspond to the edges or boundaries of the inner hole in the design. 
These regions show discretization artifacts~\cite{curvislicer}, 
  where the slicer attempts to approximate
  the circular geometry using discrete linear plastic extrusions.
  
\subsubsection*{Summary}
The key takeaway from this evaluation is that
  checking \invariant with \tool can localize
  problematic parts in 3D models arising due to different
  types of errors.
\tool's heatmap visualization
  provides an actionable diagnostic tool for the
  3D printing workflow, 
  enabling users to address slicing challenges through multiple strategies.
For locally problematic regions 
  (e.g., thin features marked in heatmaps),
  thickening those parts, adjusting parameters (minimum line width, resolution),
  or changing to a smaller nozzle can help mitigate the problem.

We note that goal for \tool is not to replace mesh repair tools and other
  debugging strategies --- rather it can serve
  as a guide to help users identify the root cause of
  a failed slicing attempt by localizing the error-prone
  parts of their model.
 While there is no single tool that can detect all kinds of problems,
 \tool is the first tool that covers a much wider range of errors
 than existing mesh or design level tools.
Since \tool reasons about \gcode,
  it can detect errors that arise due to the effect of
  slicing parameters and other slicer-level factors.

\subsection{Running Times}
\label{subsec:runtimes}
\begin{wrapfigure}{r}{0.35\textwidth}
    \vspace{-30pt}
    \centering
    \includegraphics[width=\linewidth]{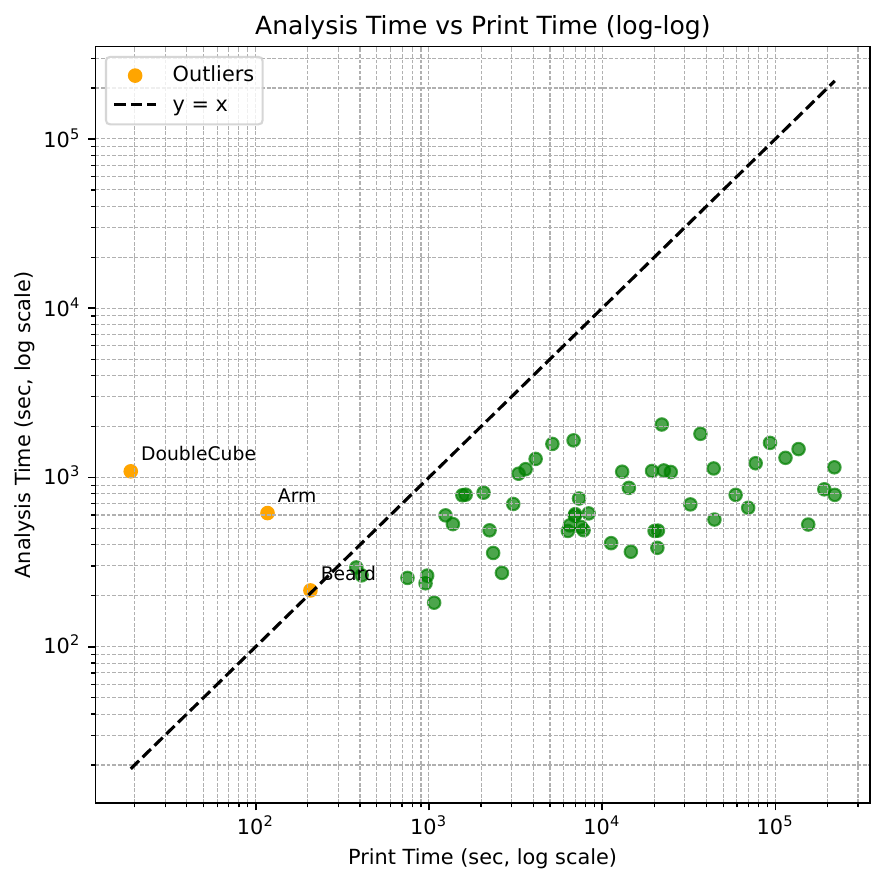}
    \caption{Comparing \tool's running time to estimated print times across all 56 benchmarks. Plots are shown in log scale, time is in seconds.}
    \vspace{-22pt}
    \label{fig:runtimes}
\end{wrapfigure}

We ran each benchmark three times, and report the running times as mean $\pm$ 95\% confidence interval.
Across all 56 benchmarks, the running times span $182\pm 2.3$s for the fastest model (\texttt{Frame}) 
  to $2055\pm 9.2$s for the slowest (\texttt{Man}).
A per-model breakdown of the running times is in \autoref{tab:model-param1} and \autoref{tab:model-param2}
  in \autoref{sec:appexdixparams}.
Since no directly comparable prior tool exists, 
  we compare the running time against the printing time
  estimates provided by the Cura slicer in \autoref{fig:runtimes}.
For most models, \tool completes its invariant checks much faster than the actual print,
  especially for larger ones like \texttt{Batman} (see \autoref{tab:model-param1}). 
Three of the small models \texttt{Arm}, \texttt{DoubleCube}, \texttt{Beard} print too quickly for
  analysis time to be faster.
However, the printing time estimates typically exclude 
  ancillary delays (bed heating, nozzle cooling, etc.) which
  would only add to the actual print times.

\section{Differential Testing of Fabrication Tools}
\label{sec:eval}
\label{sec:difftest}

This section evaluates our second claim that by comparing
  \prog programs,
  \tool finds differences between slicers and
  can evaluate the efficacy of
  mesh repair tools.
For both cases, we validated the differences highlighted
  in the heatmaps by manually performing a visual comparison of the \gcode and the model.

\subsection{Comparing Slicers using \glitchfinder}
\label{subsec:slicer-comp}

Slicers can produce different \gcode for the same model because slicing is fundamentally a path planning problem.
They use heuristics to optimize for
continuous extrusion, fewer unnecessary movements, and fabrication time.
These choices can lead to different outcomes depending on the slicer and model.
Some slicers are also tailored for specific printers, limiting their generality.
Even with the same slicer, different settings (e.g., seam placement) can change start and end points for layers, resulting in different \gcode.
We therefore used \tool  to compare the 
  behaviors of two widely used slicers, 
  Cura-5.3.1 and PrusaSlicer-2.7.4, running on MacOS M1, 
  on a set of problematic 3D models. 
The goal is to 
  identify differences in slicing outcomes 
  and understand how each slicer handles
  specific geometries. 

\subsubsection{Benchmarks and setup}
We used the 50 problematic models from ~\autoref{subsec:claim1}, 
  as well as two additional models
  (\texttt{Haut}, \texttt{DrillJig})
  from PrusaSlicer's issues
  that can be sliced correctly in Cura but not in PrusaSlicer (52 models in total). 
We imported each model into both slicers, 
  generated the \gcode, and ran \tool on it.
We configured both slicers with the same settings.

\subsubsection{Results}
For 12 of the 52 models, 
  Cura and PrusaSlicer
  produced similar \gcode
  based on visual inspection.
The remaining 40 models ---
  including all models from Section~\ref{subsubsec:small-feature} and ~\ref{subsubsec:flipped-normal}, 
  as well as 21 models from ~\ref{subsubsec:notwatertight} ---
  demonstrate differences in slicing behaviors between the two slicers.
Specifically, PrusaSlicer fails to generate \gcode for 2 models 
  (\texttt{GhostMask} and \texttt{DoubleCube}), 
  resulting in no heatmap for these cases. 
For 2 models both slicers resulted in incorrect slicing but
  in different, non-overlapping parts of the model.
Cura produces \gcode that 
  more accurately represents the original model for 16 of the other 36 models, 
  while PrusaSlicer produces more accurate \gcode than
  Cura on the rest of the 20 (out of 36) models.
We show detailed results on
  10 representative models, including the CAD design,
  rendered \gcode from Cura and PrusaSlicer
  and the corresponding heatmaps.
For 5 of these
  Cura-produced G-code more faithfully represents the original model compared to PrusaSlicer, 
  while the other 5 are the opposite. 

\begin{table}
\footnotesize
\centering
\caption{
Comparison of slicing results between Cura and PrusaSlicer for five representative models
  where Cura outperforms PrusaSlicer. 
The leftmost column shows each model's CAD design in OpenSCAD, 
  followed by \gcode visualizations from Cura (middle-left) 
  and PrusaSlicer (middle-right), 
  and heatmaps in the rightmost column. 
While Cura produces better results overall compared to PrusaSlicer, 
  only the first two models are sliced correctly,
  and the remaining three models showed slicing errors (Section~\ref{subsec:claim1}), 
  demonstrating that both slicers struggle with these particular geometries despite Cura's relative advantage.
}
\begin{tabular}{c|c|c|c}
{Model} & 
{G-code (Cura)} & 
{G-code (Prusa)} & 
{Heatmap} \\
\hline
\includegraphics[width=0.2\textwidth]{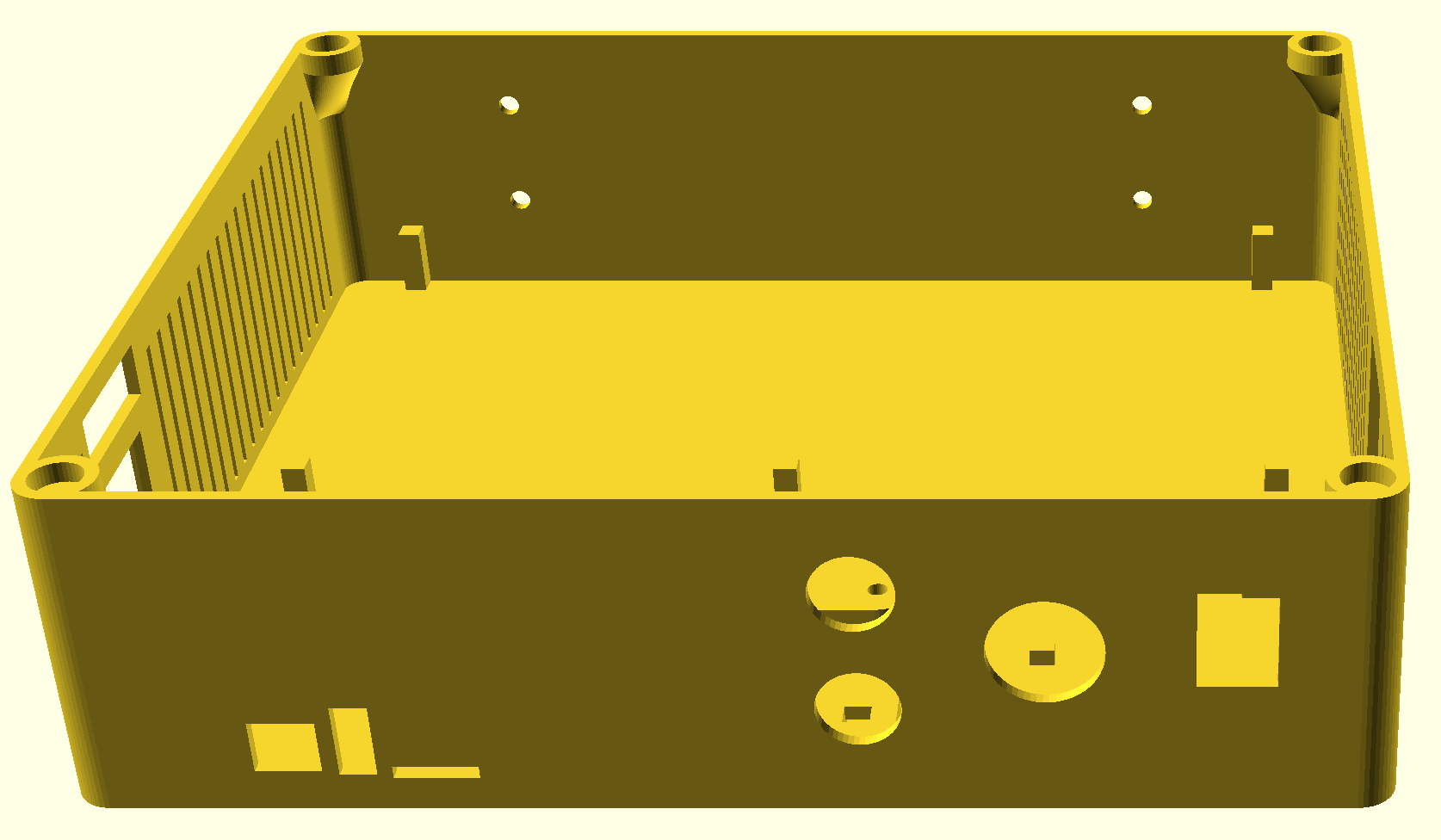} & \includegraphics[width=0.2\textwidth]{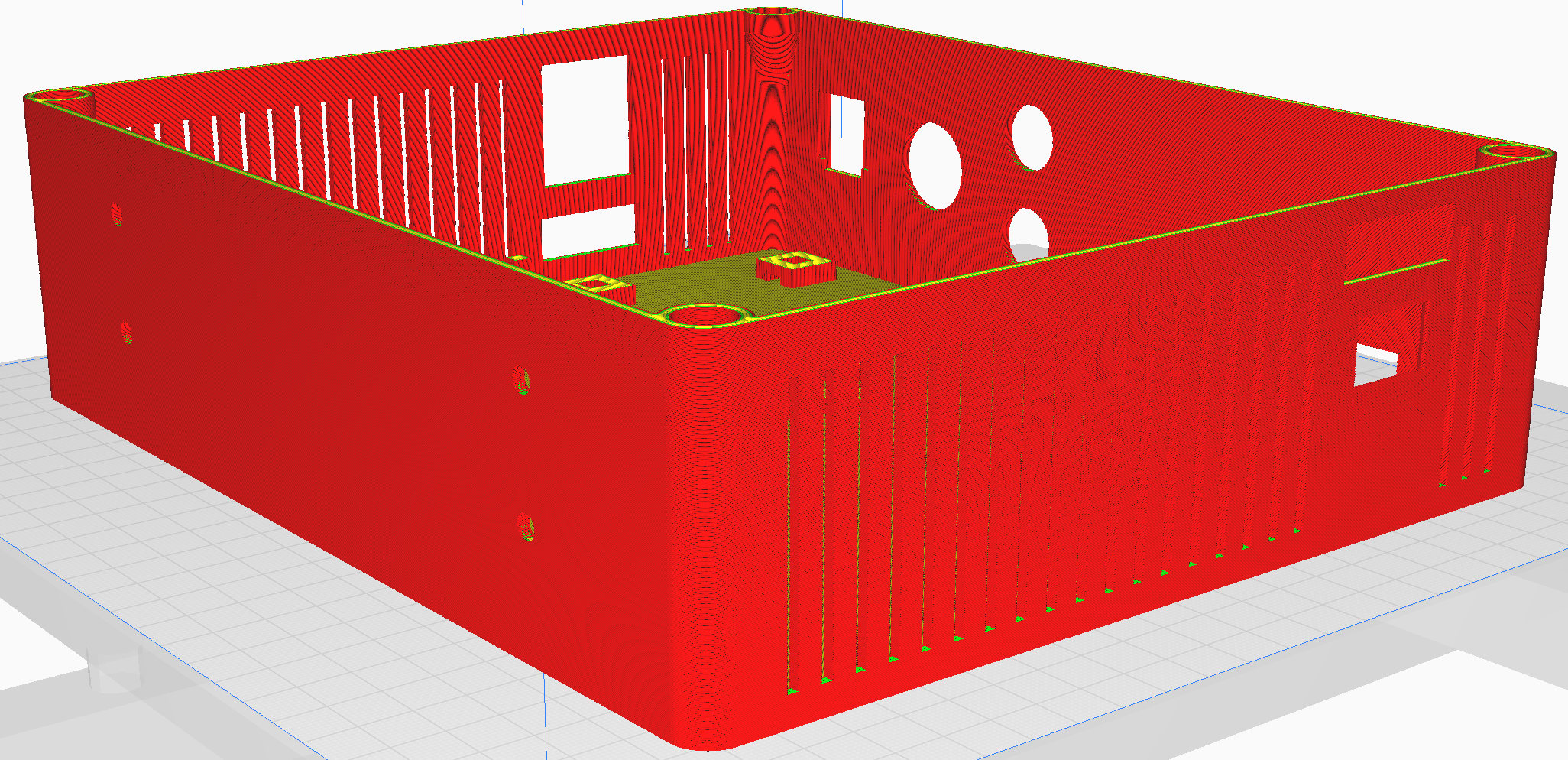} & \includegraphics[width=0.2\textwidth]{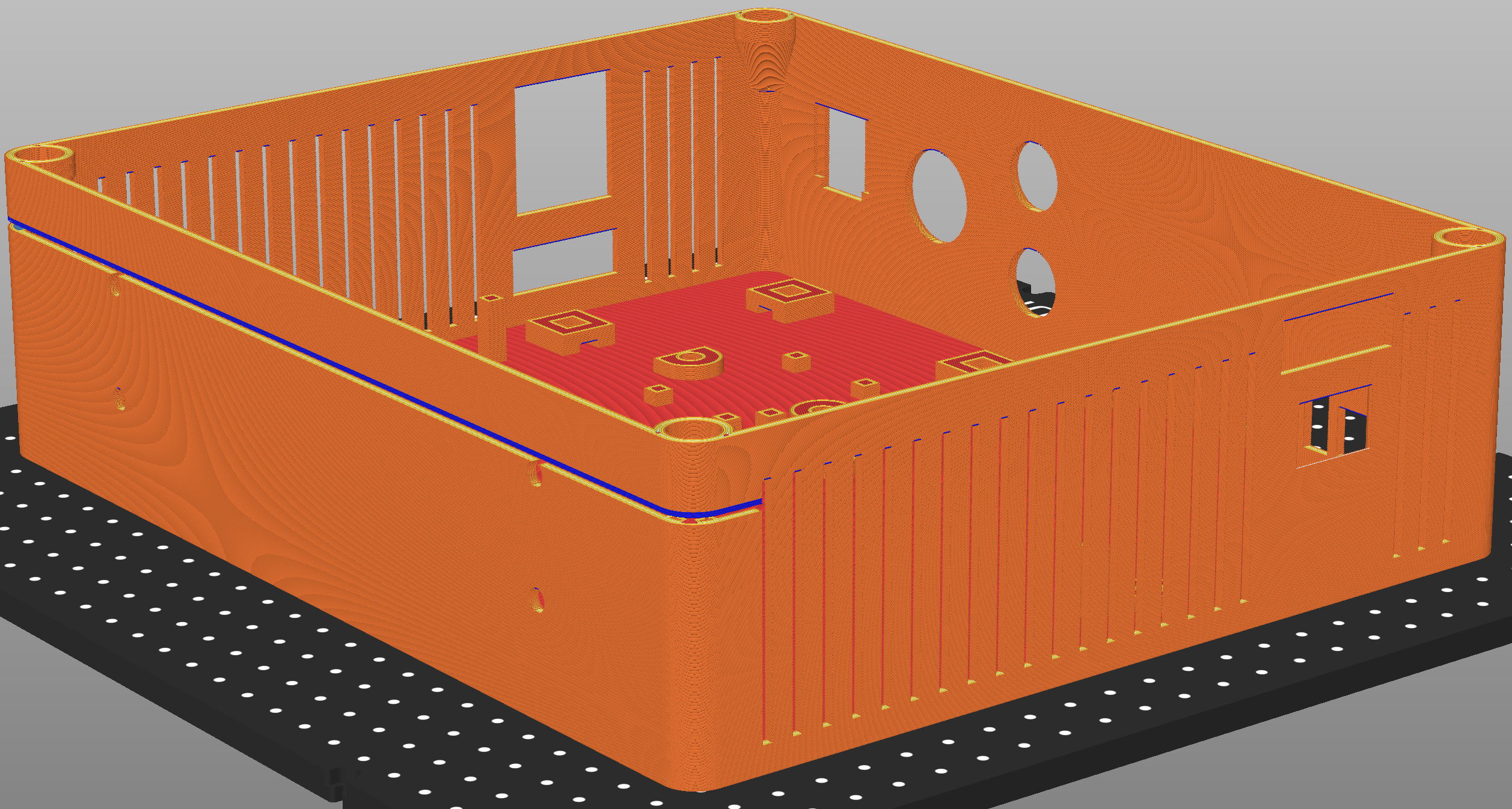} & \includegraphics[width=0.2\textwidth]{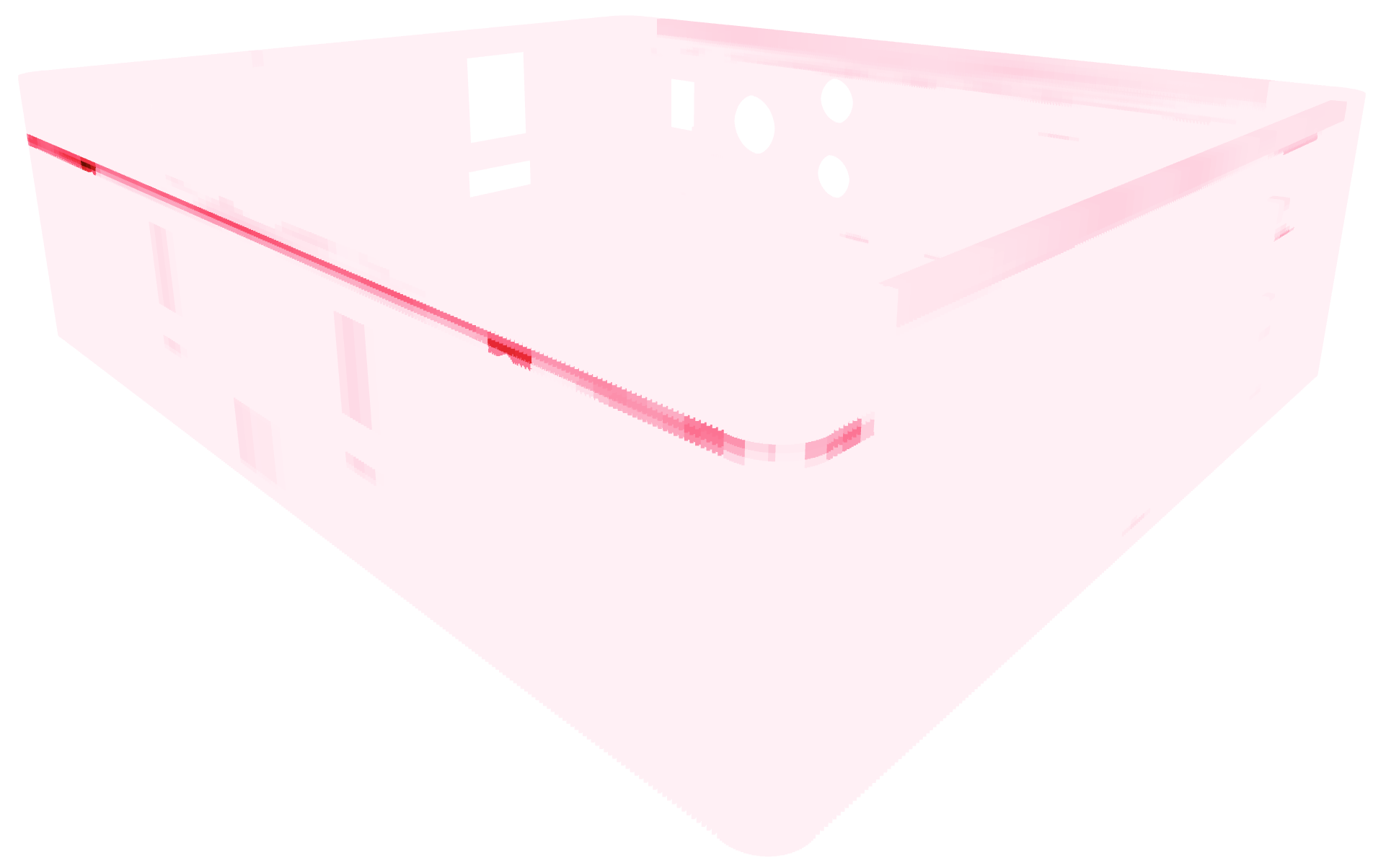} \\
\includegraphics[width=0.1\textwidth]{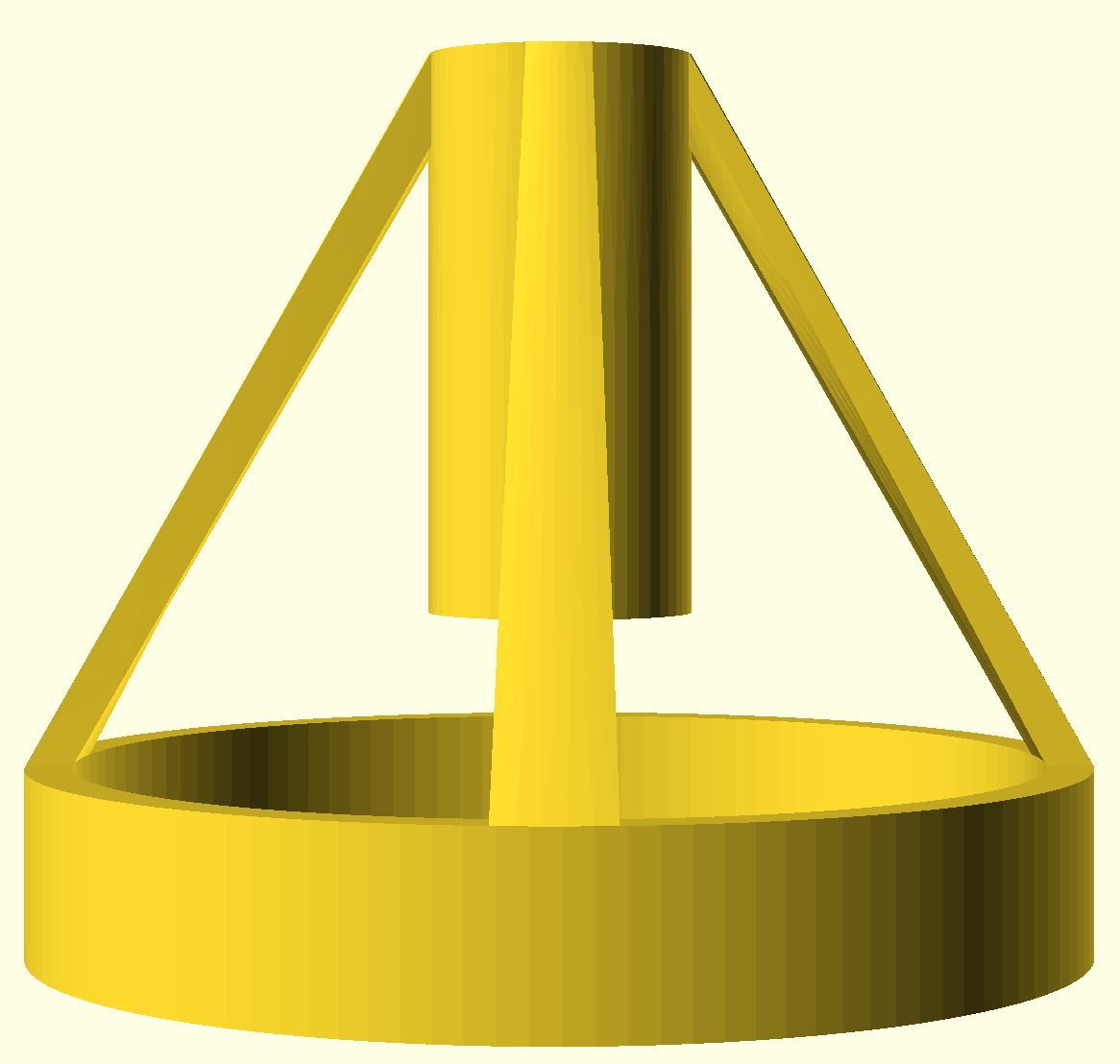} & \includegraphics[width=0.1\textwidth]{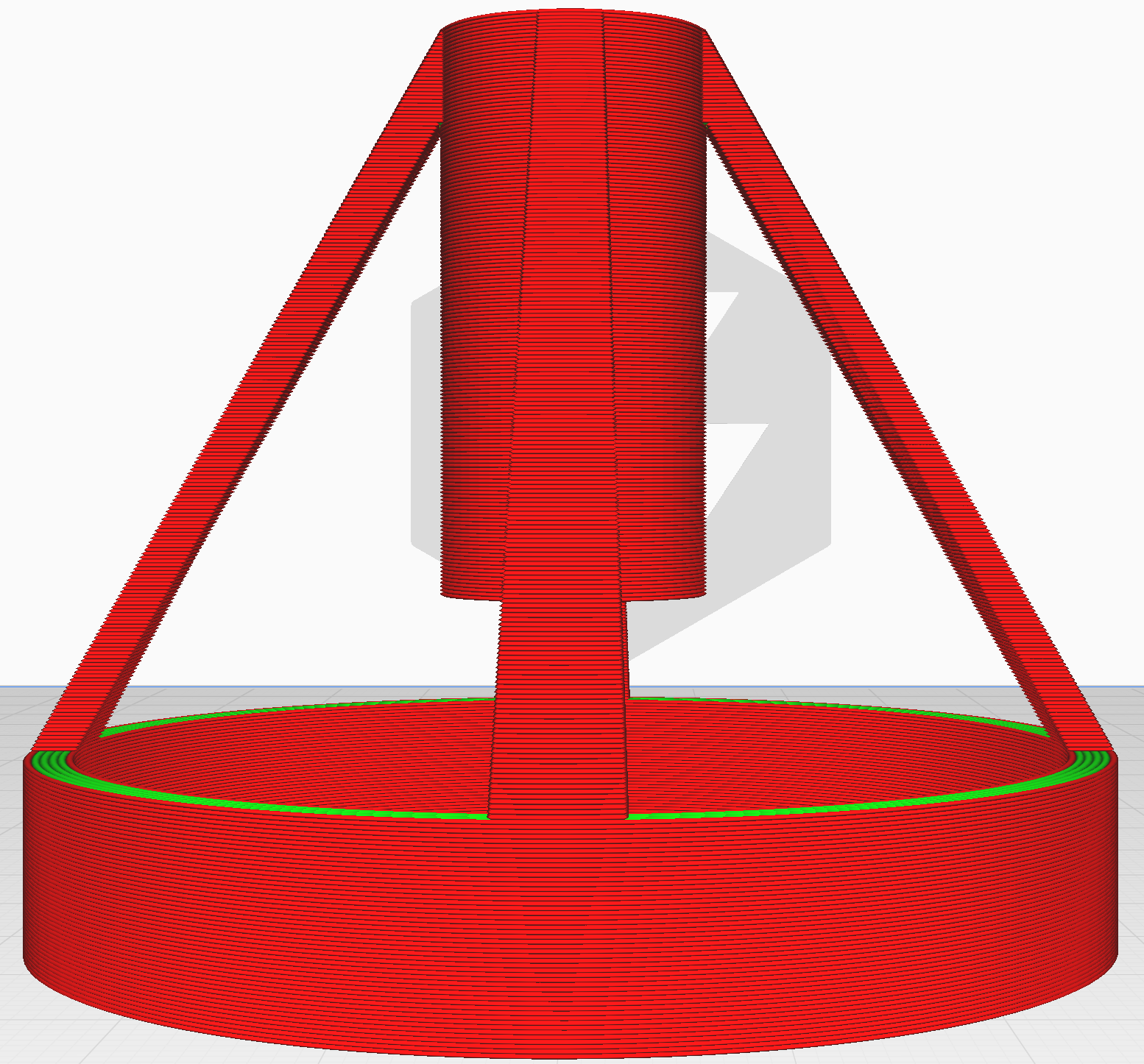} & \includegraphics[width=0.1\textwidth]{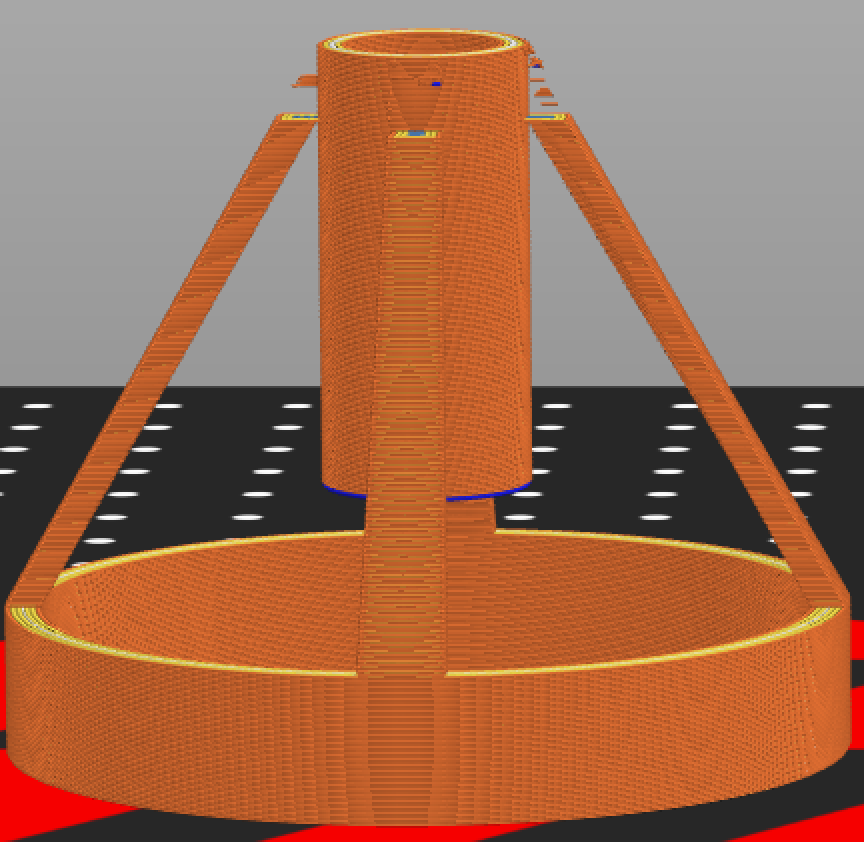} & \includegraphics[width=0.1\textwidth]{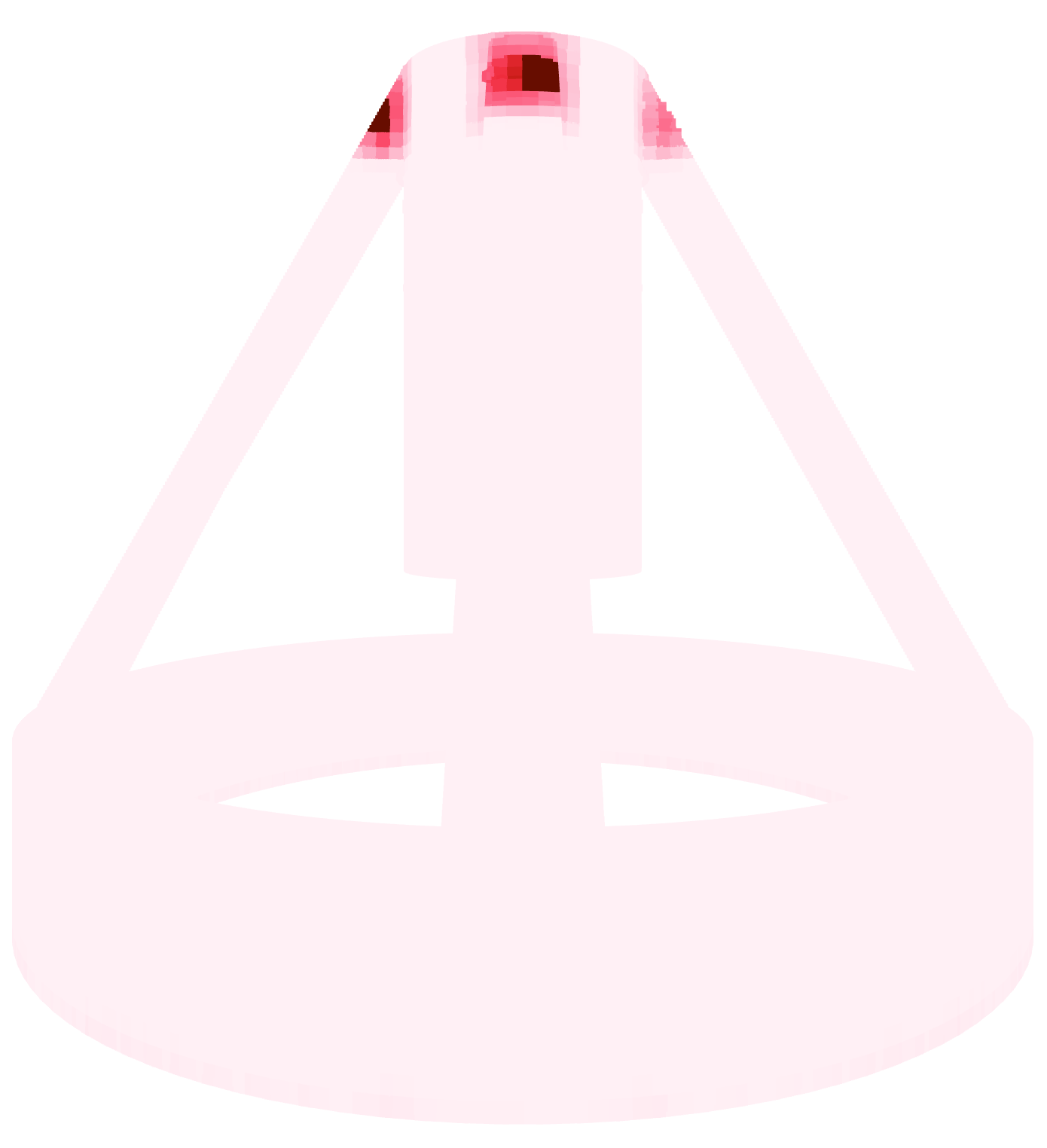} \\
\includegraphics[width=0.2\textwidth]{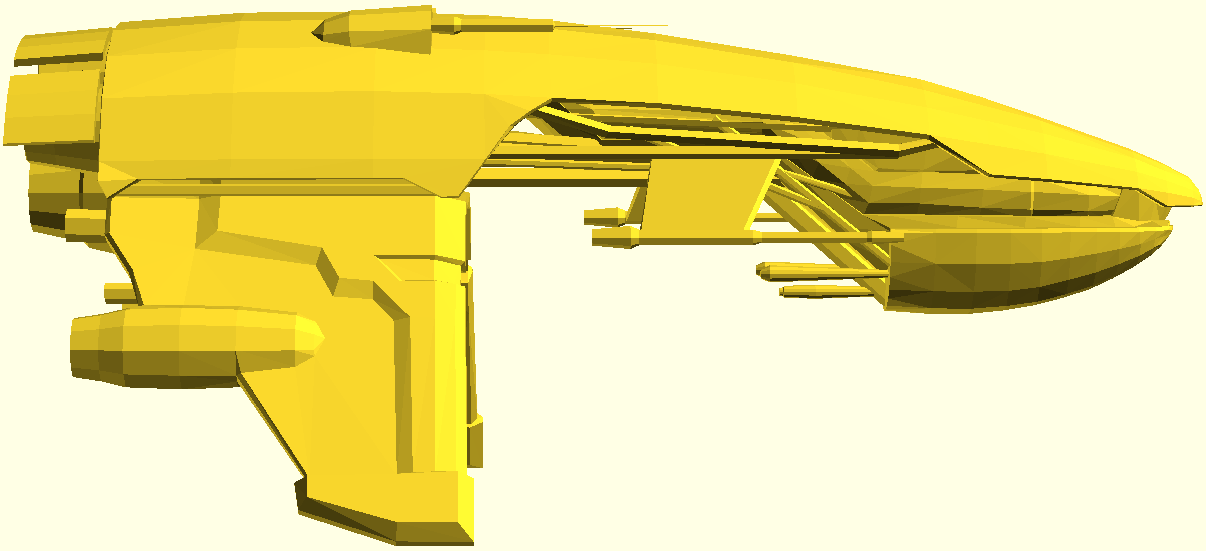} & \includegraphics[width=0.2\textwidth]{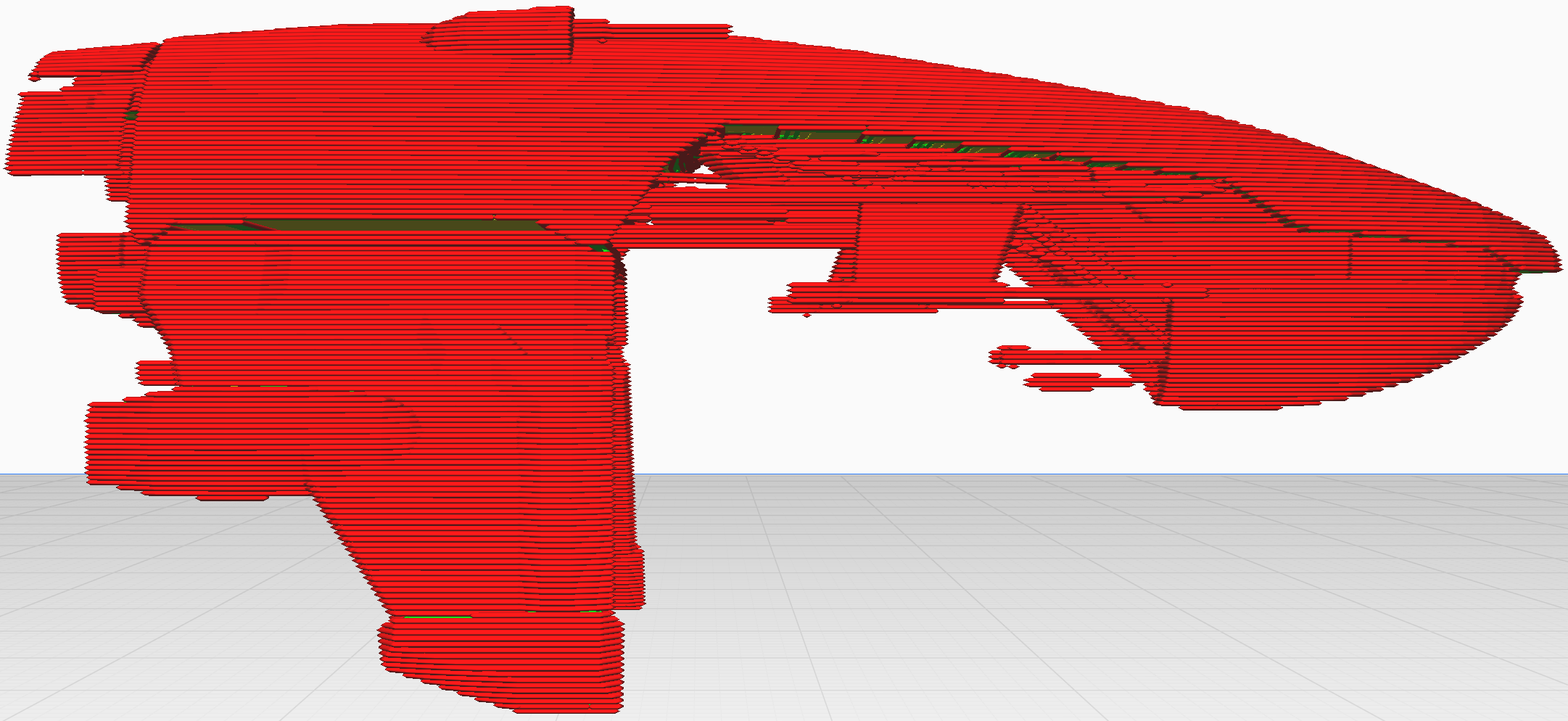} & \includegraphics[width=0.2\textwidth]{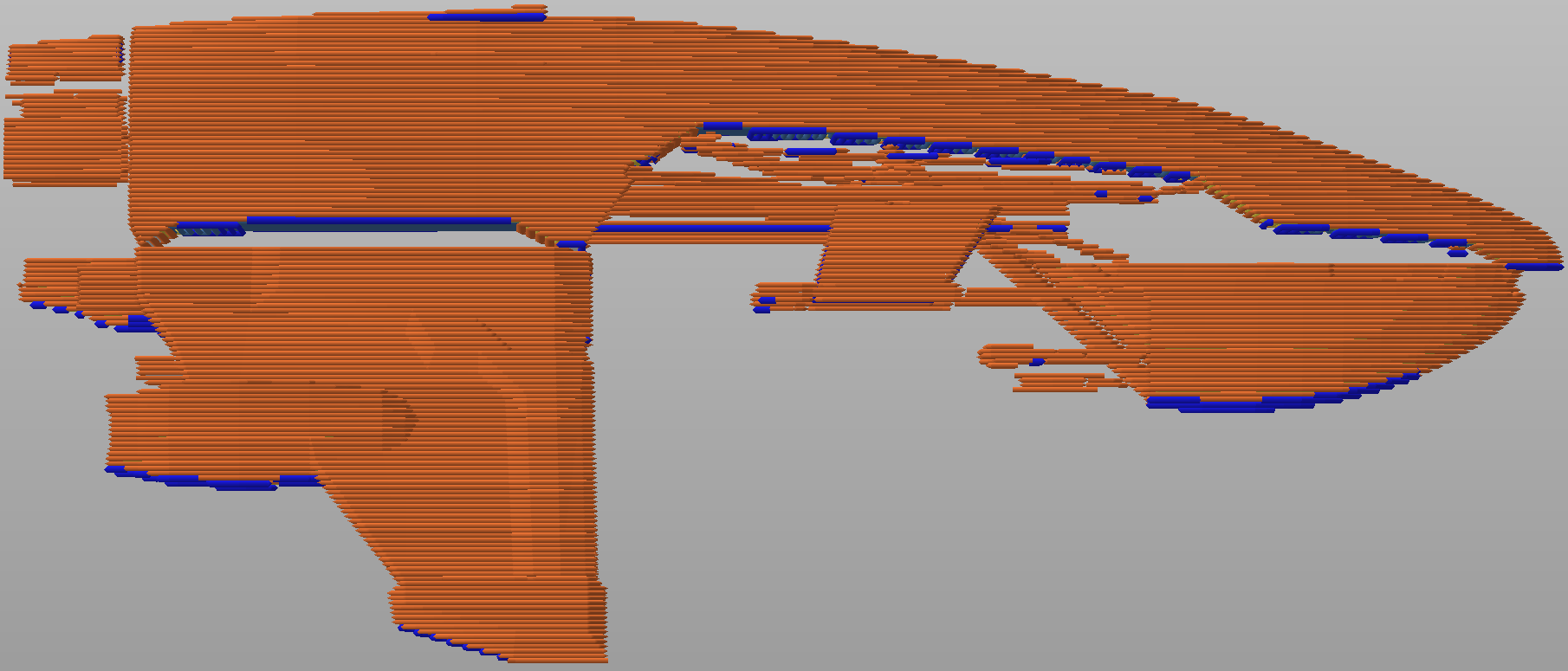} & \includegraphics[width=0.2\textwidth]{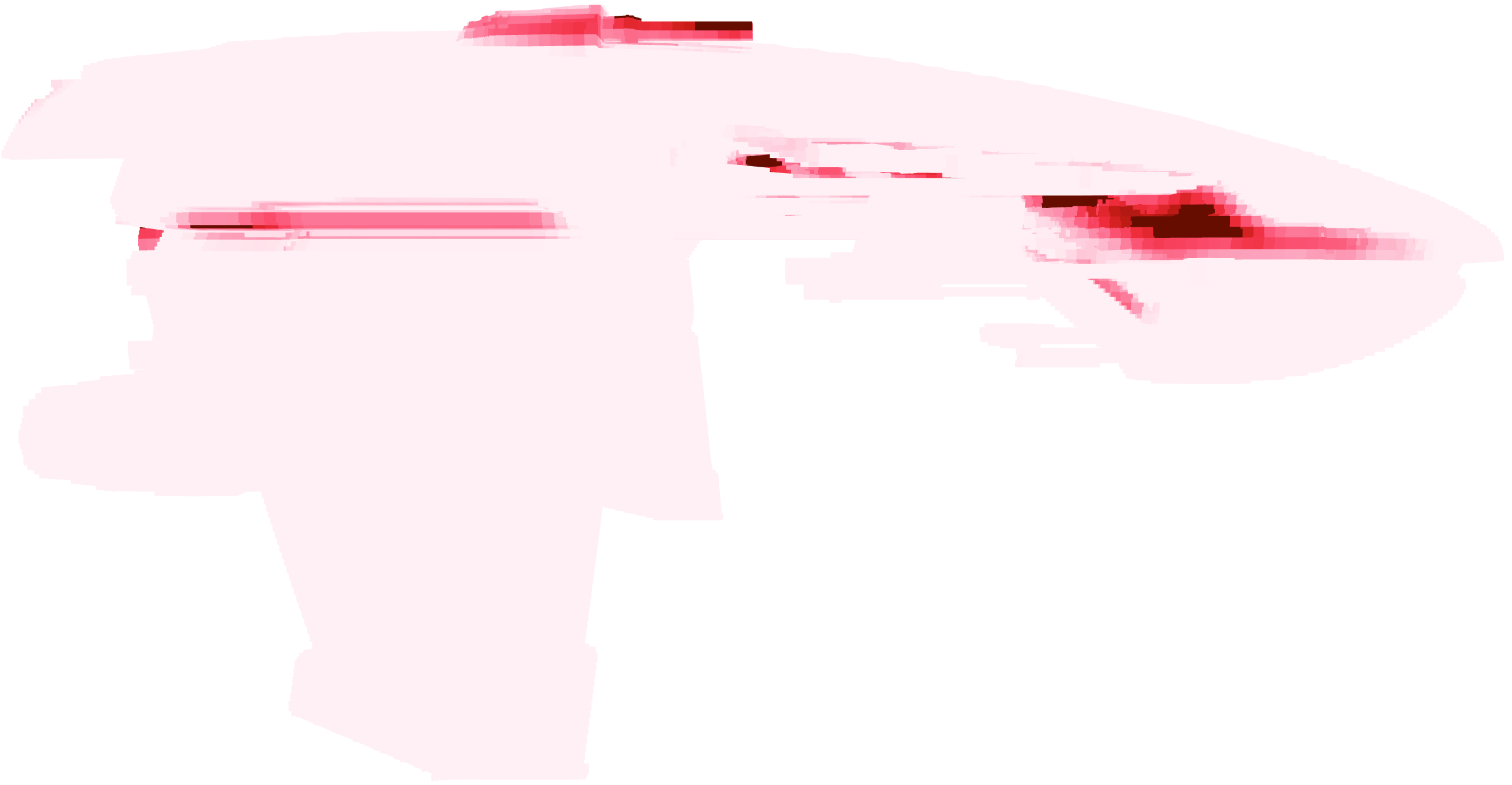} \\
\includegraphics[width=0.1\textwidth]{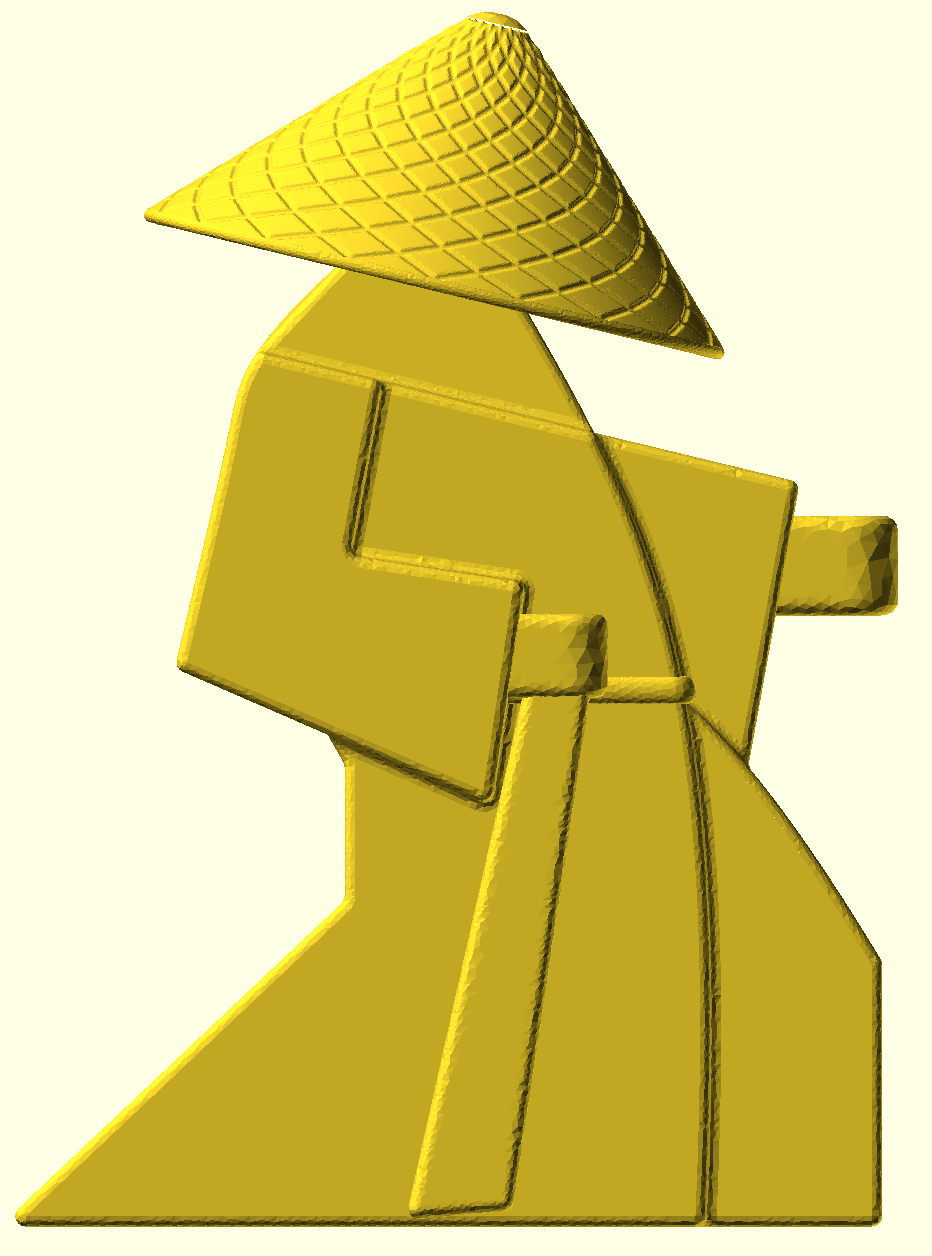} & 
\includegraphics[width=0.1\textwidth]{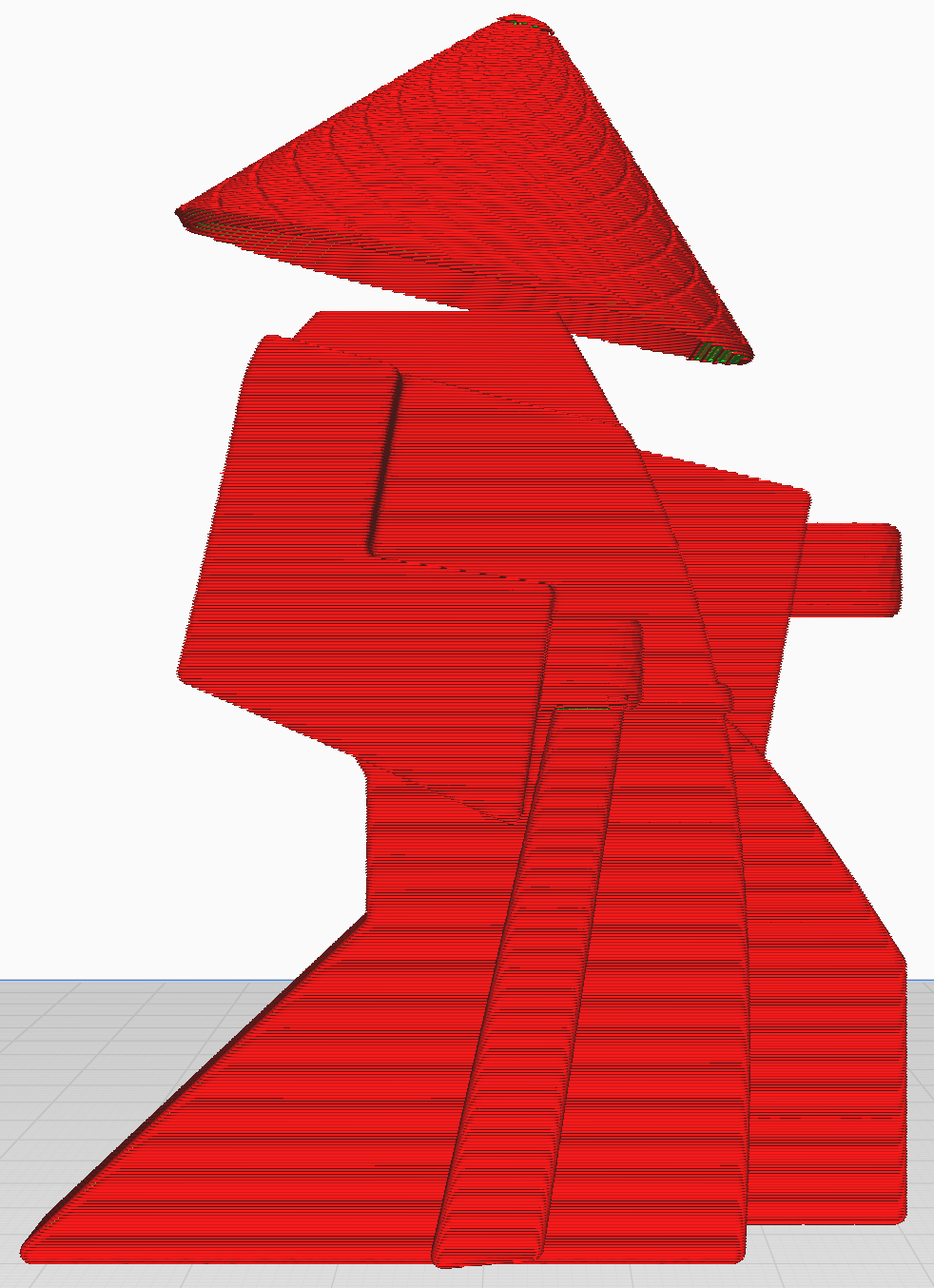} & \includegraphics[width=0.1\textwidth]{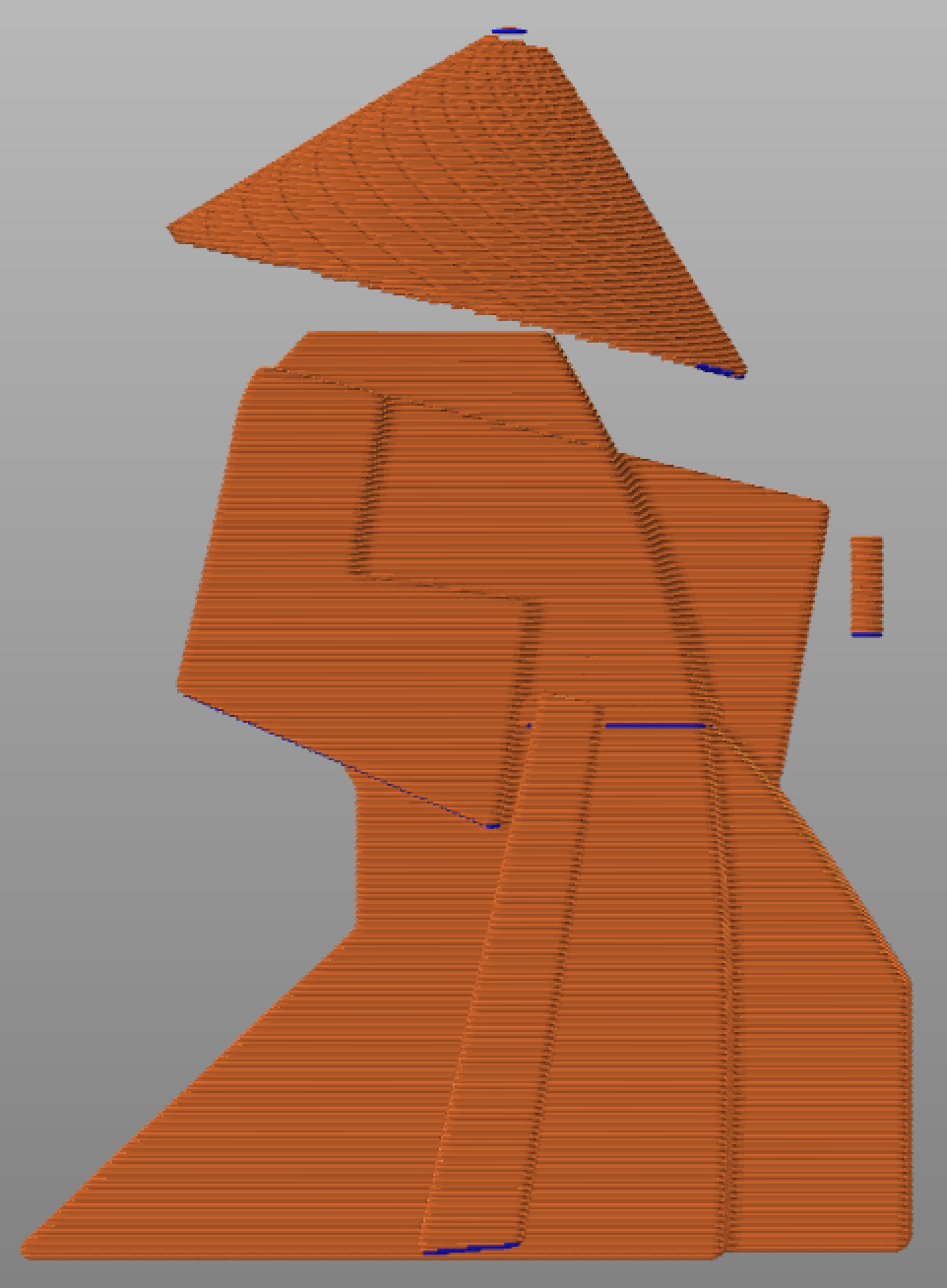} & \includegraphics[width=0.1\textwidth]{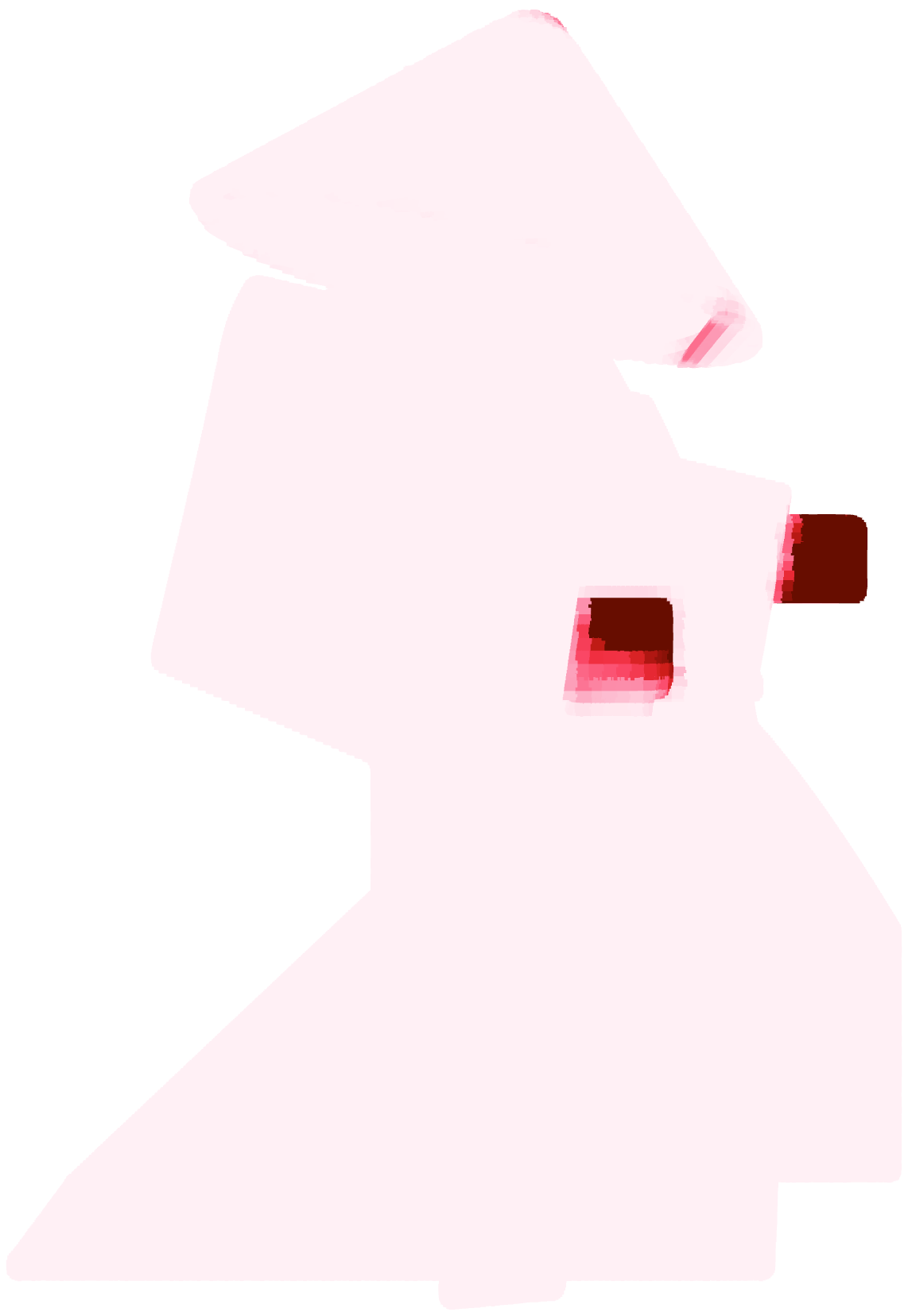} \\
\includegraphics[width=0.2\textwidth]{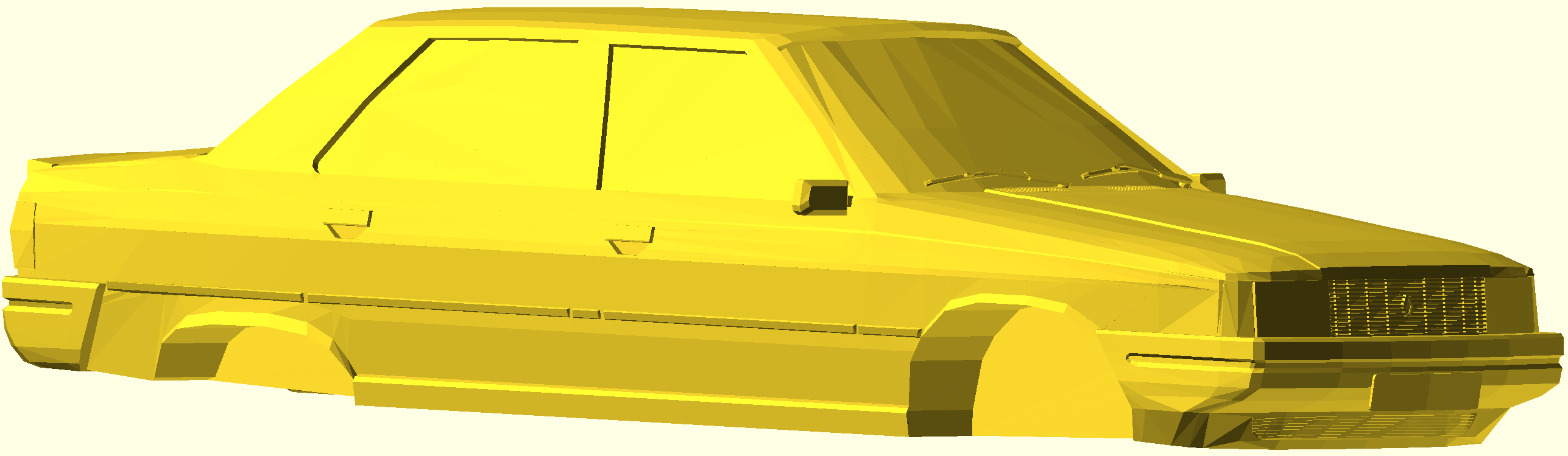} & \includegraphics[width=0.2\textwidth]{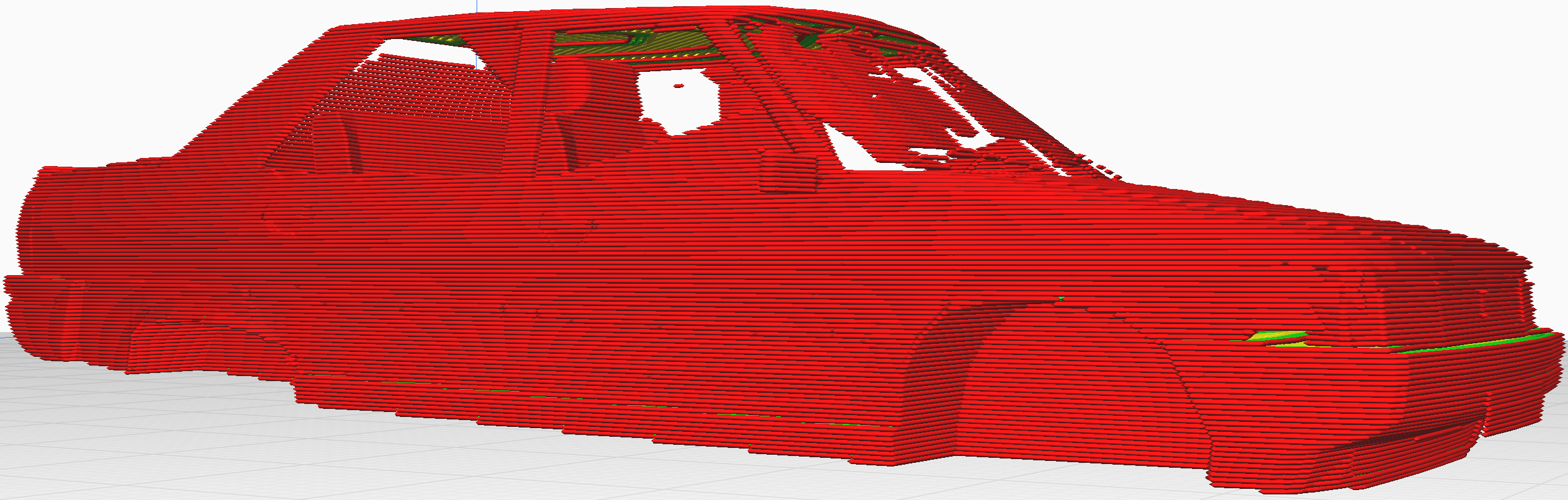} & \includegraphics[width=0.2\textwidth]{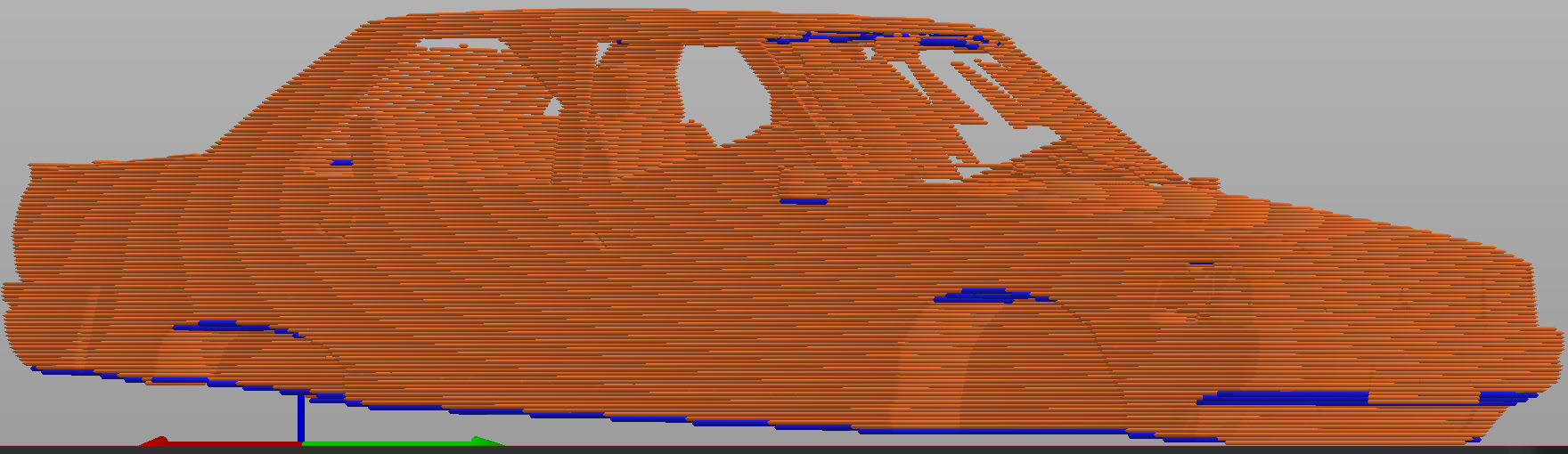} & \includegraphics[width=0.2\textwidth]{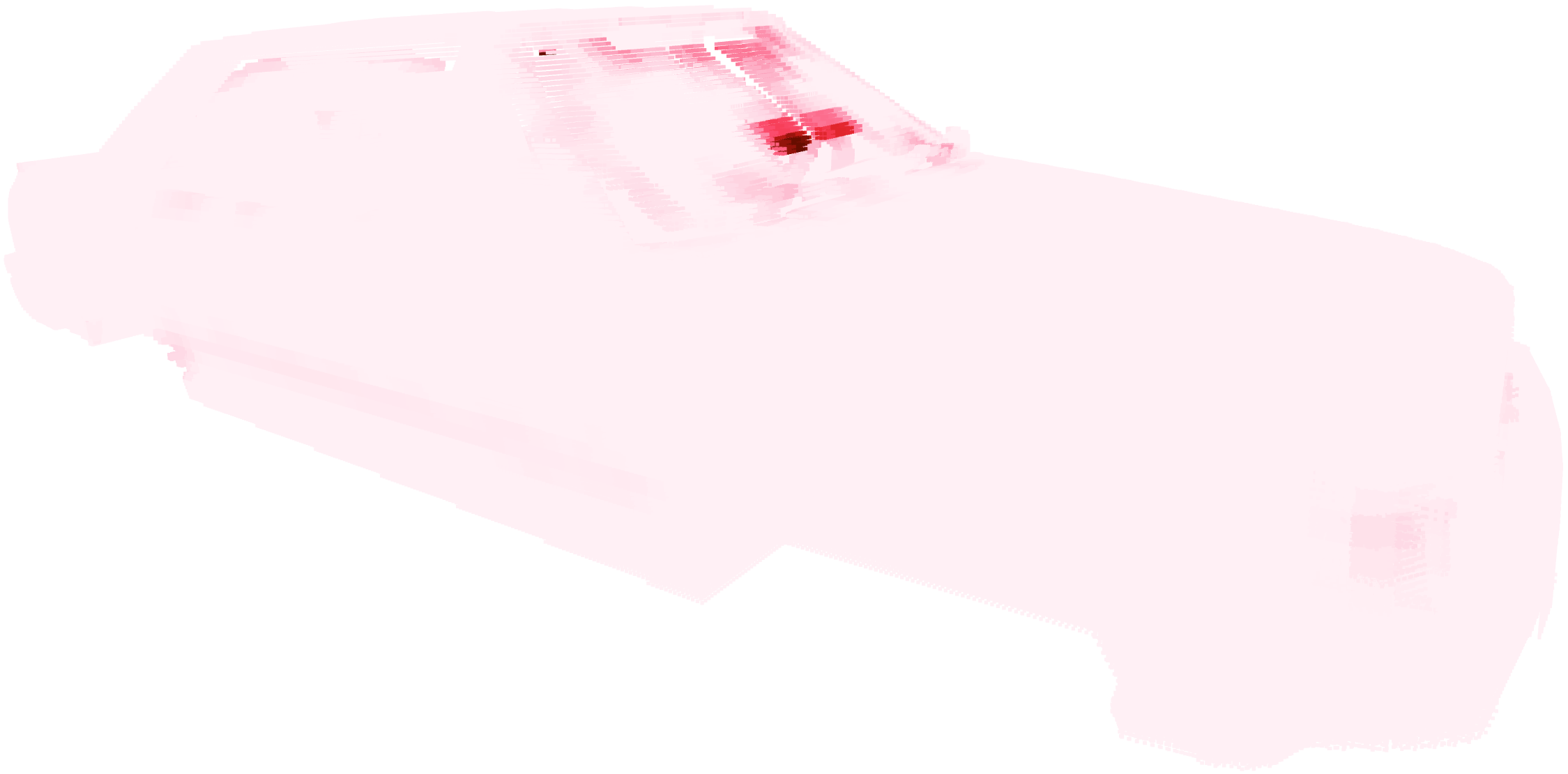}
\end{tabular}
\label{tab:cura-better-than-prusa}
\end{table}

\begin{table}[htbp]
\footnotesize
\centering
\caption{
Comparison of slicing results between Cura and PrusaSlicer for five representative models
  where PrusaSlicer outperforms Cura. 
The leftmost column shows each model's geometry in OpenSCAD, 
  followed by G-code visualizations from PrusaSlicer (middle-left) 
  and Cura (middle-right), 
  with heatmaps in the rightmost column. 
PrusaSlicer achieves completely correct slicing for only the first and last models, 
  while producing more accurate results than Cura for the remaining three cases.
}
\begin{tabular}{c|c|c|c}
{Model} & 
{G-code (Prusa)} & 
{G-code (Cura)} & 
{Heatmap} \\
\hline \\
\includegraphics[width=0.18\textwidth]{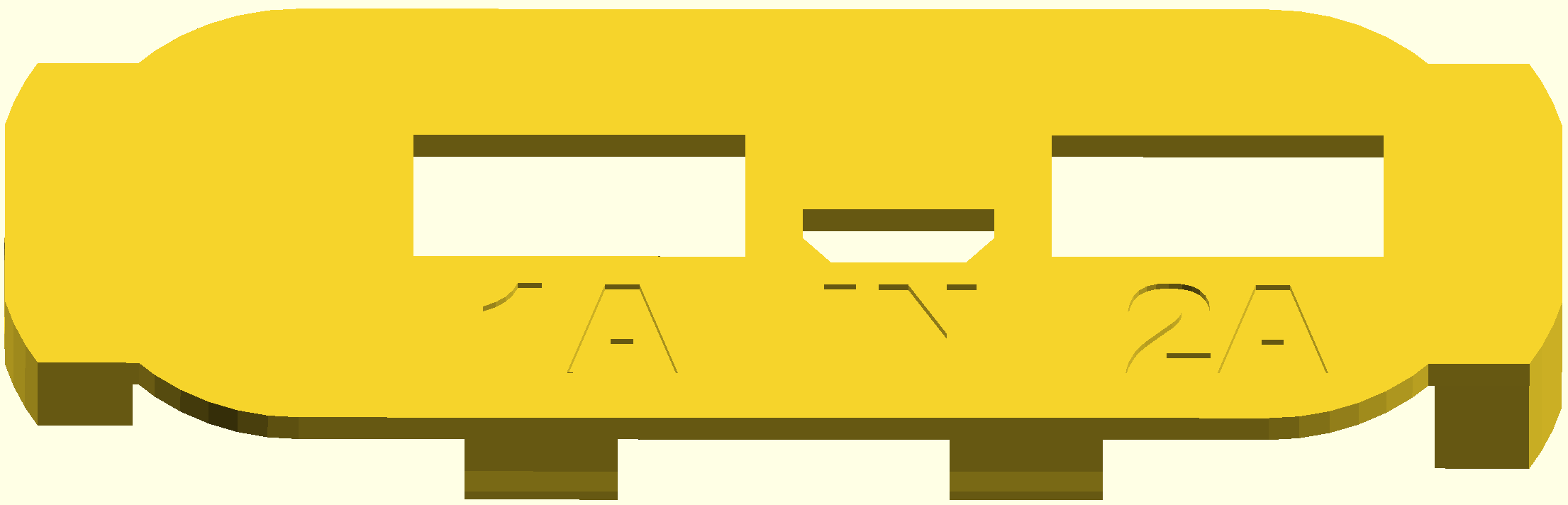} & \includegraphics[width=0.18\textwidth]{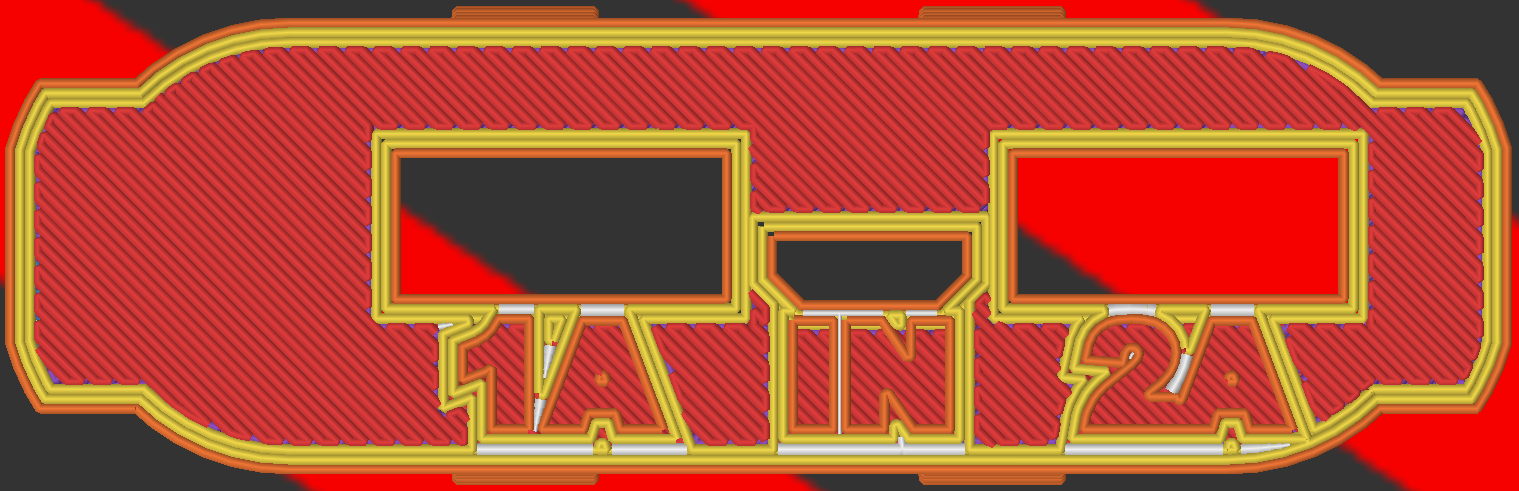} & \includegraphics[width=0.18\textwidth]{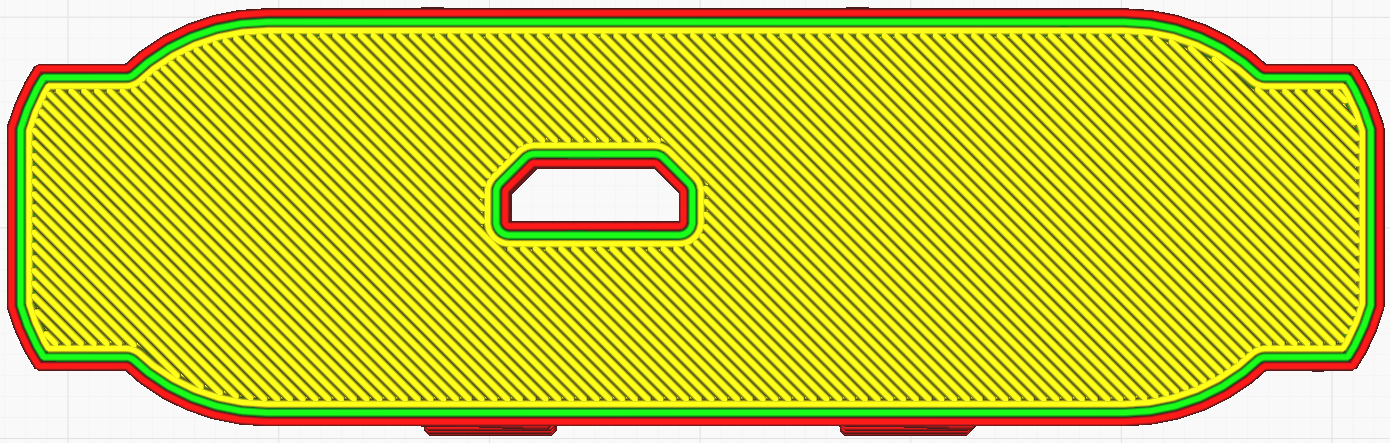} & \includegraphics[width=0.18\textwidth]{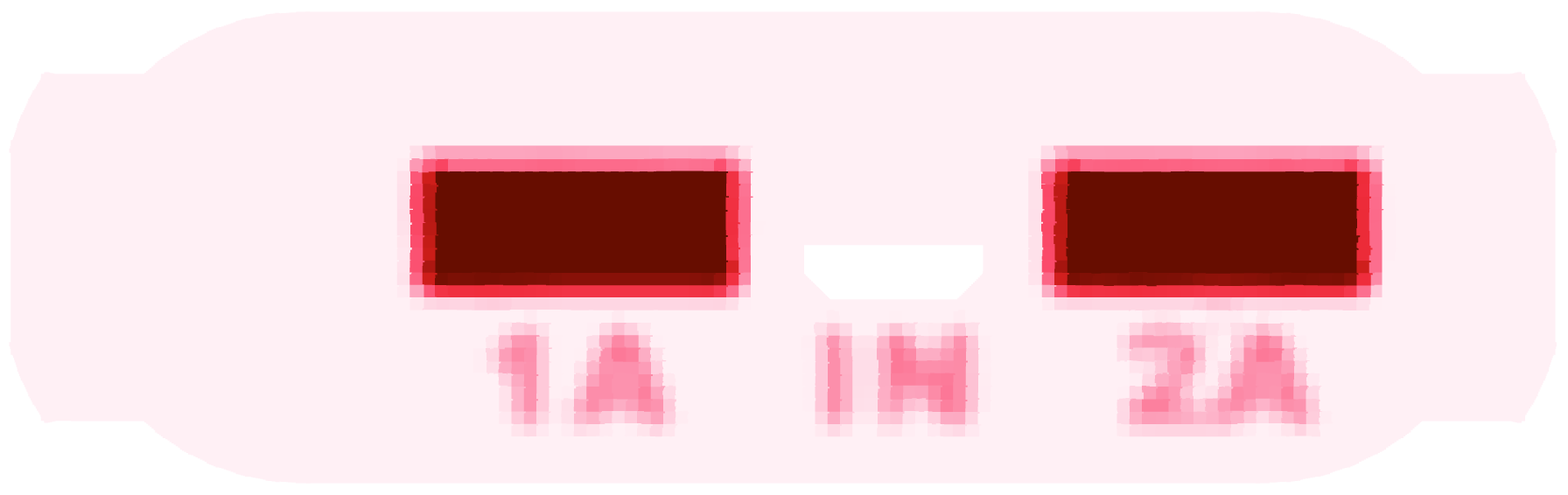} \\
\includegraphics[width=0.1\textwidth]{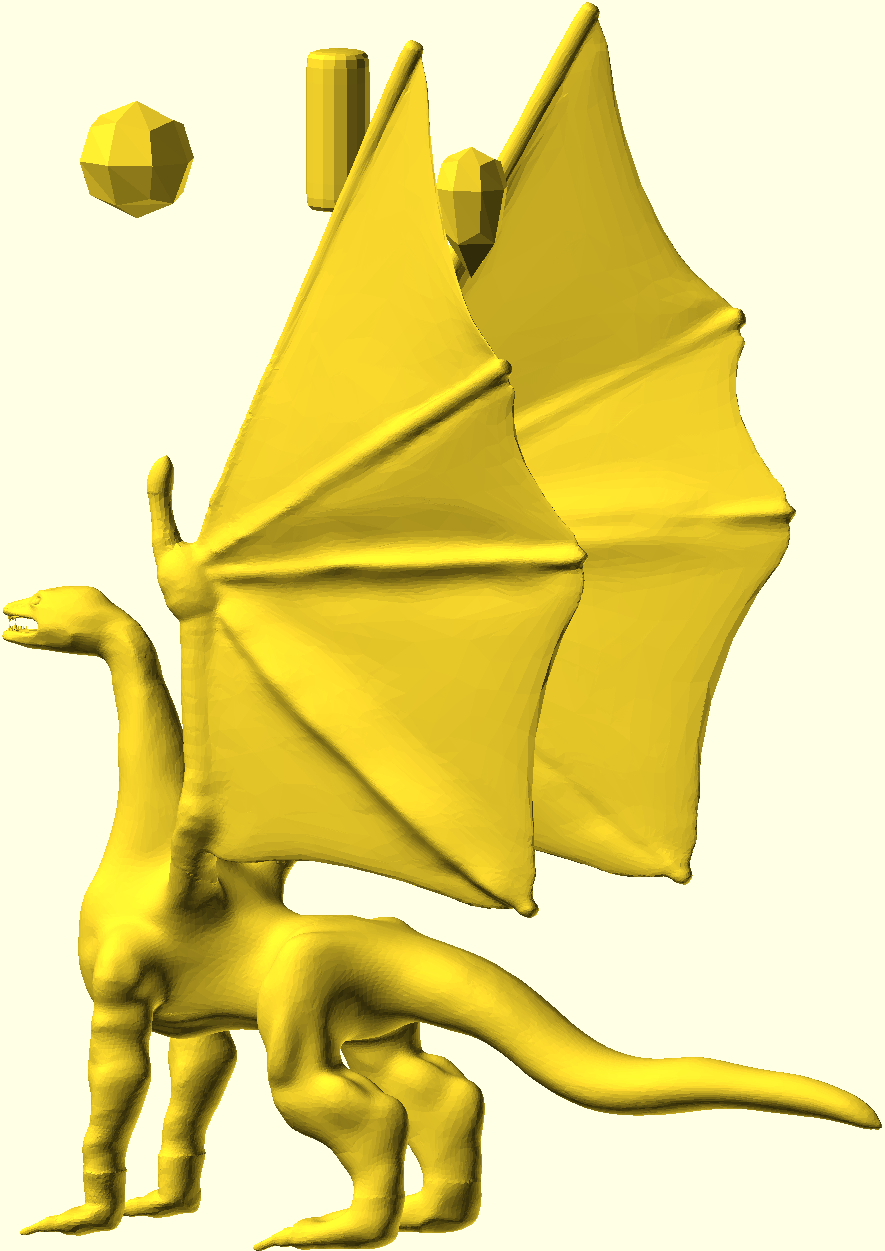} & \includegraphics[width=0.1\textwidth]{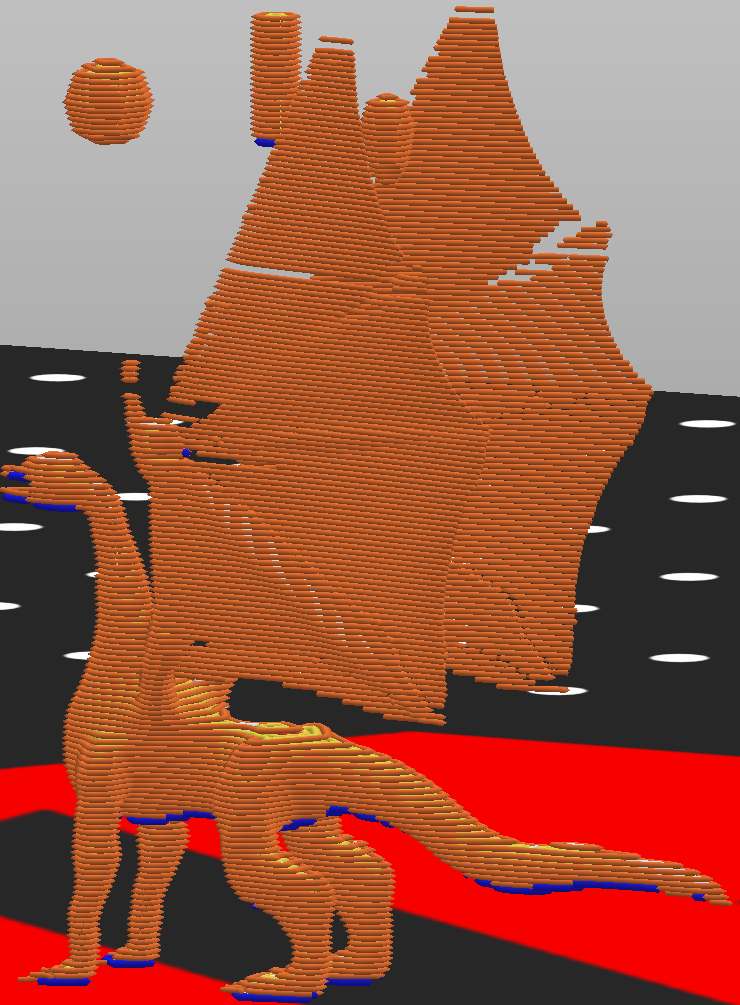} & \includegraphics[width=0.1\textwidth]{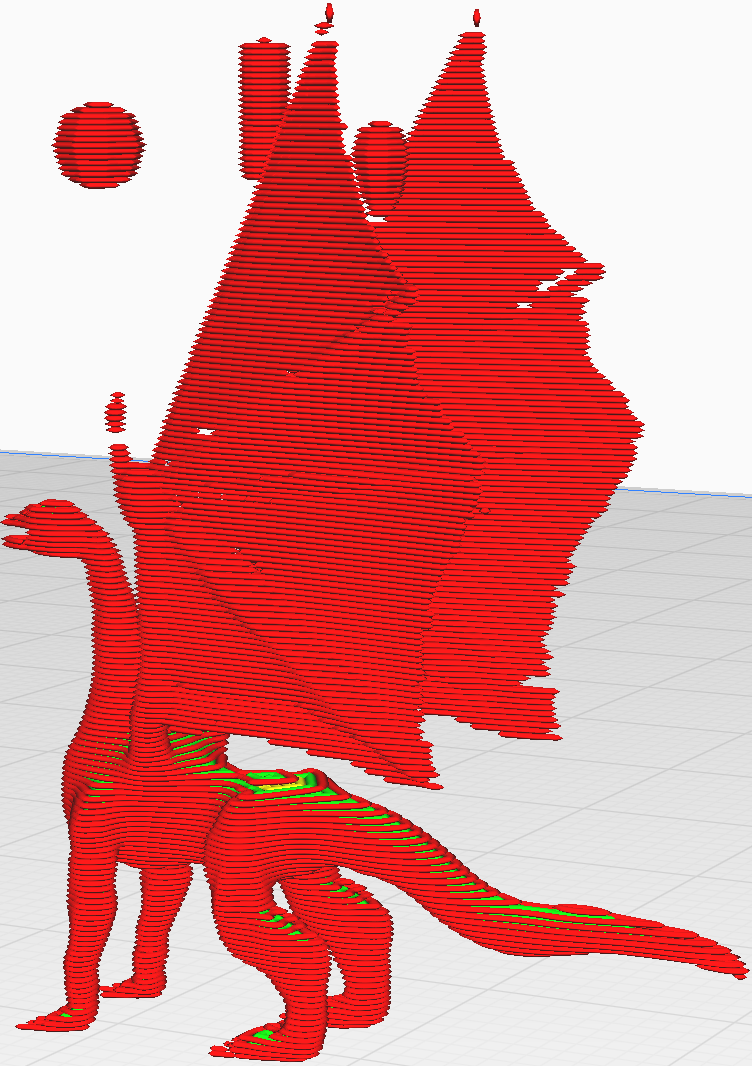} & \includegraphics[width=0.1\textwidth]{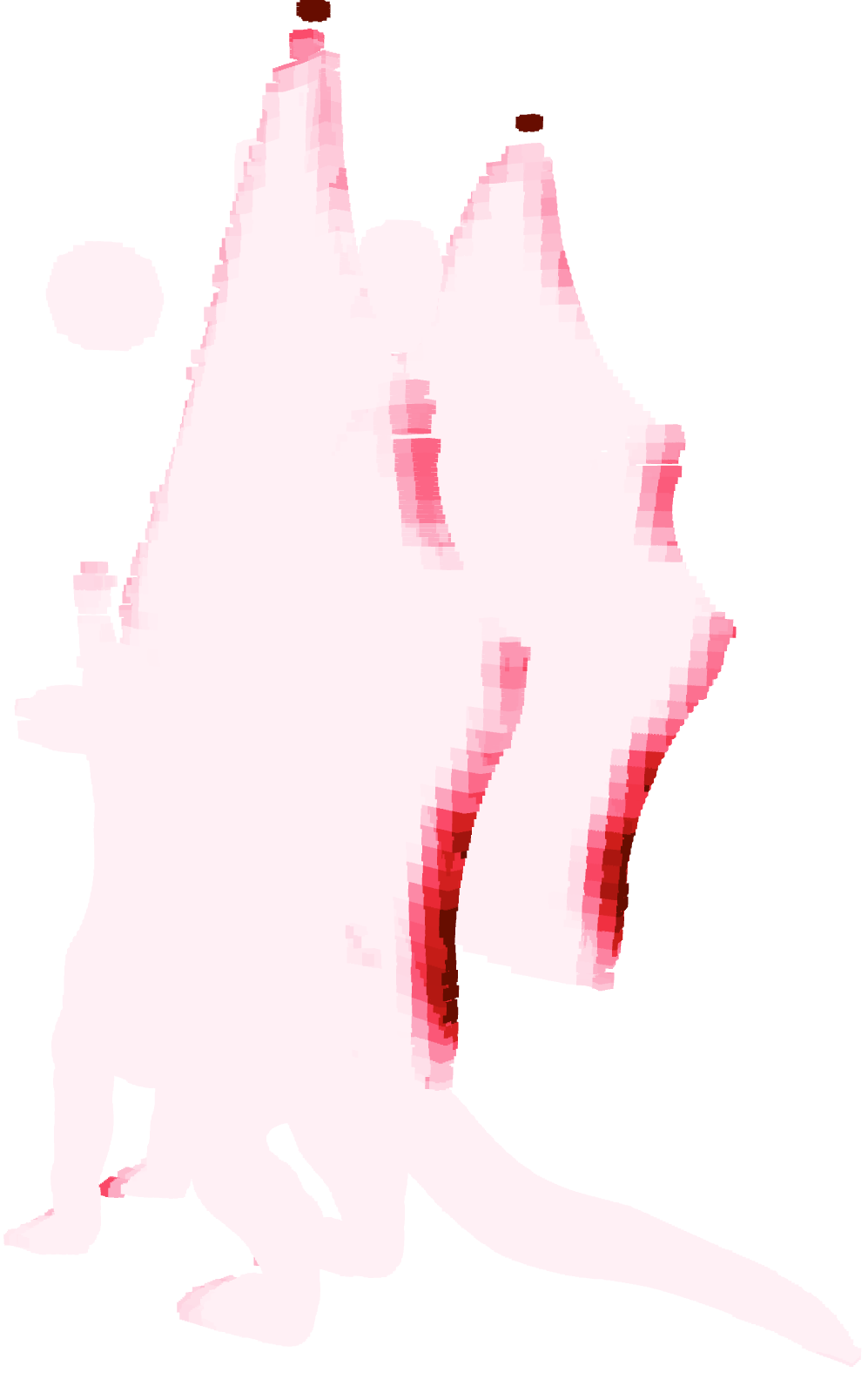} \\
\includegraphics[width=0.1\textwidth]{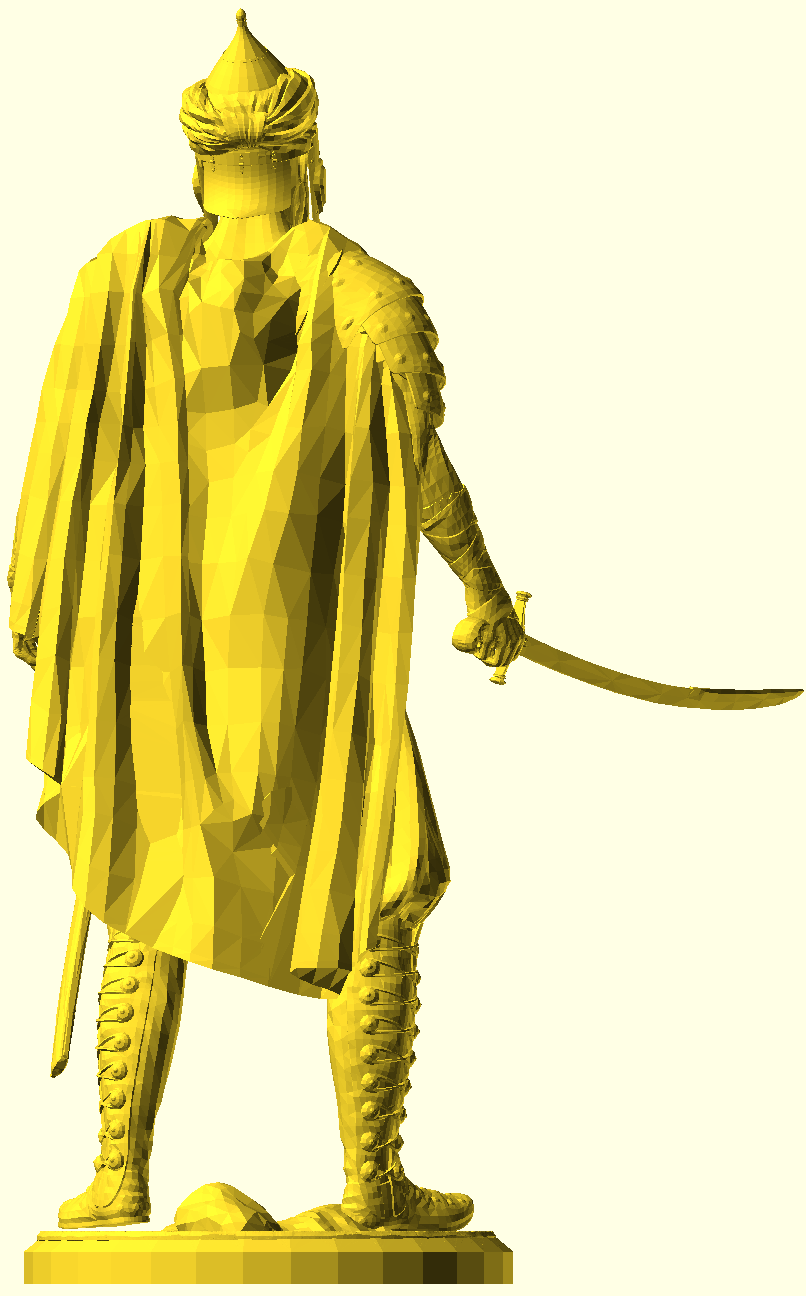} & \includegraphics[width=0.1\textwidth]{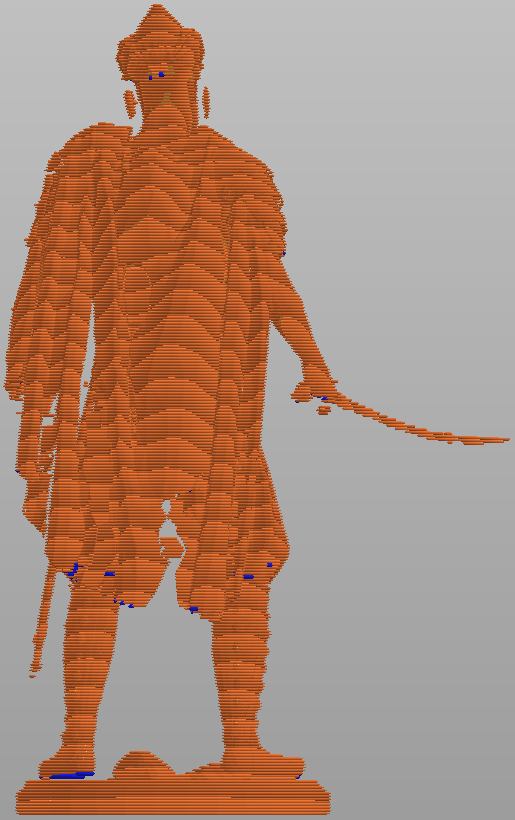} & \includegraphics[width=0.1\textwidth]{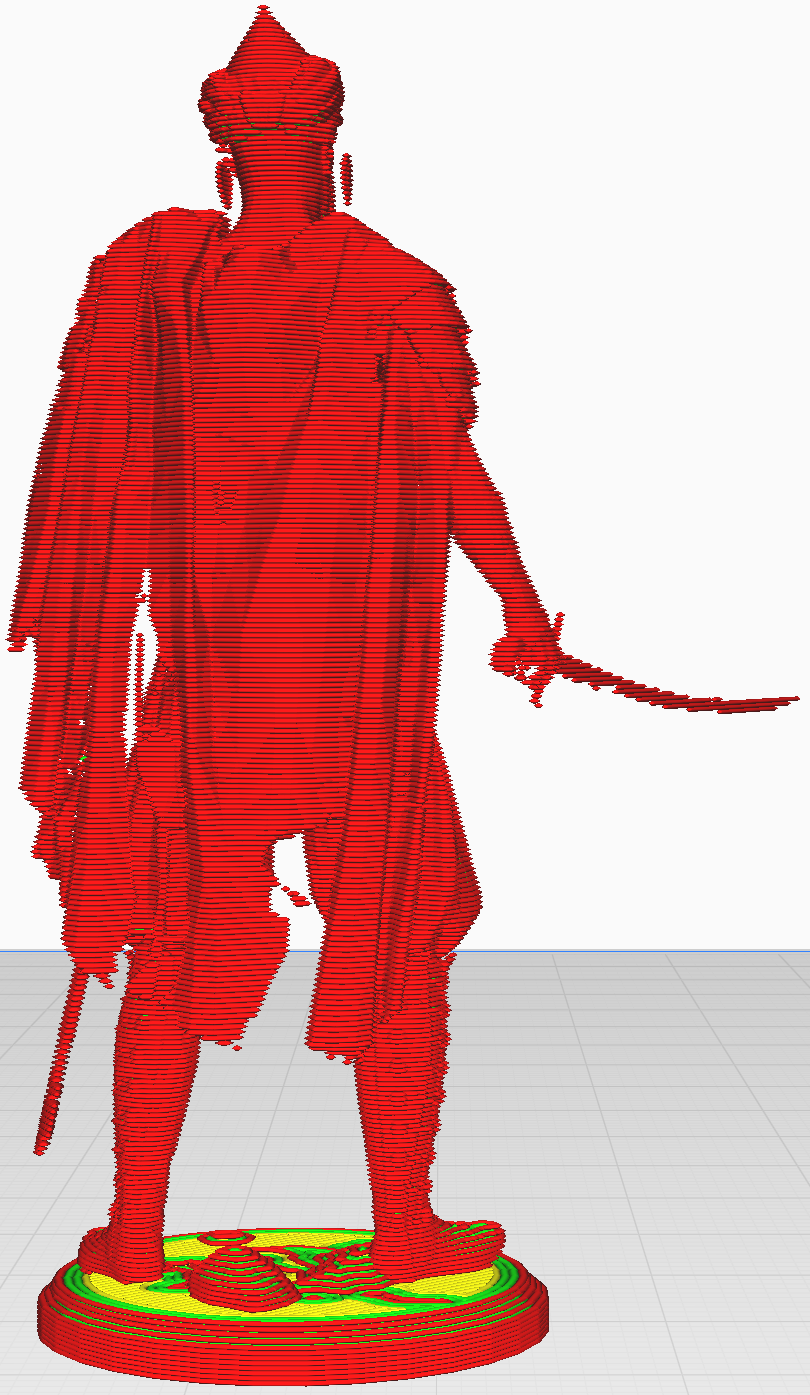} & \includegraphics[width=0.1\textwidth]{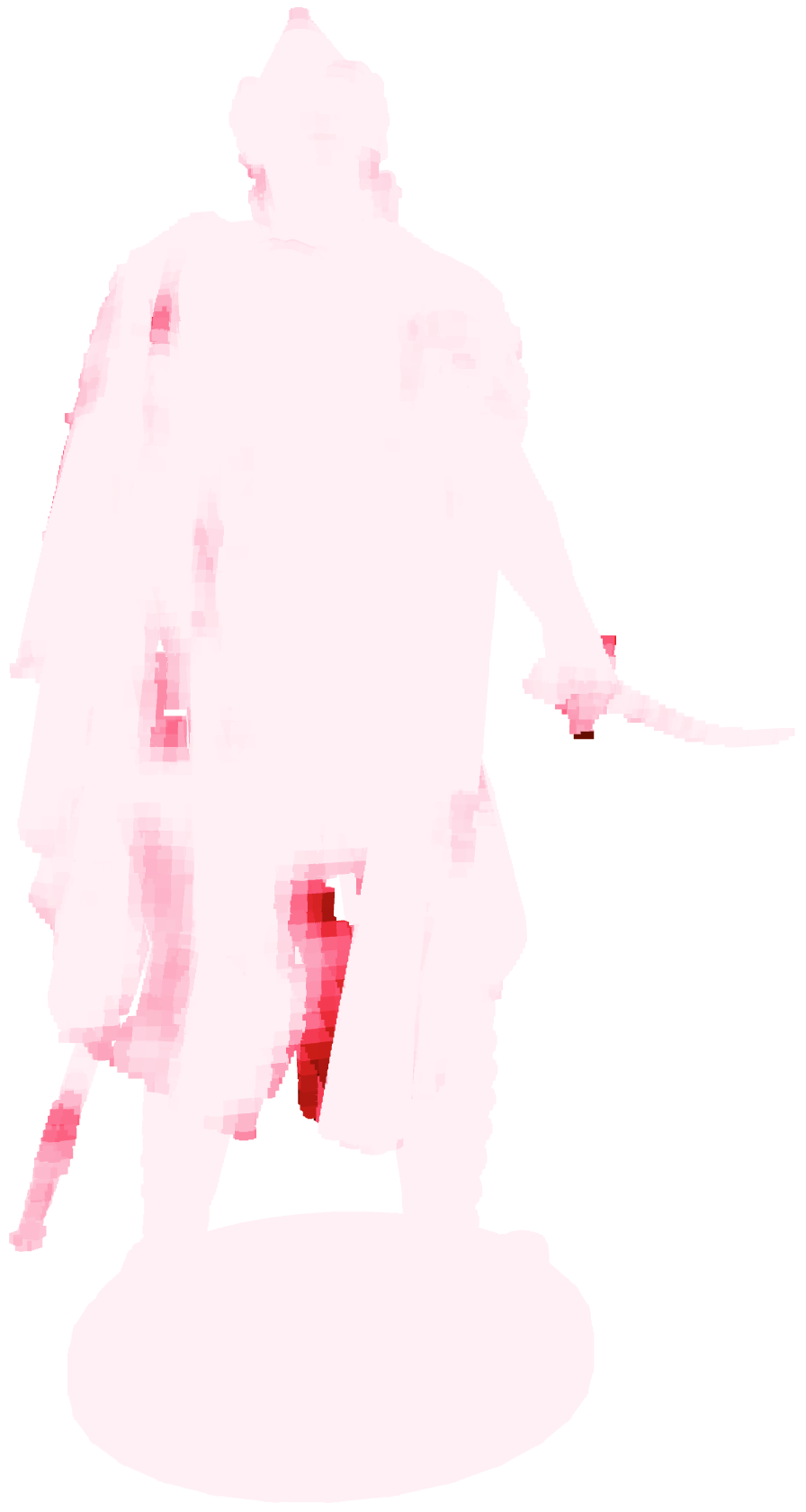} \\
\includegraphics[width=0.15\textwidth]{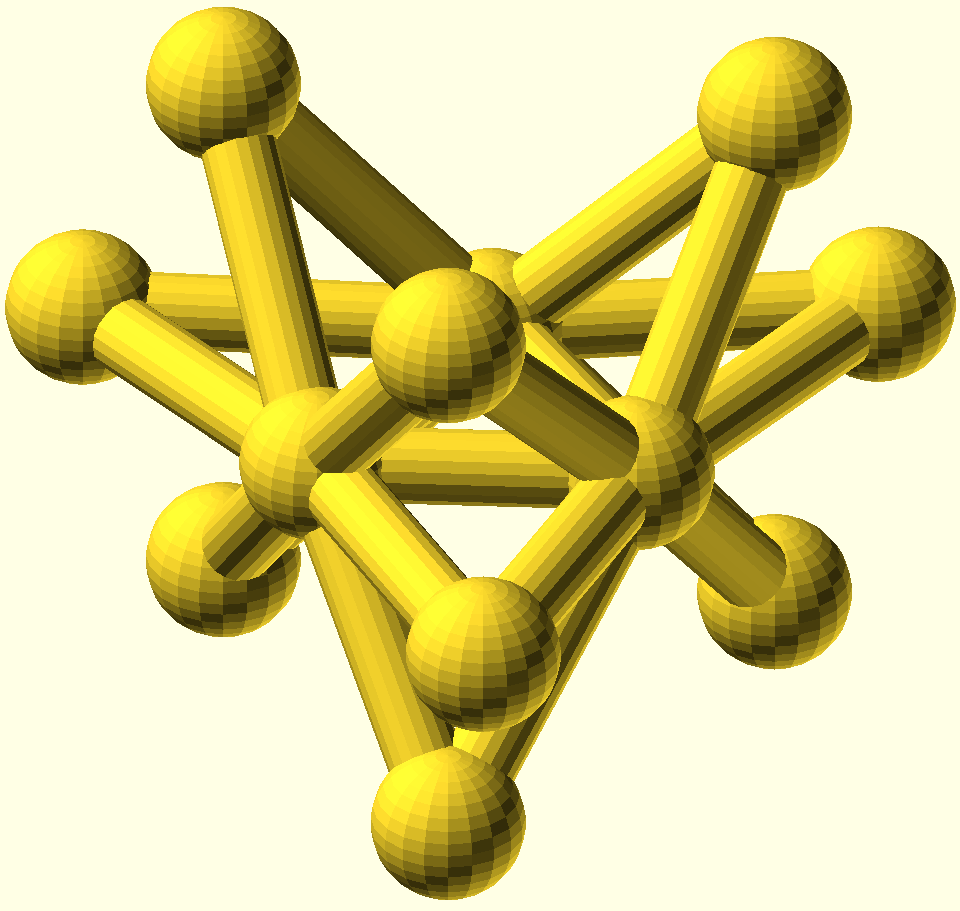} & \includegraphics[width=0.14\textwidth]{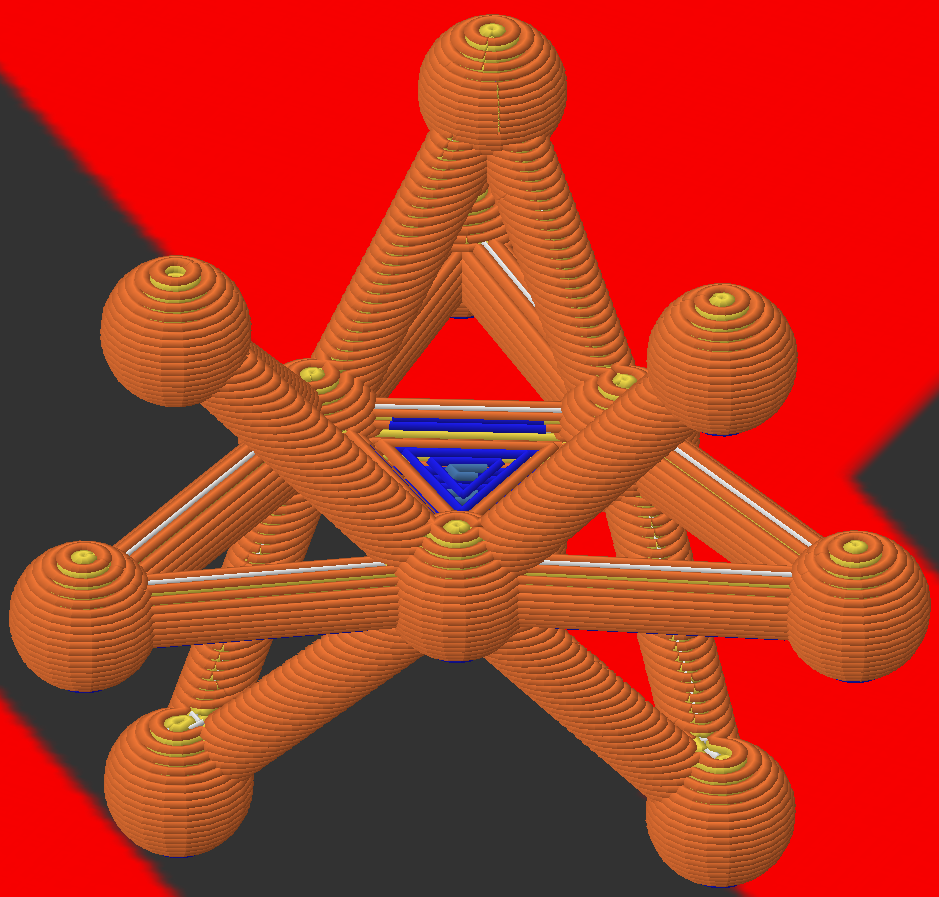} & \includegraphics[width=0.16\textwidth]{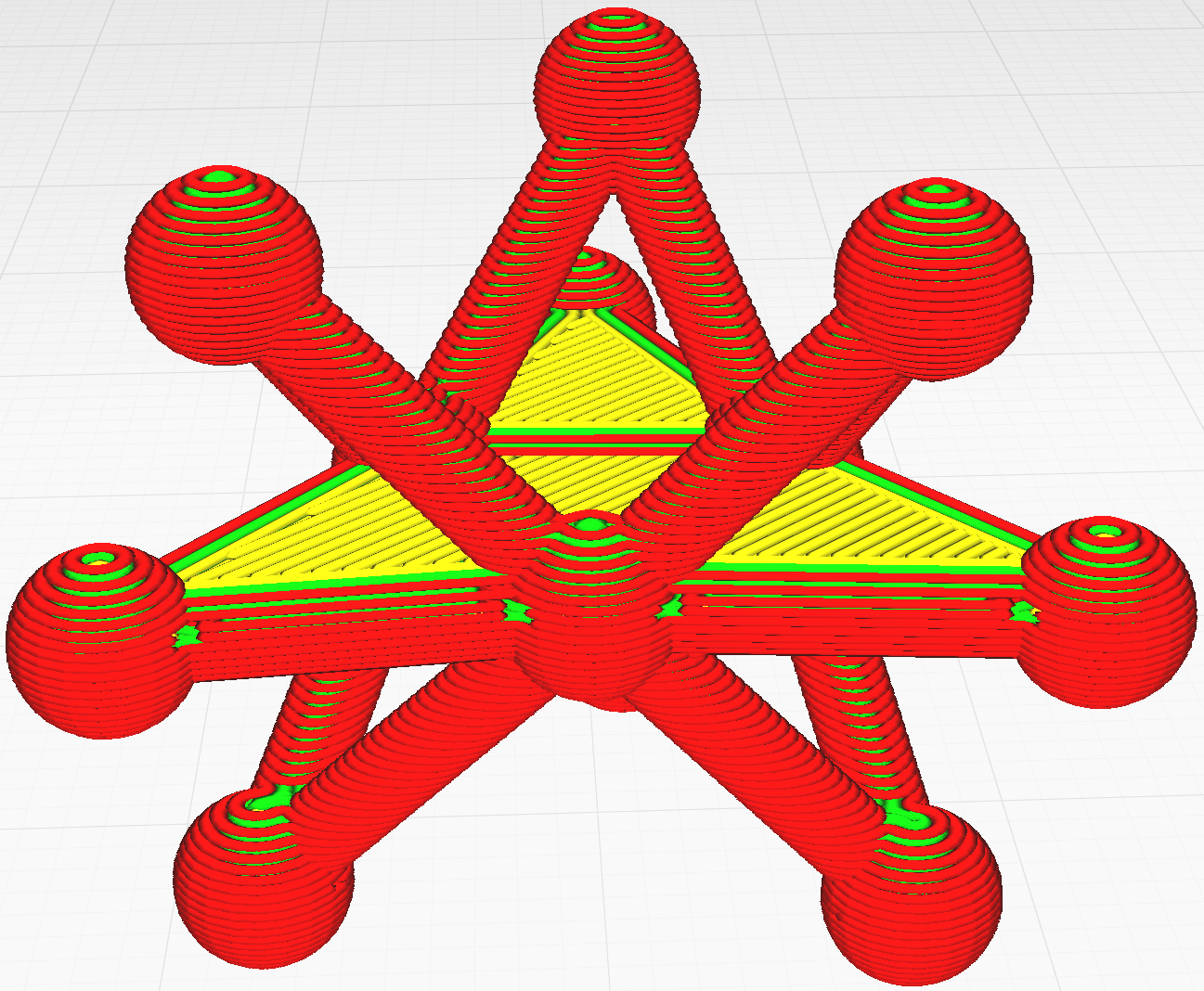} & \includegraphics[width=0.2\textwidth]{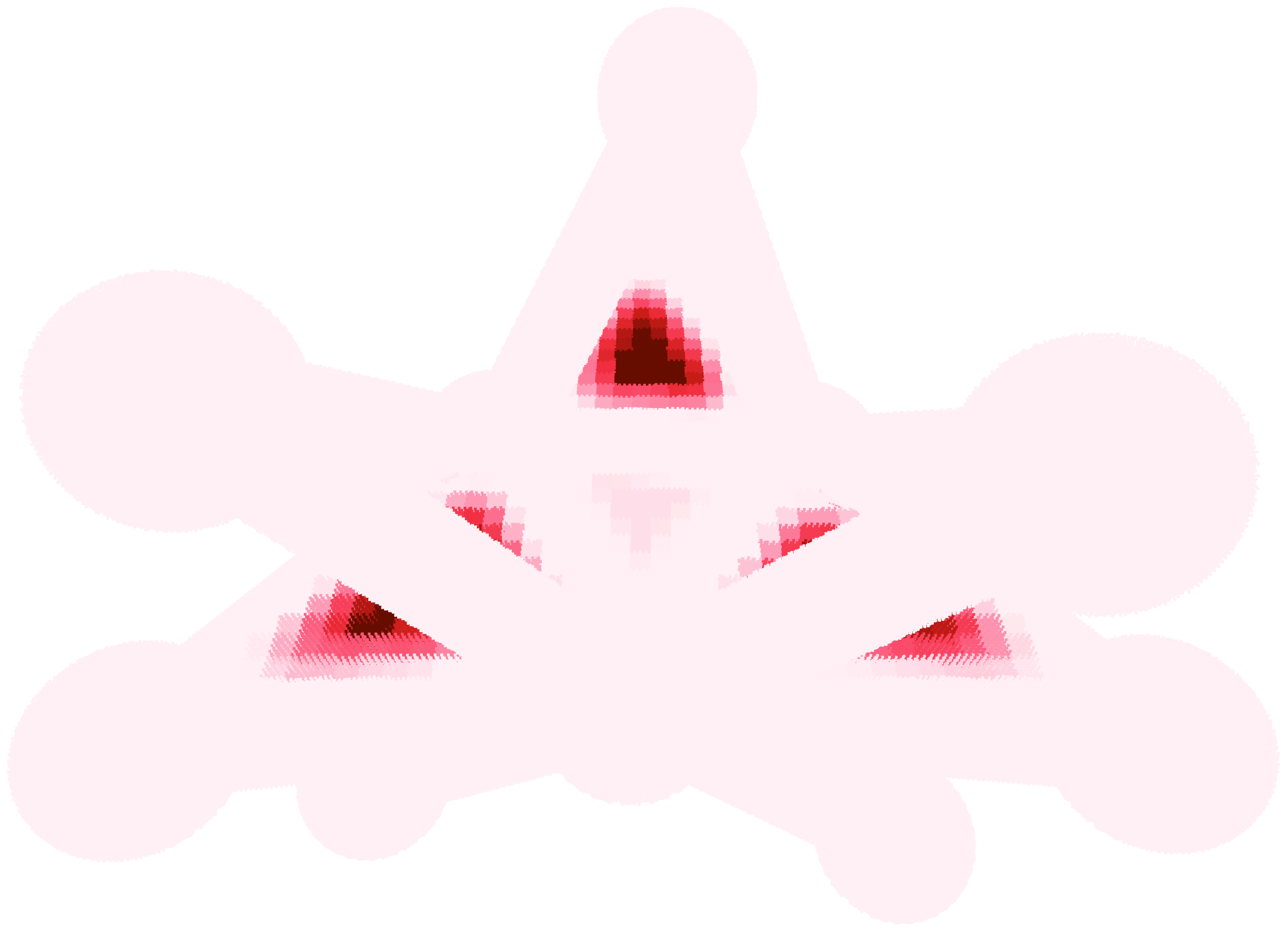} \\
\includegraphics[width=0.15\textwidth]{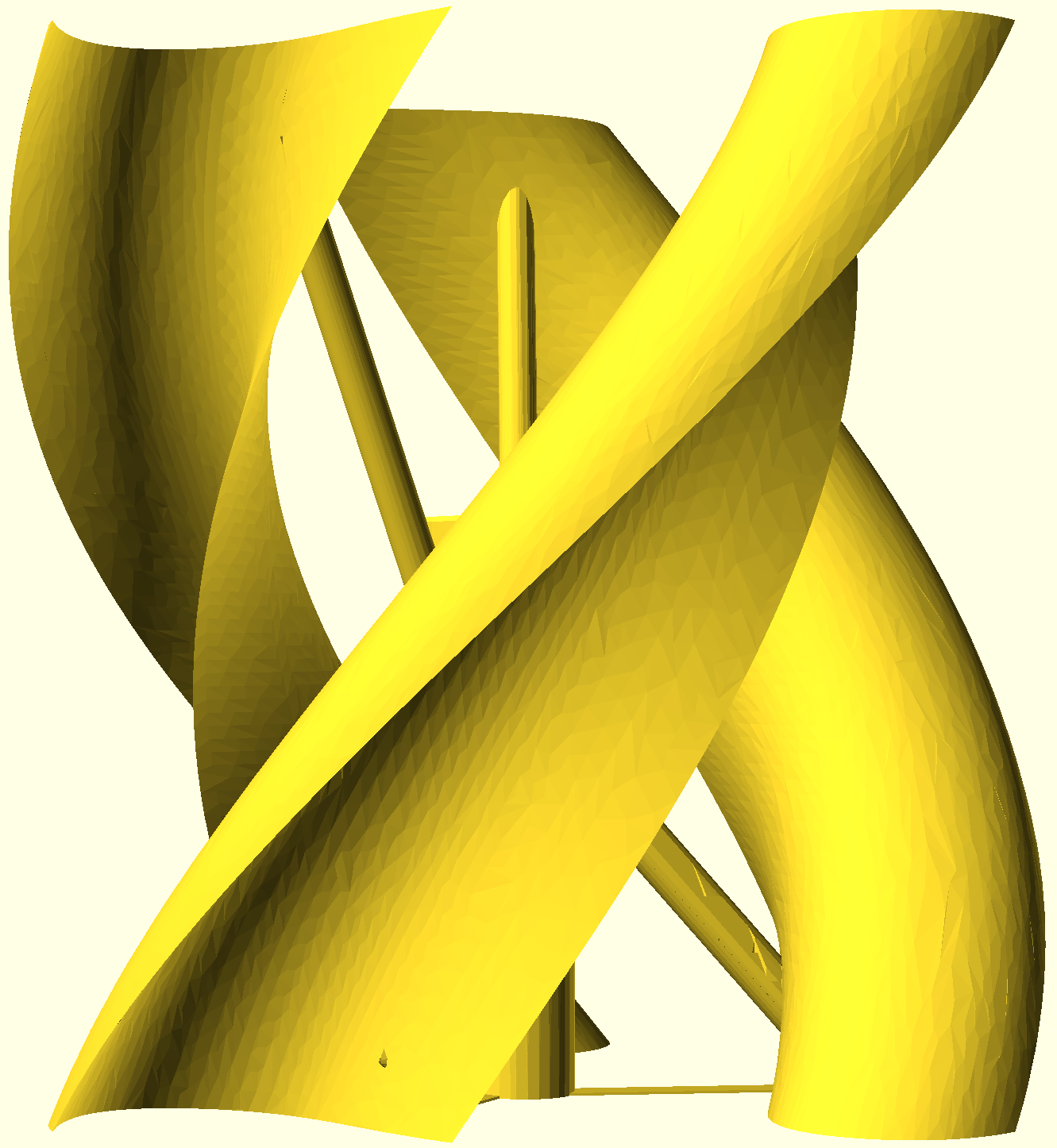} & \includegraphics[width=0.14\textwidth]{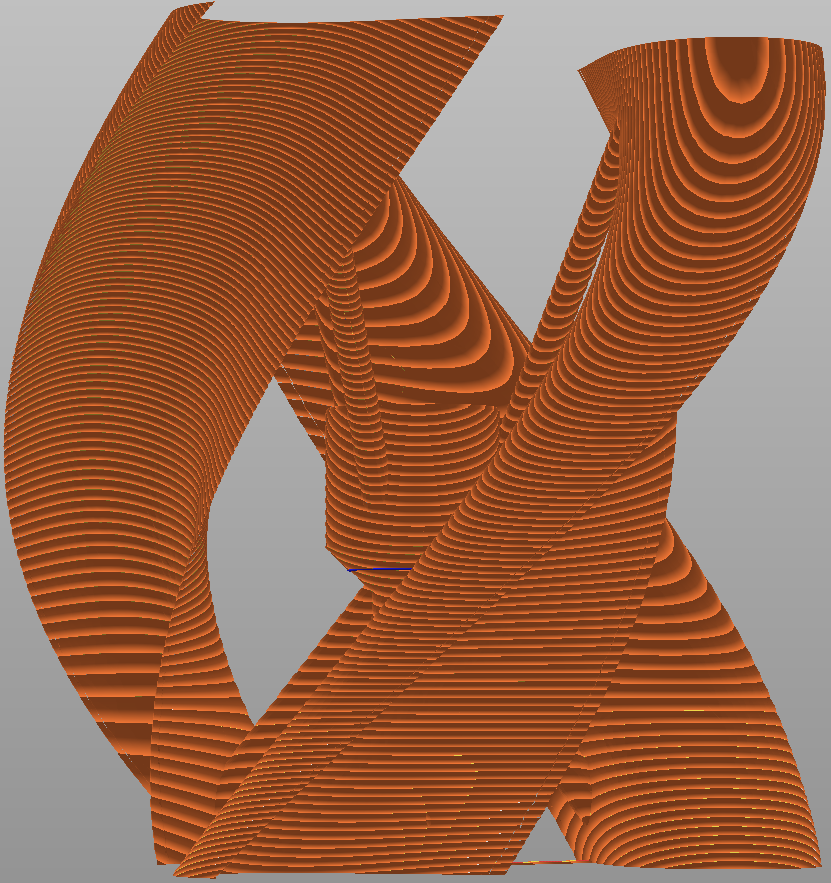} & \includegraphics[width=0.14\textwidth]{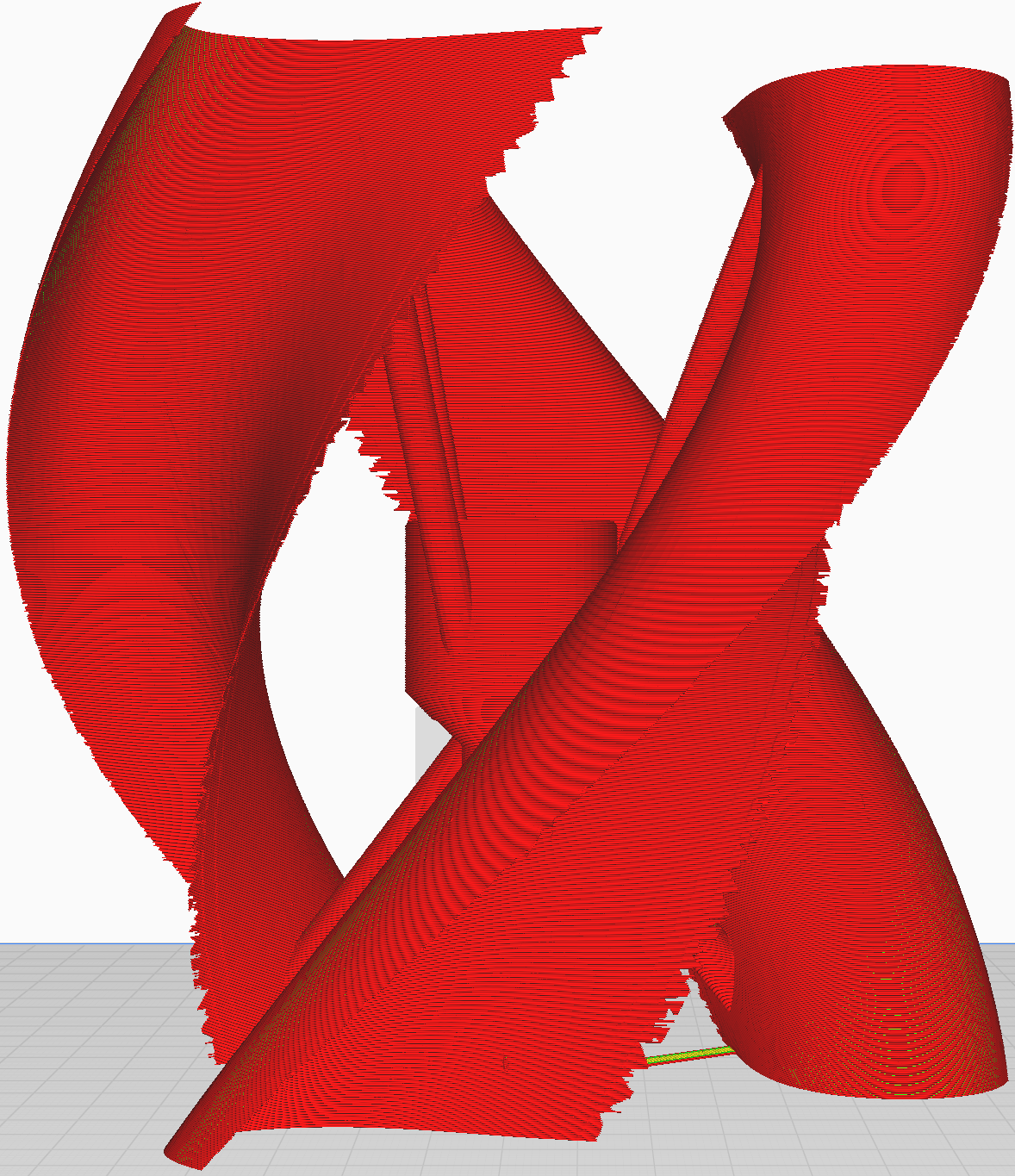} & \includegraphics[width=0.13\textwidth]{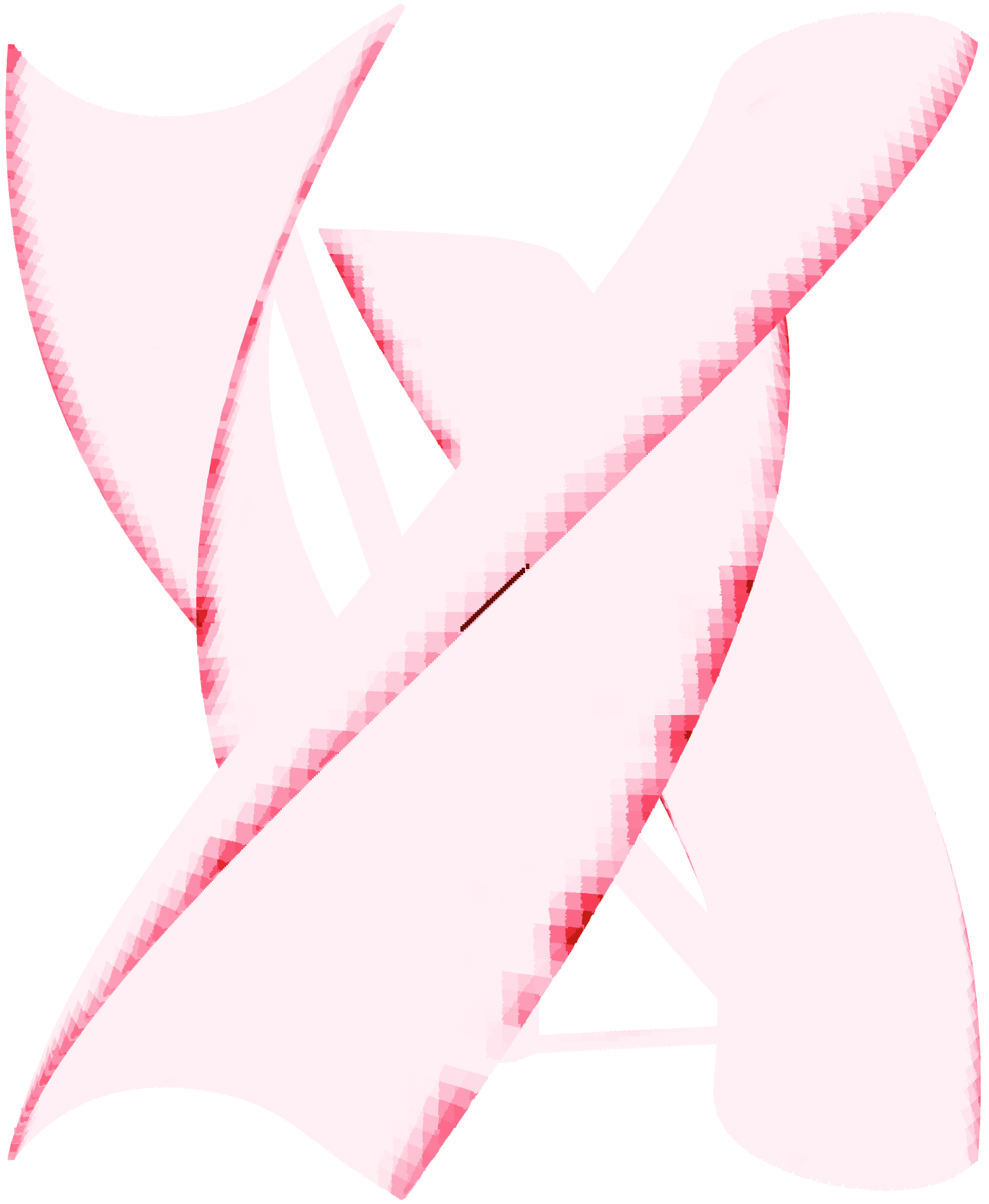}
\end{tabular}
\label{tab:prusa-better-than-cura}
\end{table}

\begin{table}[h]
\footnotesize
 \caption{Comparing Meshmixer and MeshLab on 37 benchmarks that had no suggested fix on their corresponding Github issue. Note that for 3 models not accounted for in this table, Meshmixer showed a combination of partial improvement with newly introduced defects.}
    \centering
    \begin{tabular}{c|c|c|c|c}
        Mesh Tool & Complete resolution & Partial improvement & Not fixed       & New slicing defects \\
        \hline
        MeshLab   &  5.4\% (2/37)       & 5.4\% (2/37)        & 78.3\% (29/37)  & 10.8\% (4/37) \\
        Meshmixer &  8.1\% (3/37)       & 8.1\% (3/37)        & 45.9\% (17/37)  & 29.7\% (11/37) \\
    \end{tabular}
    \label{tab:mesheval}
\end{table}
The first 2 models included in Table~\ref{tab:cura-better-than-prusa} are \texttt{Haut} and \texttt{DrillJig}. 
Like the 
  3\textsuperscript{rd} 
  and 
  4\textsuperscript{th} model in the  table, 
  both these models exhibit non-watertight mesh issues.
Among the {21} non-watertight models that exhibit differences in slicing behavior, 
  Cura outperforms PrusaSlicer in 14 cases,
  including a model for which PrusaSlicer fails to generate G-code.
PrusaSlicer produces G-code that more precisely represents the model in 9 cases;
  for example, 
  the last 2 models in Table~\ref{tab:prusa-better-than-cura}.
Although Cura might appear to have an advantage in handling non-watertight models, 
  only 1 out of 5 models with flipped normal issues is handled better in Cura. 
  
The remaining models are all 
  handled better by PrusaSlicer --- it resolved the issues 
  and generated \gcode, 
  as demonstrated by the first model in Table ~\ref{tab:prusa-better-than-cura}.
Among the 12 models with small features, 
  excluding 1 model for which PrusaSlicer fails to generate G-code 
  and 2 models where the outcome is difficult to determine, PrusaSlicer performs better in 7 cases.
Two of these cases are highlighted as the 
  2\textsuperscript{nd} 
  and 
  3\textsuperscript{rd} models 
  in Table~\ref{tab:prusa-better-than-cura}. 
In contrast, 
  Cura generates better slicing results in only 2 cases, 
  one of which is showcased as the final model in Table~\ref{tab:cura-better-than-prusa}.
  
Overall, we conclude that Cura and PrusaSlicer are good at handling different
  kinds of problematic models --- 
  Cura was able to handle the
  non-watertight models better than PrusaSlicer but was outperformed
  by the latter on the flipped normal models and small-feature models.

\subsubsection{Summary}
This study shows how \tool can be used to
  differentially test slicers across a diverse set of 3D models.
The output heatmaps correctly highlighted the discrepancies between the two programs in all cases.
\tool can be used by developers to
   compare slicers.
For users of slicers, \tool can act as a guide for
deciding which slicer to select for their model.
\tool can empower both developers
  to improve their software and
  users to achieve optimal slicing results.
Indeed, users can compare \gcode produced by the same slicer with different settings to evaluate the effects of configuration settings on slicing quality.

\subsection{Comparing Mesh Repair Tools using \glitchfinder}
\label{subsec:mesh-tool-comp}
It is common in the fabrication community to use
  mesh repair tools to fix meshes and re-slice when
  the original model fails to slice.
We compare two popular mesh repair tools:
  MeshLab~\cite{meshlab} and Meshmixer2017~\cite{meshmixer} to
  evaluate their effectiveness in fixing problematic
  3D models that fail to slice correctly in the Cura slicer. 

\subsubsection{Benchmarks and setup}
We selected 37 models from the 50-model benchmark suite in \autoref{sec:invcheck}  
  for which a viable repair solution was not already
  suggested by someone on the corresponding GitHub issue. 
These models
  cover all three defect categories
  (small features, flipped normals, and not-watertight models)
  in Section~\ref{subsec:claim1}.
For each model,
  we used both Meshmixer and MeshLab to repair the mesh.
We used Meshmixer's ``inspector'' feature that can auto-repair all errors.
MeshLab has many mesh repair functionalities of which we applied
  5 fixes to each model:
  removing duplicate faces,
  removing duplicate vertices,
  removing unreferenced vertices,
  removing zero area faces, and
  repairing non-manifold edges.
We then sliced both repaired meshes with Cura and compare
  the \gcode against the original (uncorrected) \gcode
  using \tool.
We define 4 outcomes:
\begin{itemize}
\item \textbf{Complete resolution}: 
All original slicing errors are corrected with no residual defects.
\item \textbf{Partial improvement}: 
A subset of the original slicing errors persists.
\item \textbf{Not fixed}: 
The repaired mesh has identical slicing
failures as the original mesh.
\item \textbf{New slicing defects}: 
New errors emerged that were not present for the original mesh.
\end{itemize}

\subsubsection{Results}
\autoref{tab:mesheval} summarizes the results.
For both tools, many models remained unfixed for slicing purposes.
Notably, for Meshmixer,
  8.1\% (3 models) showed the combination of
  partial improvement with newly introduced defects.
For both tools,
  all successfully repaired models (complete or partial fixes) were non-watertight, 
  suggesting that non-watertight models are a primary source of repairable slicing errors.
Meshmixer also tends to
  change geometrically valid models, 
  particularly those containing fine features
  (e.g., \texttt{CresGaruru} and \texttt{BearingInsert}).
In these cases, Meshmixer 
  not only failed to resolve the original slicing defects, 
  but also either introduced extraneous features or,
  more often, 
  removed features from the original model,
  thus ultimately producing worse slicing outcomes.
For both tools, we show two representative cases each: 
  one demonstrating successful repair 
  (\autoref{tab:meshlab-fixed}, \autoref{tab:meshmixer-fixed})
  and one showing worsened slicing after repair
  (\autoref{tab:meshlab-worse}, \autoref{tab:mieshmixer-worse}). 

\begin{table}
\footnotesize
\centering
\caption{
Comparison of slicing outcomes before and after MeshLab's repair. 
Columns show: 
  (1) original CAD model, 
  (2) G-code from defective mesh, 
  (3) G-code from repaired mesh, and 
  (4) heatmap. 
MeshLab successfully resolved all mesh defects, 
  producing a fully correct slicing result.
}
\begin{tabular}{c|c|c|c}
{Model} & 
{G-code (Original)} & 
{G-code (MeshLab)} & 
{Heatmap} \\
\hline
\\
\includegraphics[width=0.2\textwidth]{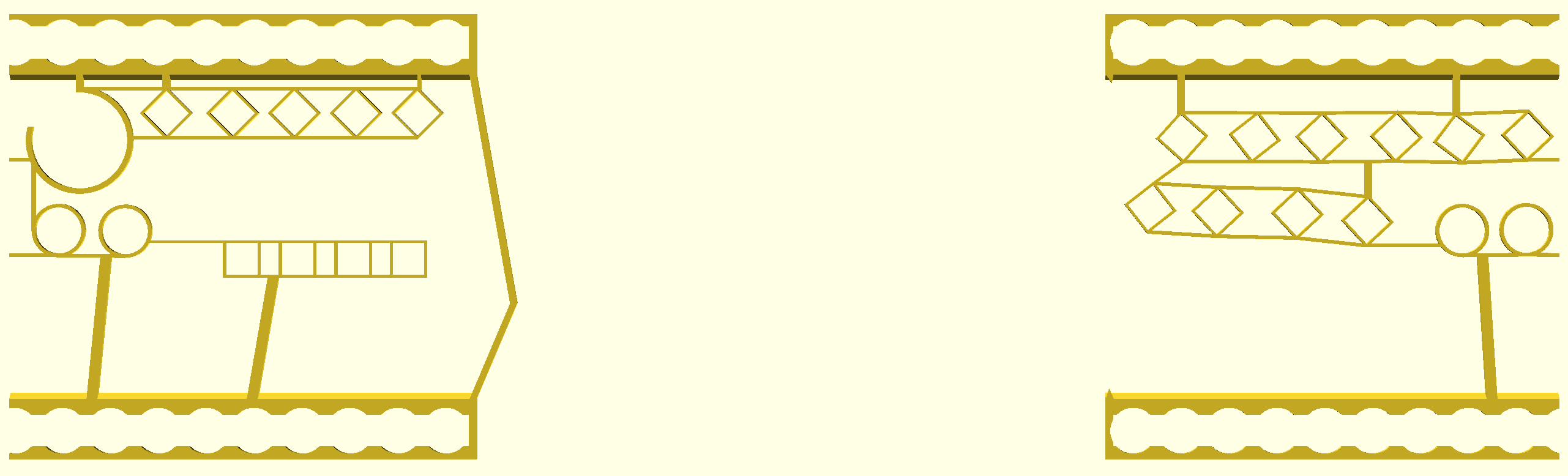} & \includegraphics[width=0.2\textwidth]{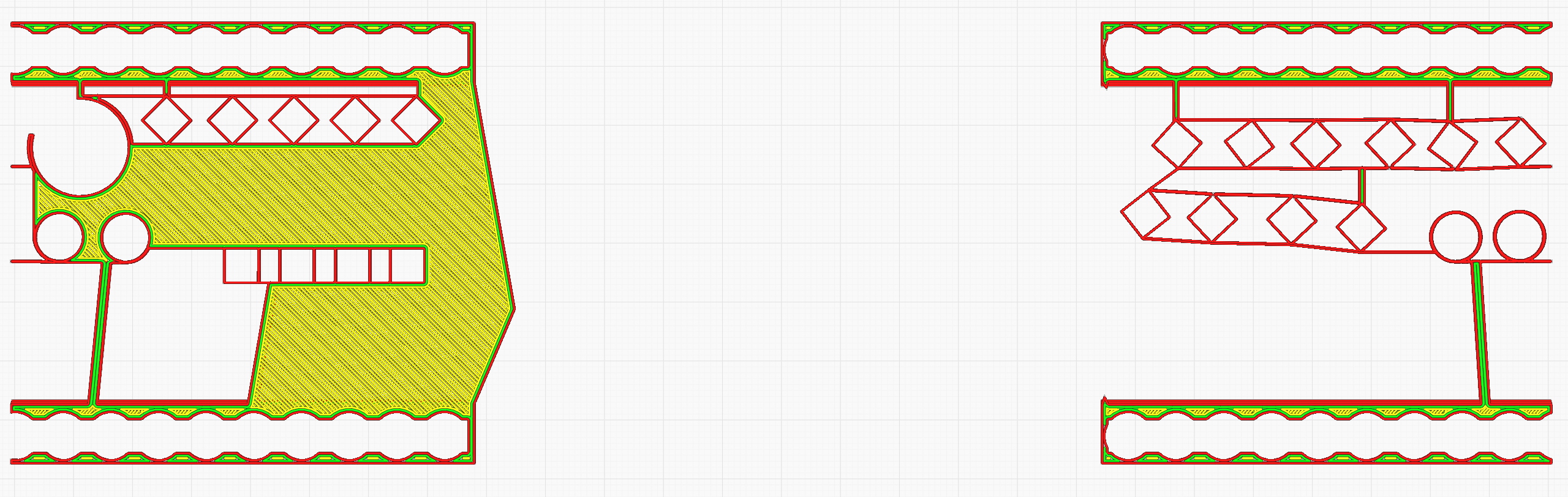} & \includegraphics[width=0.2\textwidth]{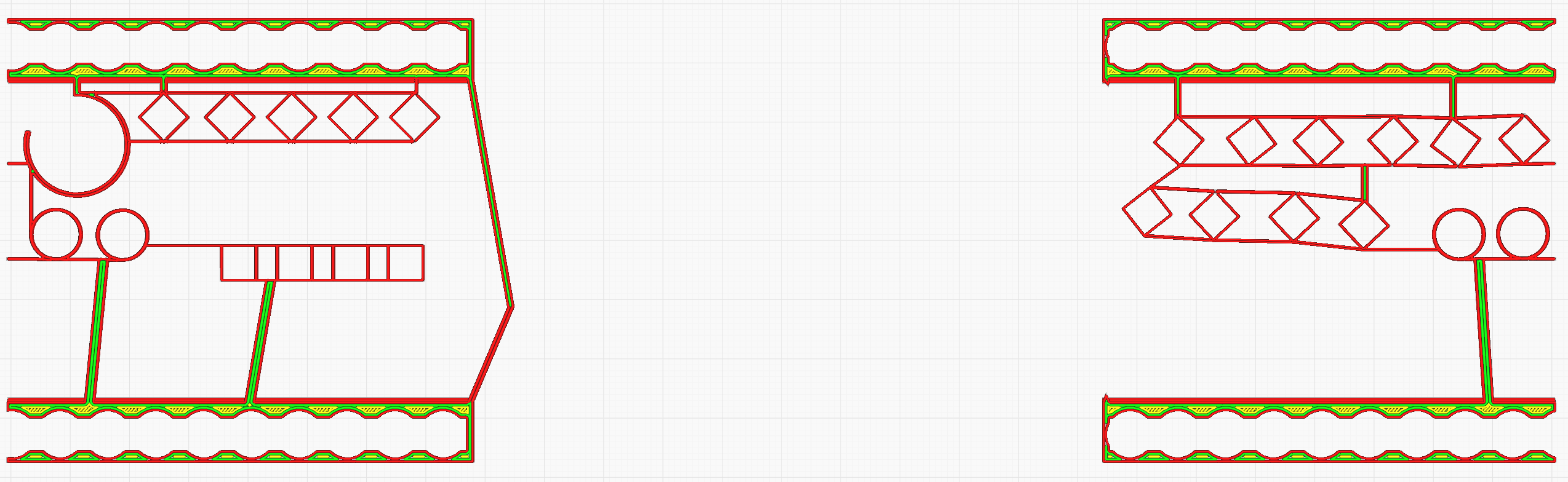} & \includegraphics[width=0.2\textwidth]{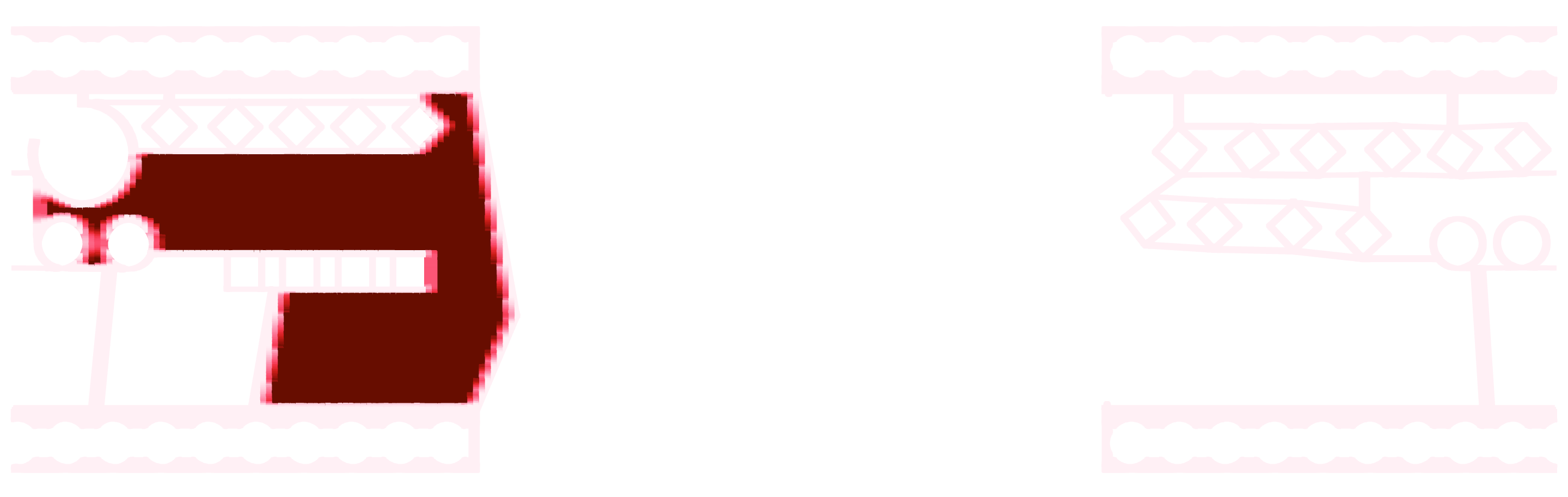}
\end{tabular}
\label{tab:meshlab-fixed}
\end{table}

\begin{table}
\footnotesize
\caption{
Comparison of slicing outcomes before and after MeshLab's repair. 
Columns show: 
  (1) original CAD model, 
  (2) G-code from defective mesh, 
  (3) G-code from repaired mesh, and 
  (4) heatmap. 
MeshLab introduced critical errors in this case, 
  incorrectly filling all designed holes in the model. 
The original G-code only exhibited 
  partial slicing defects at a single circular corner (top-left) feature.
}
\centering
\begin{tabular}{c|c|c|c}
{Model} & 
{G-code (Original)} & 
{G-code (MeshLab)} & 
{Heatmap} \\
\hline
\\
\includegraphics[width=0.1\textwidth]{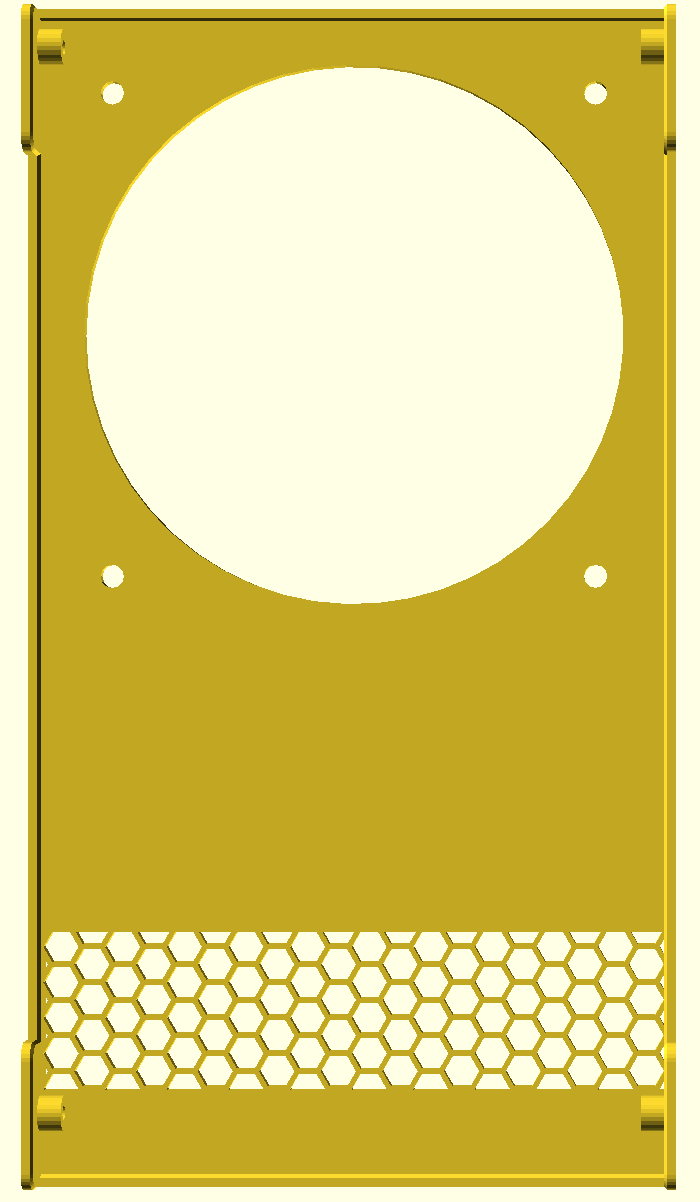} & \includegraphics[width=0.1\textwidth]{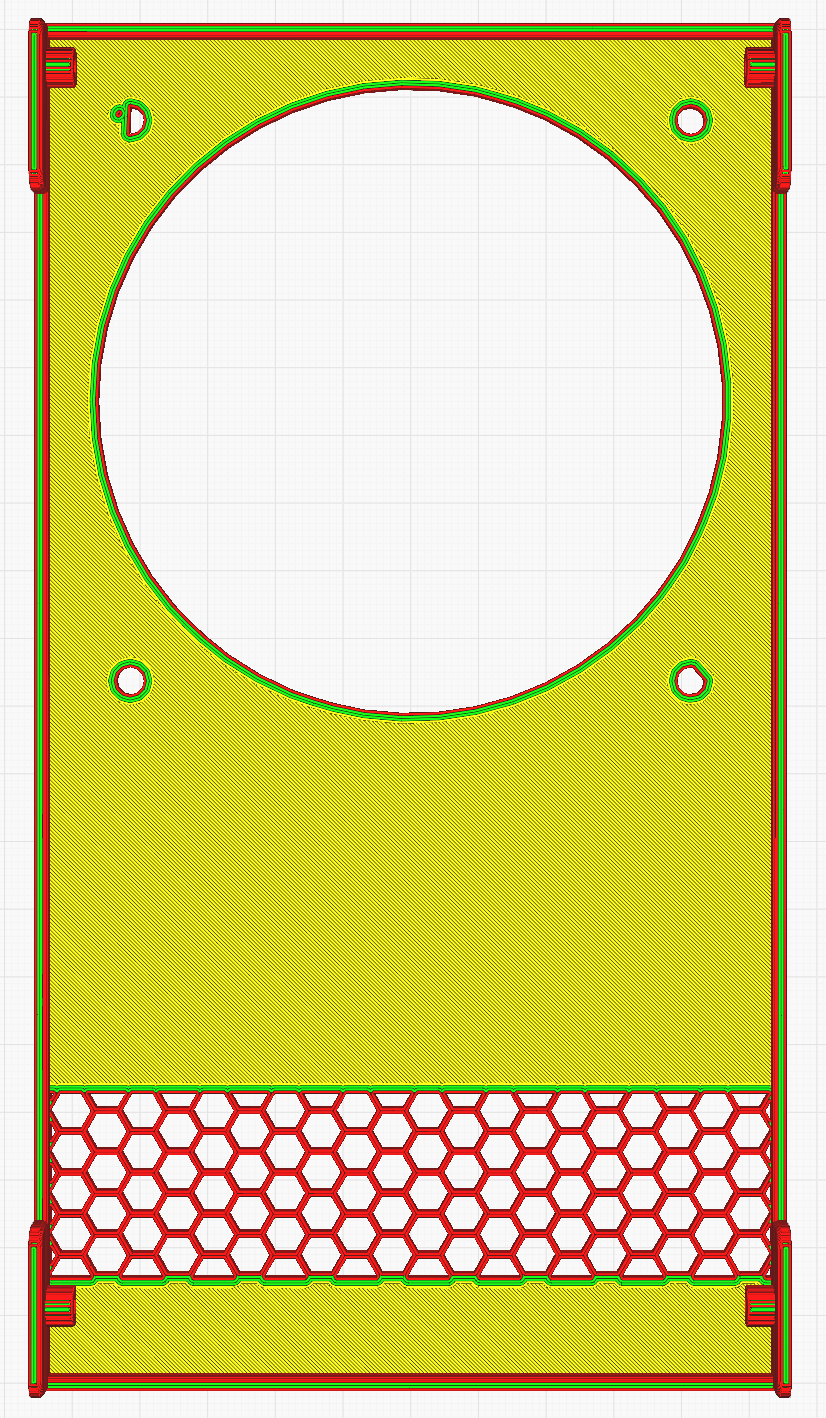} & \includegraphics[width=0.1\textwidth]{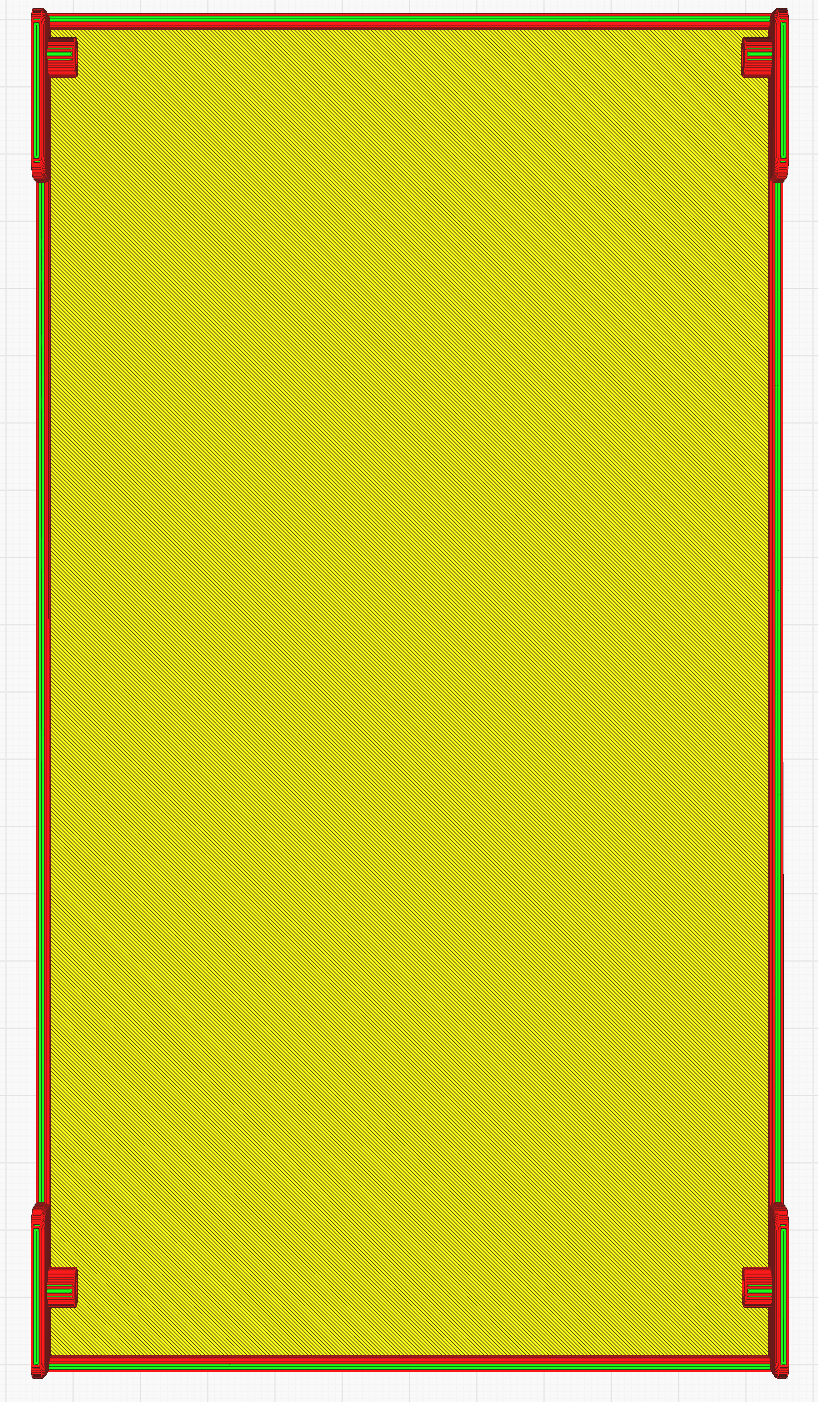} & \includegraphics[width=0.1\textwidth]{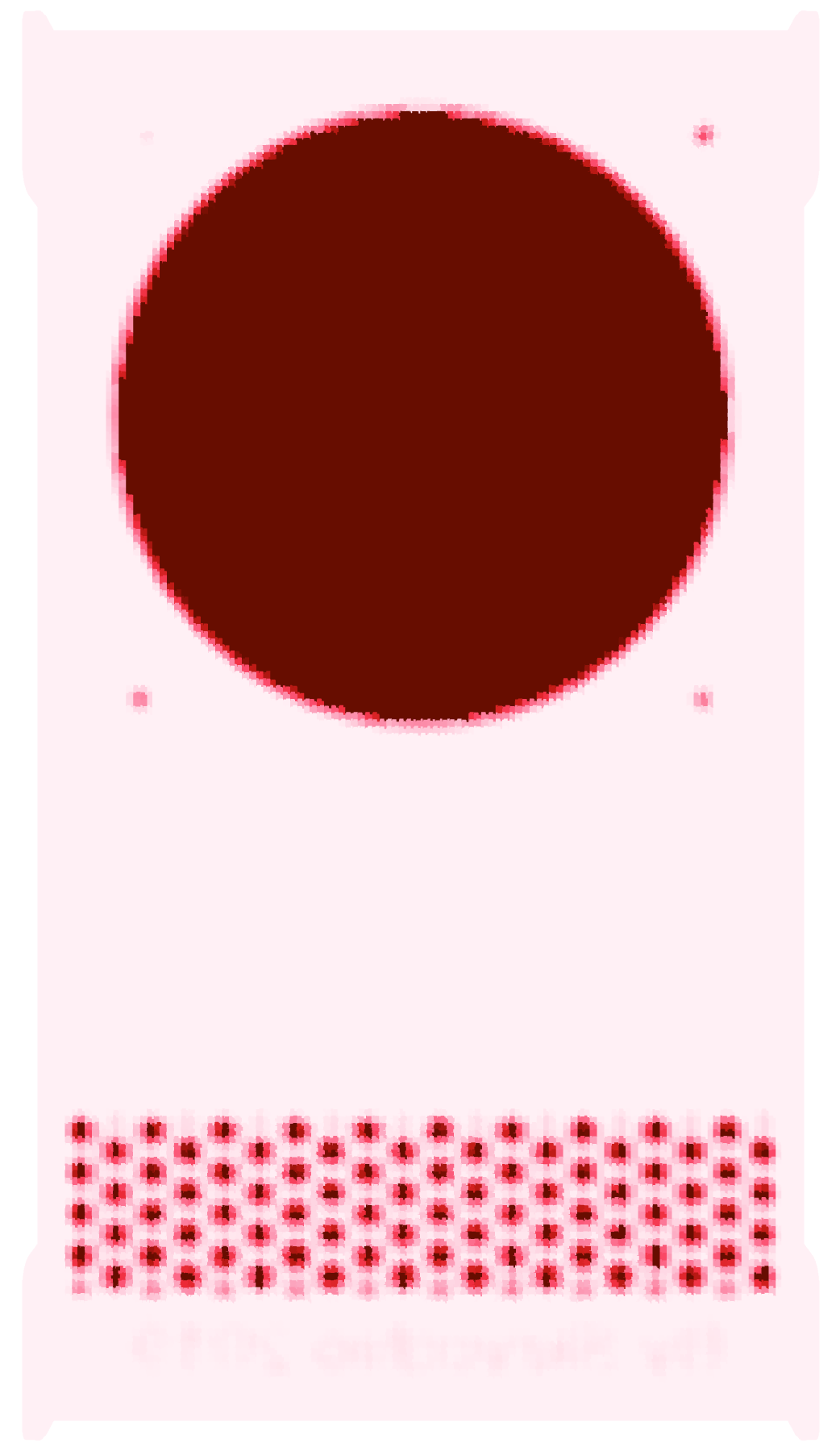} \\
\end{tabular}
\label{tab:meshlab-worse}
\end{table}

\begin{table}[htbp]
\footnotesize
\caption{
Comparison of slicing outcomes before and after Meshmixer's repair. 
Columns show: 
  (1) original CAD model, 
  (2) G-code from defective mesh, 
  (3) G-code from repaired mesh, and 
  (4) heatmap. 
Meshmixer successfully addressed all geometric flaws, 
  resulting in a correct slicing.
}
\centering
\begin{tabular}{c|c|c|c}
{Model} & 
{G-code (Original)} & 
{G-code (Meshmixer)} & 
{Heatmap} \\
\hline
\\
\includegraphics[width=0.15\textwidth]{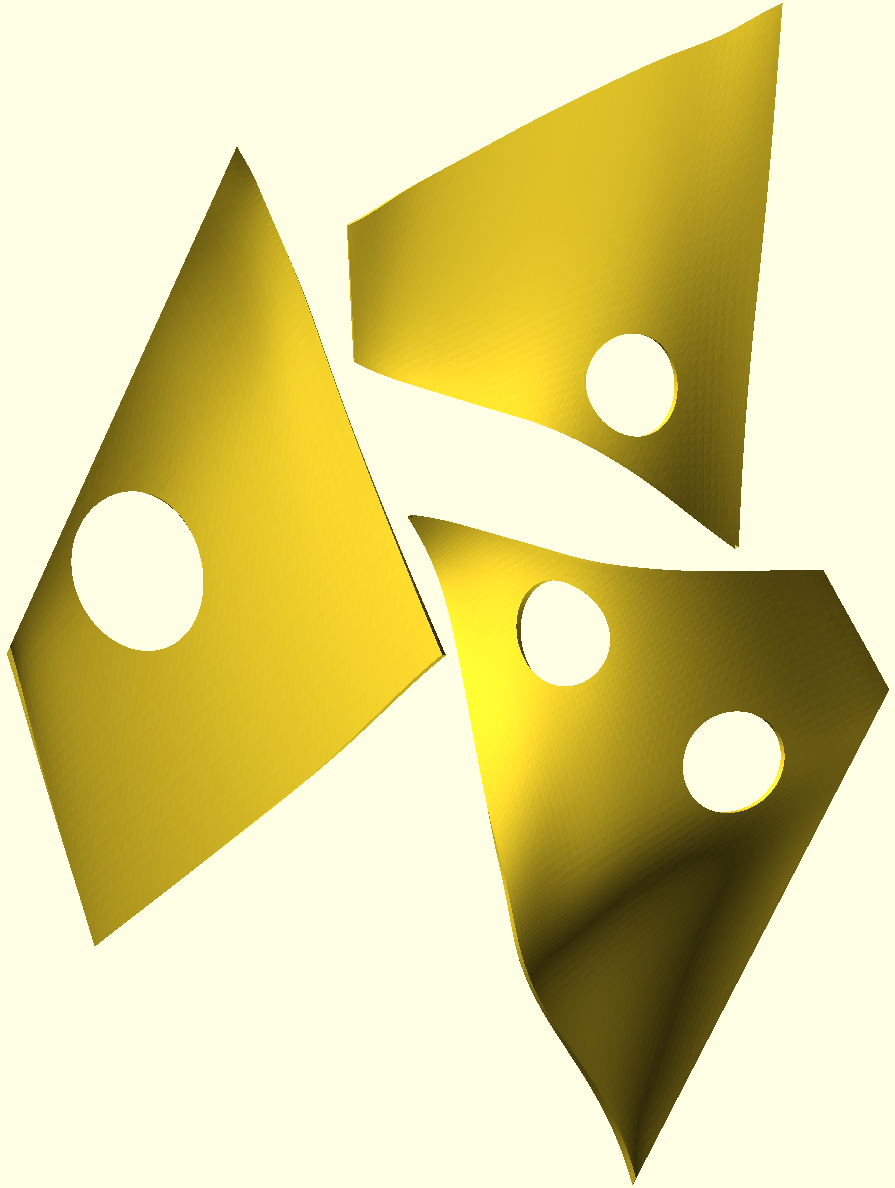} & \includegraphics[width=0.15\textwidth]{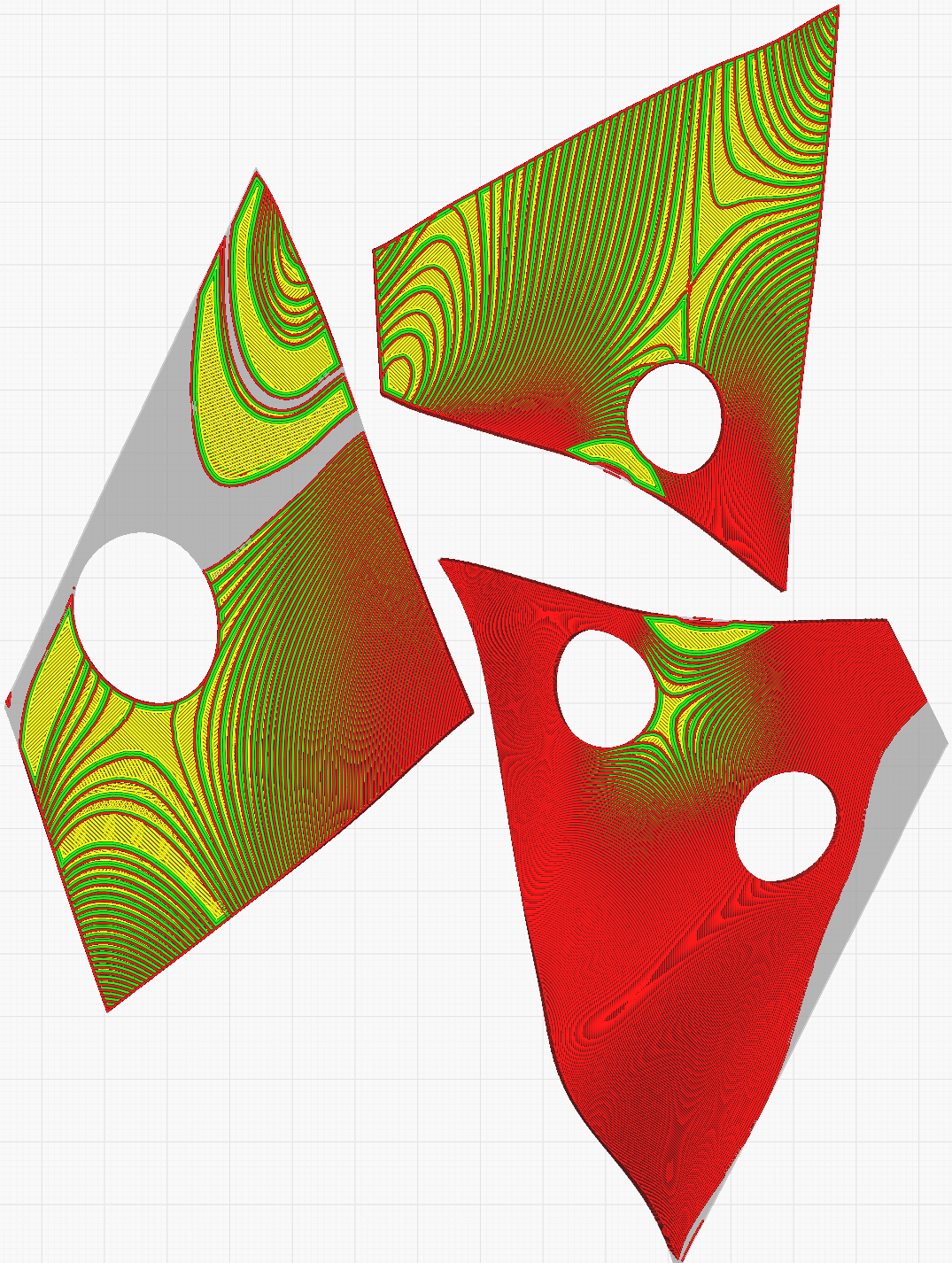} & \includegraphics[width=0.15\textwidth]{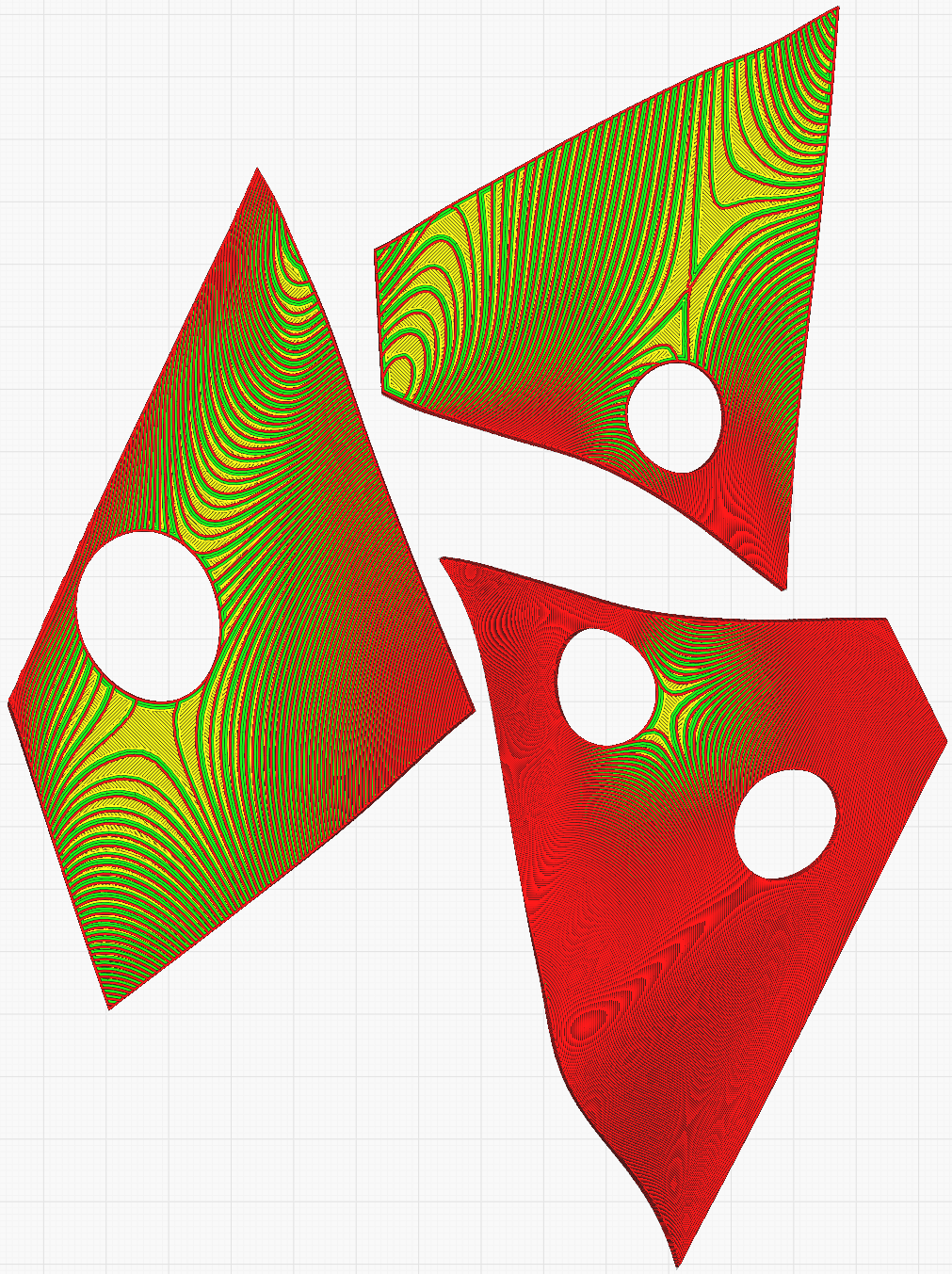} & \includegraphics[width=0.15\textwidth]{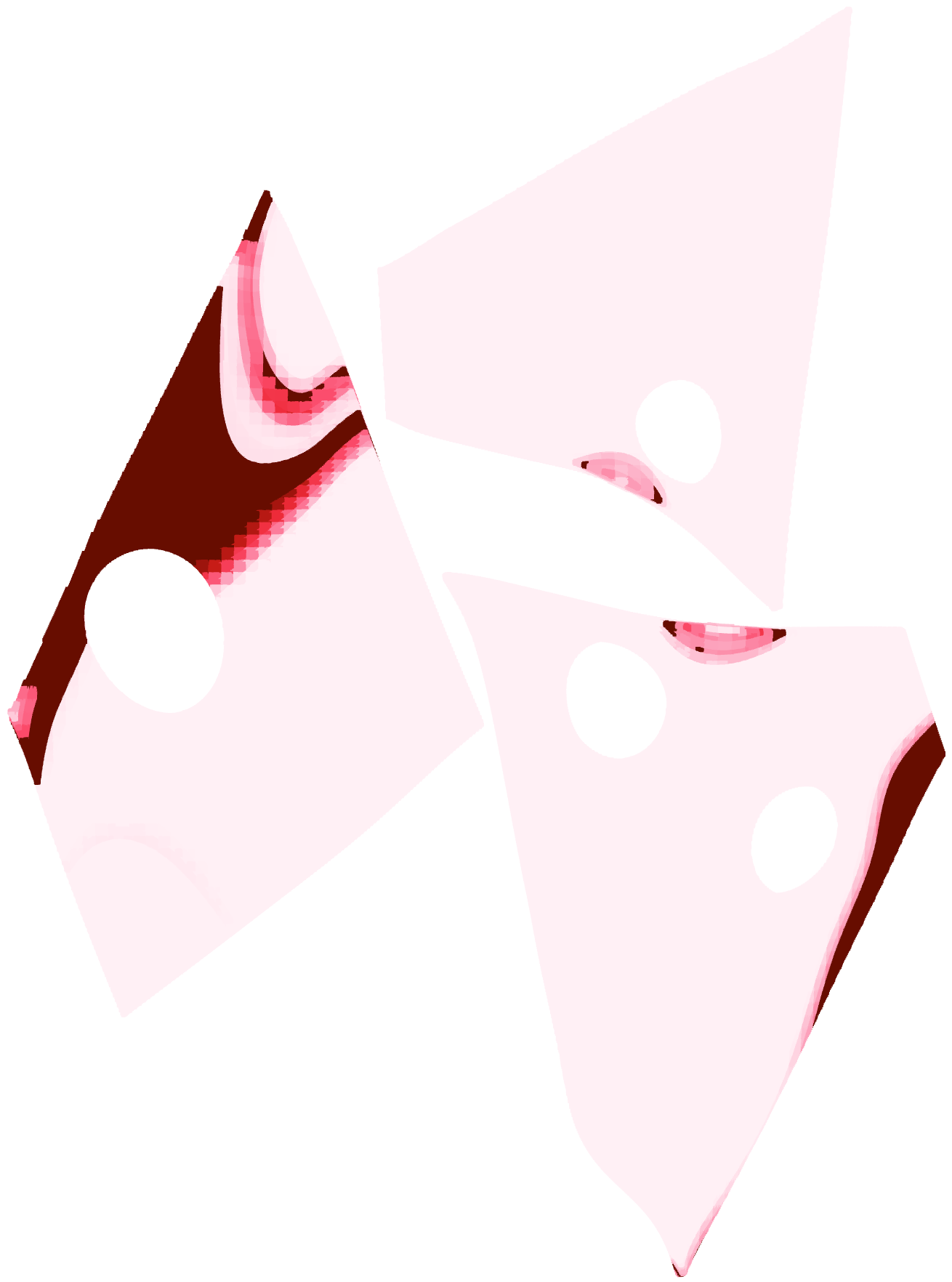} 
\\
\end{tabular}

\label{tab:meshmixer-fixed}
\end{table}

\begin{table}
\footnotesize
\caption{
While manually corrected for manifold errors by the user, 
  this model still showed slicing failures in fine features (e.g., thin cape).
Meshmixer further degraded results by removing certain mesh elements at the model's foot and ankle  area, which can be clearly seen in the heatmap.
}
\centering
\begin{tabular}{c|c|c|c}
{Model} & 
{G-code (Original)} & 
{G-code (Meshmixer)} & 
{Heatmap} \\
\hline
\includegraphics[width=0.2\textwidth]{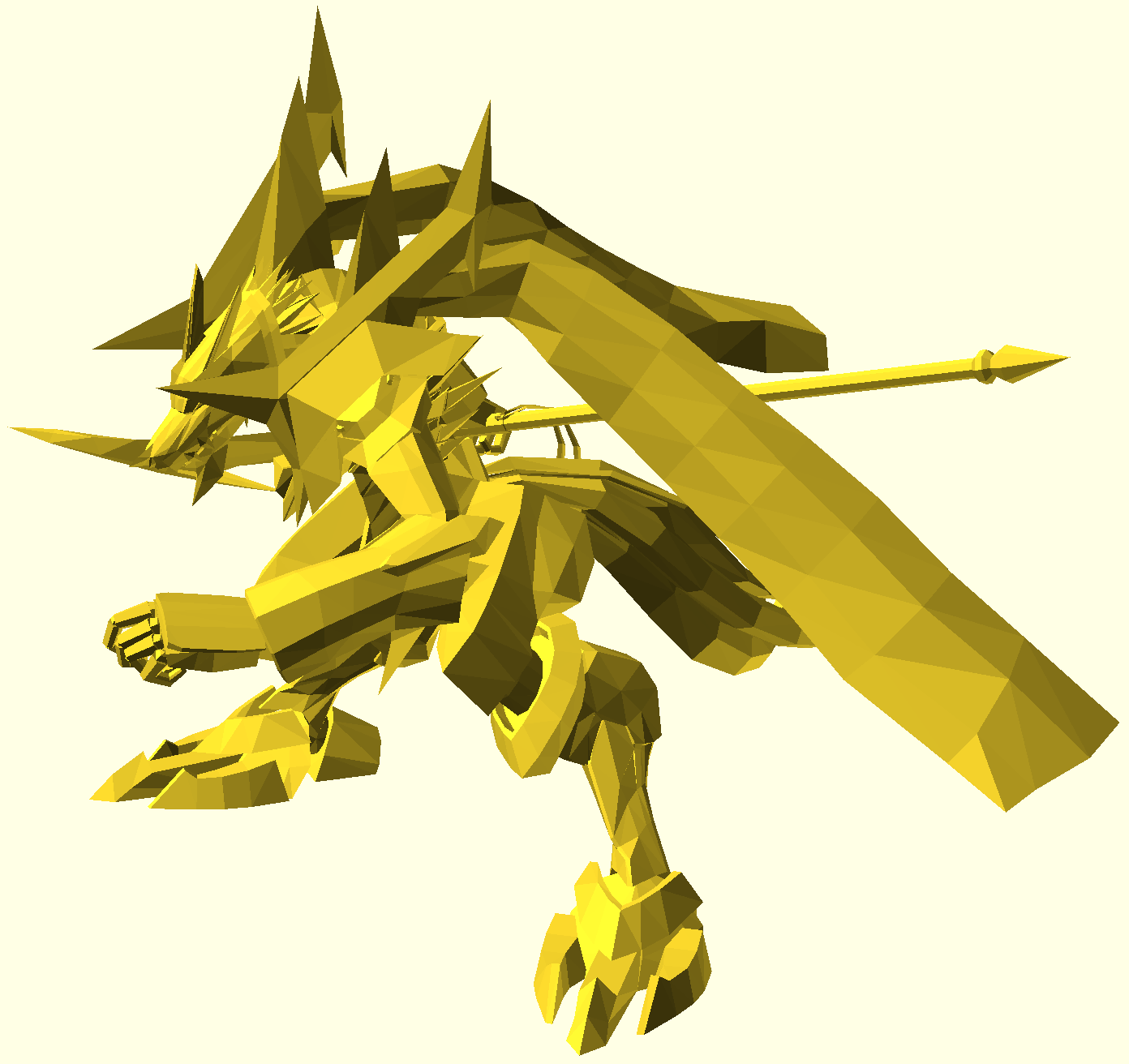} & \includegraphics[width=0.22\textwidth]{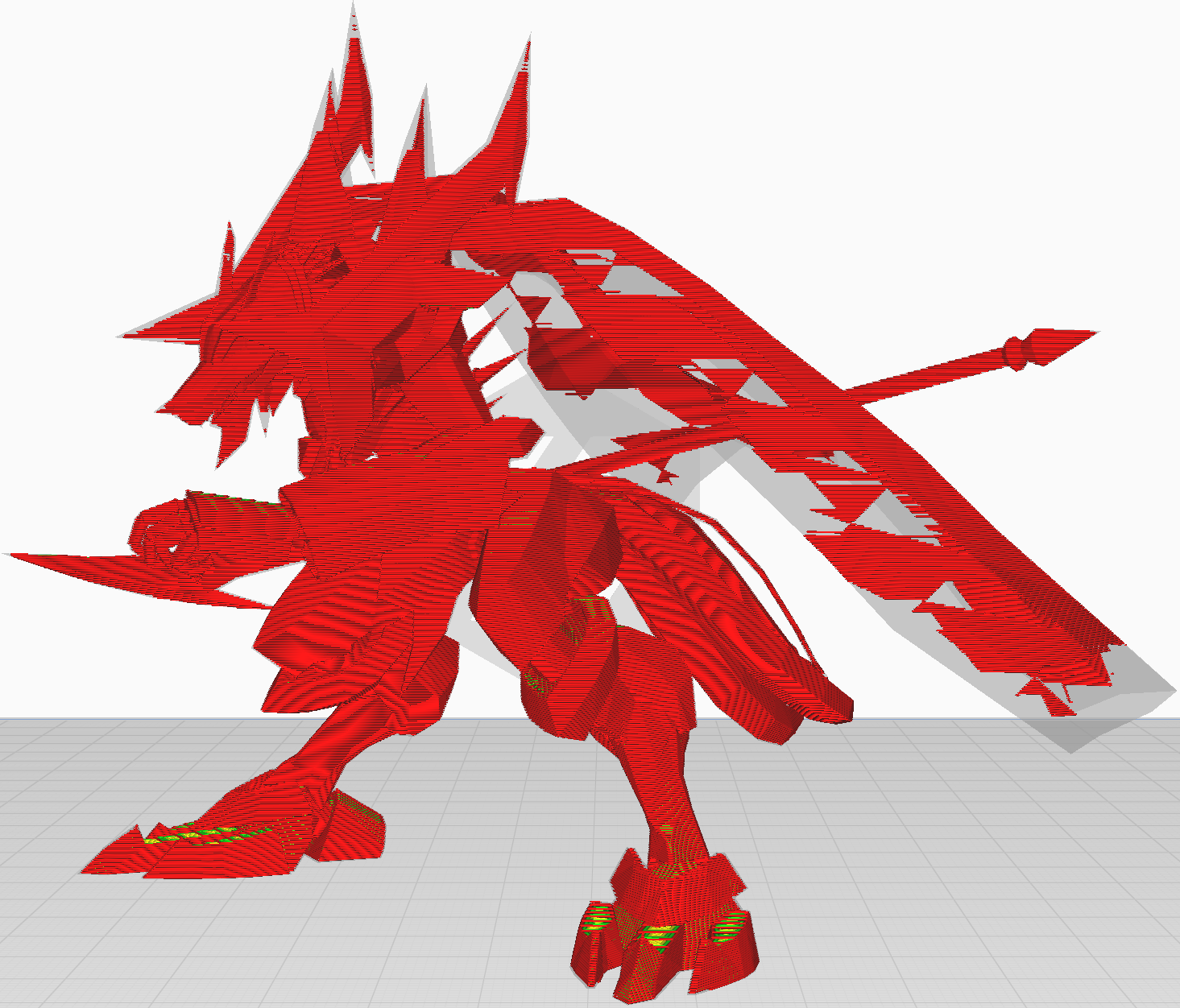} & \includegraphics[width=0.22\textwidth]{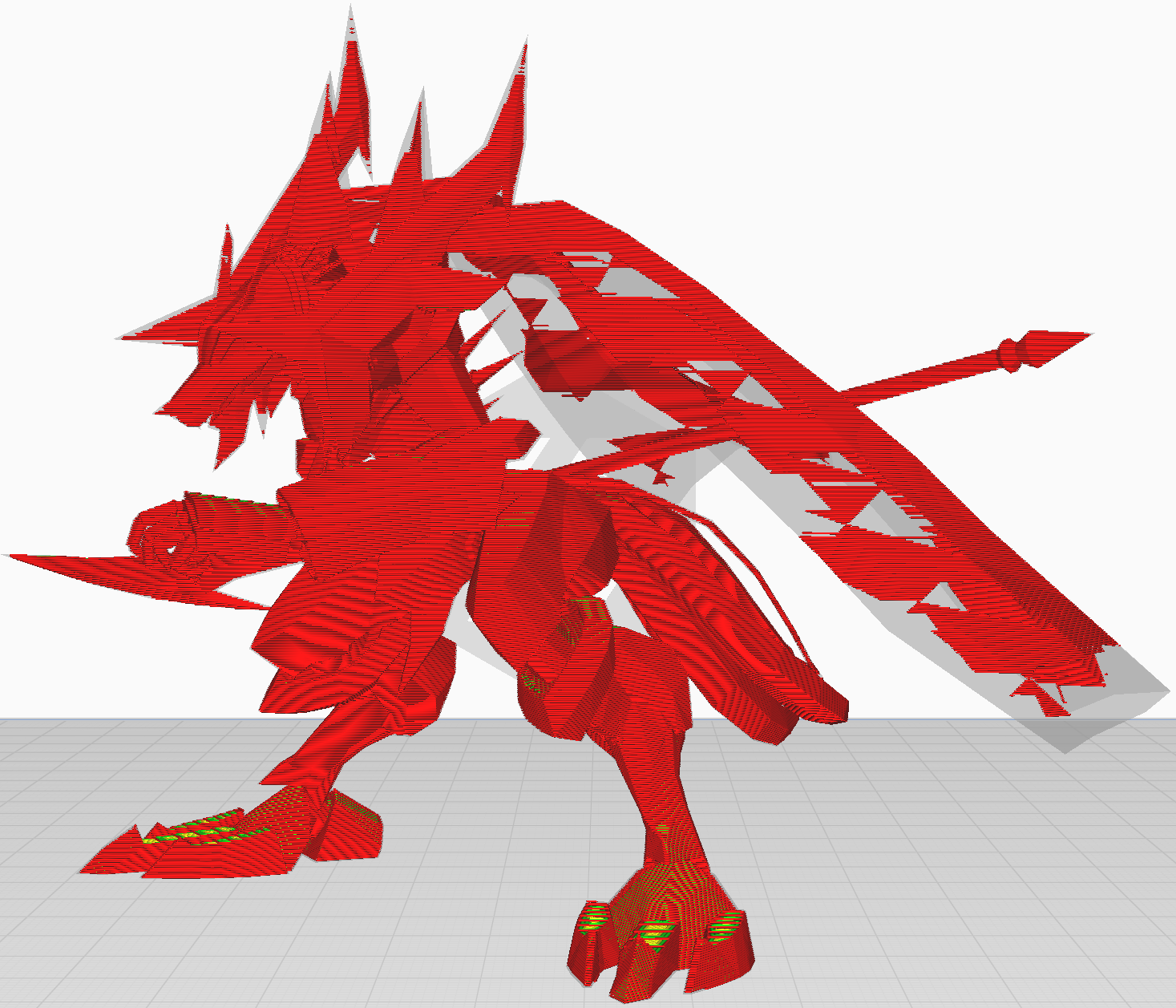} & \includegraphics[width=0.2\textwidth]{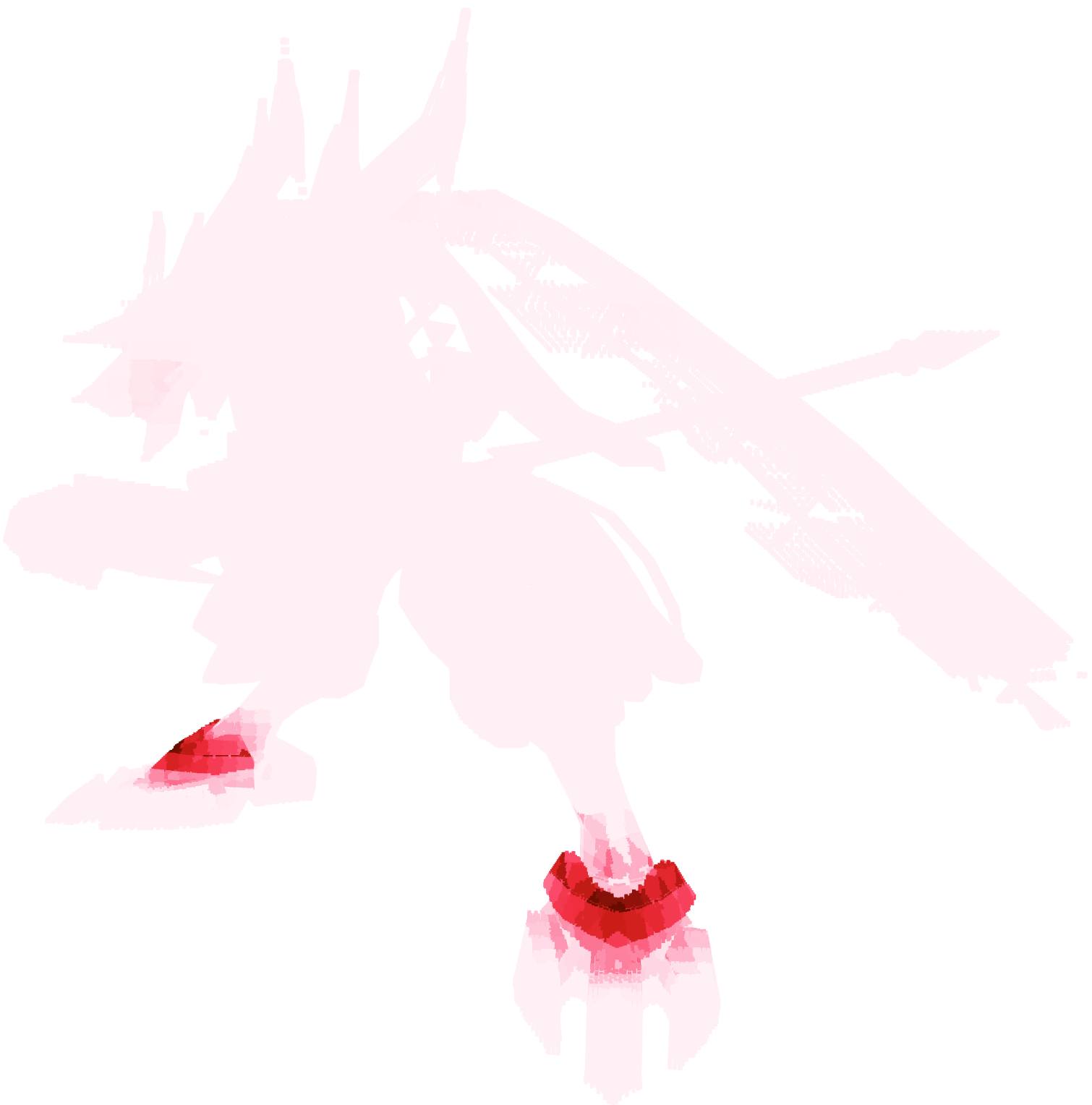}
\\
\end{tabular}

\label{tab:mieshmixer-worse}
\end{table}

\subsubsection{Summary}
This showed yet another
  application of \tool --- by enabling \gcode comparison
  it allowed us to compare the efficacy of popular mesh repair tools
  when used in the context of slicing models for 3D printing.
We found that both mesh repair tools tend to have 
  many ``false negatives'', i.e., 
  they do not successfully fix the models even when
  they generate an output.

\section{Related Work}
\label{sec:related}

\textbf{Programming languages for fabrication} Prior work has used program synthesis, term rewriting, and
  decompilation techniques to reconstruct CAD models from polygon meshes~\cite{carton2021parallel, snapl, altidor, Jones2021, csgnet, shape,  szalinski, inverse, babble}, provided a formal semantics
  for CSG and mesh~\cite{reincarnate, sherman} which has been used
  to develop semantics preserving compilers and decompilers~\cite{reincarnate}, and laid the foundation for developing tools like debuggers and analyzers for toolpaths~\cite{taxon}.
Vespidae~\cite{vespidae} allows users to develop and visualize custom toolpaths.
Imprimer~\cite{imprimer} has explored literate programming in the context of CNC milling. 
In concurrent work,
\citet{nfm25} presented the first
  mechanized semantics of a subset of \gcode and
  a formally verified interpreter
  in Why3 and Rocq.
Our definition of state, $\sigma$, is similar to
  the machine's physical state defined by \citet{nfm25} although
  they do not ``lift'' the \gcode commands
  by modeling them as cuboids.
However, none of these formally reason about \gcode for checking invariants of models or differential testing of fabrication tools.

\textbf{Understanding \gcode.} GSim (\citet{tsoutsou-17}) decompiles \gcode  to obtain an approximation of the input as a CSG model.
We implemented \citet{tsoutsou-17}'s approach and found that their method for reconstructing
   \gcode using CSG does not scale for comparing \gcode programs (\autoref{subsec:cloud}).
Recent work~\cite{jignasu2023foundational, BADINI2023278, makatura2023large} has used
  large language models (LLMs) for debugging and 
  comprehension of \gcode and for CAD/CAM more broadly.
\citet{garcia} developed an algorithm for constructing point clouds from
  scanned objects and compiling them directly to toolpaths.
We start with \gcode (which is closer to toolpaths)
  and reconstruct a point cloud modeling the lines in the \gcode.
\citet{e7f3508655234bc787b90338c07b2332} used machine learning to reverse engineer
  the toolpath used to manufacture a 3D object using fiber reinforced ABS filament.
\citet{Baumann2017} developed a simulation
  technique based on contour detection
  for reverse engineering STL meshes from \gcode.
Prior work has studied techniques for optimizing the orientation of a model when 3D printing
  in order to increase mechanical strength~\cite{umetani, DELFS2016314},
  and developed methods for analyzing the quality of 3D printed objects~\cite{luban}.
While in this work we targeted uniform, planar slicing~\cite{3dp-overview, DOLENC1994119},
  other slicing techniques~\cite{curvislicer, mit-class} are witnessing adoption in many tools~\cite{LIM2016216}.

\textbf{Studying slicers.} 
\citet{Sljivic2019} and \citet{Ariffin2018} compared slicers by analyzing printed objects,
  while \citet{bryla2021mex} examined how filament, slicer, and printer together affect output.
These works studied the manufactured results rather than the \gcode itself.
\citet{machines9080163} conducted a \textit{syntactic} comparison of \gcode from two slicers~\cite{cura, prusaslicer},
  examining how identical settings yield different artifacts.
In contrast, we use \tool to \textit{semantically} analyze slicer behavior, highlighting differences in how each handles geometric challenges.
Recent work~\cite{nadyachi2025} explores interactive tuning via parameter adjustment.

\section{Conclusion}
\label{sec:conclusions}
In this work, we lay the groundwork for formal reasoning
  of toolpaths in the form on \gcode.
We defined semantics for linear-motion \gcode and developed \tool, a static analysis tool for comparing \gcode programs.
We showed two key applications of \tool: checking model invariants and differential testing
  of slicers and mesh repair tools and demonstrated both on real-world models and popular tools.
By operating directly on \gcode, \tool enables rigorous analysis of slicer behavior
  and can be integrated into CI pipelines to monitor tooling quality.
Future directions include supporting additional slicing strategies and
  exploring further applications of high-level geometry reconstruction from \gcode.

\begin{acks}
Thanks to the anonymous reviewers for their feedback. We are grateful to Zachary Tatlock for sharing valuable, insightful comments about the work and to Amy Zhu for reading an earlier draft.
\end{acks}

\section*{Data Availability Statement}
The code and data used in this paper are available on Zenodo~\cite{heartifact2025}.
The code is developed on GitHub~\cite{glitch,glitchscripts}.
The benchmarks we used are all public, though we do not own them~\cite{trickymodelslink}.
\autoref{tab:model-param1}, \autoref{tab:model-param2} in the
  submitted supplementary material (\autoref{sec:appexdixparams}) have
links to all benchmarks and they are included as part of the artifact.

\section{Parameters for running \tool}
\label{sec:appexdixparams}

The parameters used in our evaluation are shown in 
\autoref{tab:model-param1} and \autoref{tab:model-param2}.
The value of Sampling Gap depends on the size of the smallest feature in the model.
Selecting an appropriate Unit Box Size is crucial,
  as it helps isolate errors and ensures the heatmaps accurately highlight problematic regions.
The right value will obtain the ``right'' number of errors -- a number which  does not overwhelm the user but still provides ``actionable'' information,
  a key criteria for success in traditional
  static analysis tools.
We list the values that we found worked best for each model.

\begin{table}[h]
\footnotesize
\centering
\caption{Parameters used for all 58 benchmarks in our evaluation: 50 models with errors, 6 error-free Voron-0 models, and 2 additional models we used for comparing Prusa and Cura slicers (\autoref{subsec:slicer-comp}). 
All dimensions, sampling gaps, and unit box sizes are given in millimeters (mm).
Running time of \tool for the
  invariant checking application (\autoref{sec:invcheck}) in seconds is shown in the last column. 
\texttt{DrillJig} and \texttt{Haut} were not included in the invariant checking benchmark suite, so no runtime data are reported for them.
Times are presented as mean $\pm$ 95\% confidence interval, rounding the mean up to the next whole second and the interval up to the next-highest tenth of a second.
}
\label{tab:model-param1}
\begin{tabular}{llclr}
\toprule
{Name} & {Dimensions} & {Sampling Gap} & {Unit Box Size} & {Time (s)} \\
\midrule
Adapter~\cite{adapter} & $130.0\times 130.0\times 6.7$ & 0.5 & $1.0\times 1.0\times 1.0$ & $485\pm 8.1$\\
AirNozzle &  $19.4\times 19.3\times 31.3$ & 0.08 & $0.6\times 0.6\times 0.6$ & $530\pm 1.6$\\
Airplane~\cite{airplane} & $98.8\times 41.1\times 65.9$ & 0.5 & $1.5\times 1.5\times 1.5$ & $483\pm 2.0$\\
Amy~\cite{amy} & $76.9\times 88.3\times 5.0$ & 0.3 & $1.0\times 1.0\times 0.5$ & $607\pm 3.1$\\
Arm~\cite{arm} & $11.8\times 8.9\times 10.2$ & 0.03 & $0.2\times 0.2\times 0.2$ & $616\pm 0.2$\\
Bagon~\cite{bagon} & $60.8\times 83.2\times 112.9$ & 1.0 & $2.0\times 2.0\times 2.0$ & $564\pm 1.0$\\
Batman~\cite{batman} & $189.9\times 239.6\times 155.0$ & 1.5 & $2.0\times 2.0\times 2.0$ & $1600\pm 9.7$\\
Beard~\cite{beard} & $27.5\times 10.5\times 8.1$ & 0.05 & $0.3\times 0.3\times 0.3$ & $215\pm 1.8$\\
BearingInsert~\cite{ercf} & $40.0\times 25.0\times 7.0$ & 0.08 & $0.6\times 0.6\times 0.6$ & $785\pm 6.0$\\
Bolt~\cite{bolt} & $28.0\times 27.9\times 43.0$ & 0.1 & $0.3\times 0.3\times 0.8$ & $1119\pm 10.0$\\
Borboleta~\cite{borboleta} & $20.8\times 38.0\times 20.0$ & 0.1 & $0.3\times 0.3\times 0.3$ & $813\pm 7.0$\\
Bottlecap~\cite{bottlecap} & $33.0\times 33.0\times 10.3$ & 0.1 & $1.0\times 1.0\times 1.0$ & $358\pm 0.5$\\
Carcasa~\cite{carcasa} & $100.0\times 61.0\times 22.4$ & 0.2 & $0.8\times 0.8\times 0.5$ & $1081\pm 3.1$\\
Car~\cite{car} & $36.4\times 78.7\times 21.8$ & 0.2 & $0.8\times 0.8\times 0.8$ & $483\pm 4.4$\\
CircularHole~\cite{hole} & $110.0\times 200.0\times 12.0$ & 0.5 & $1.0\times 1.0\times 1.0$ & $409\pm 3.6$\\
Clip~\cite{clip} & $28.8\times 40.1\times 21.9$ & 0.1 & $1.0\times 1.0\times 1.0$ & $697\pm 3.8$\\
Cloud~\cite{cloud} & $41.5\times 66.8\times 15.0$ & 0.08 & $0.3\times 0.3\times 1.0$ & $791\pm 1.0$\\
CresGaruru~\cite{cresgaruru} & $138.6\times 136.8\times 98.1$ & 1.0 & $2.0\times 2.0\times 2.0$ & $364\pm 1.1$\\
DoubleCube~\cite{doublecube} & $3.6\times 3.3\times 2.8$ & 0.01 & $0.05\times 0.05\times 0.05$ & $1087\pm 2.4$\\
Dragon~\cite{dragon} & $14.5\times 29.5\times 36.3$ &
0.1 &
$0.5\times 0.5\times 0.5$ 
& $237\pm 3.1$\\
DrillJig~\cite{drilljig} & 
$57.0\times 57.1\times 48.3$ &
0.1 &
$0.8\times 0.8\times 0.8$ 
& N/A\\
DriveFrameUpper~\cite{driveFrameUpper} & 
$61.0\times 50.0\times 22.5$ &
0.2 &
$1.0\times 1.0\times 1.0$ 
& $510\pm 2.6$\\
Drum~\cite{drum} & 
$163.0\times 163.0\times 38.0$ &
1.2 &
$2.0\times 2.0\times 2.0$ 
& $1148\pm 7.0$\\
Fuselage_A~\cite{fuselage1} & 
$99.5\times 100.5\times 133.4$ &
1.5 &
$2.0\times 2.0\times 2.0$ 
& $789\pm 6.9$\\
Fuselage_B~\cite{fuselage2} & 
$120.3\times 137.2\times 85.8$ &
1.0 &
$2.0\times 2.0\times 2.0$ 
& $851\pm 4.5$\\
FaucetHead~\cite{faucetHead} & 
$35.0\times 34.9\times 42.0$ &
0.2 &
$1.0\times 1.0\times 1.0$ 
& $521\pm 2.5$\\
Ford~\cite{ford} & 
$244.7\times 153.4\times 113.1$ &
0.5 &
$2.5\times 2.5\times 2.5$ 
& $1213\pm 3.0$\\
Frame~\cite{frame} & 
$260.9\times 75.0\times 0.6$ &
0.1 &
$1.0\times 1.0\times 0.3$ 
& $182\pm 2.3$\\
GhostMask~\cite{ghostmask} & 
$150.0\times 70.6\times 162.3$ &
0.6 &
$1.5\times 1.5\times 1.5$ 
& $868\pm 2.2$\\
Gun~\cite{gun} & 
$32.0\times 27.8\times 212.1$ &
0.3 &
$0.6\times 0.6\times 1.0$ 
& $1092\pm 16.6$\\
Haut~\cite{haut} & 
$247.3\times 67.5\times 211.0$ &
0.85 &
$3.0\times 3.0\times 1.0$ 
& N/A\\
HexagonalCap~\cite{hexcap} & 
$20.9\times 24.1\times 17.2$ &
0.1 &
$0.5\times 0.5\times 0.5$ 
& $487\pm 3.9$\\
HolyGrail~\cite{holygrail} & 
$15.2\times 15.2\times 6.3$ &
0.05 &
$0.3\times 0.3\times 0.3$ 
& $294\pm 1.7$\\
Horse~\cite{horse} & 
$43.5\times 49.2\times 53.6$ &
0.1 &
$0.8\times 0.8\times 0.8$ 
& $1656\pm 25.5$\\
M2Nut~\cite{m2Nut} & 
$145.0\times 5.5\times 2.4$ &
0.1 &
$1.0\times 0.2\times 0.2$ 
& $255\pm 1.5$\\
Man~\cite{man} & 
$58.8\times 32.0\times 149.6$ &
0.3 &
$0.6\times 0.6\times 0.6$ 
& $2055\pm 9.2$\\
OmegaBoard~\cite{omega} & 
$245.0\times 191.0\times 13.2$ &
0.8 &
$1.2\times 1.2\times 1.0$ 
& $693\pm 4.0$\\
Ring~\cite{ring} & 
$90.0\times 90.0\times 10.0$ &
0.5 &
$1.0\times 1.0\times 1.0$ 
& $384\pm 3.1$\\
Radenci~\cite{radenci} & 
$250.3\times 250.0\times 150.0$ &
0.5 &
$3.0\times 3.0\times 3.0$ 
& $789\pm 7.3$\\
\bottomrule
\end{tabular}
\end{table}

\begin{table}[h]
\footnotesize
\centering
\caption{Continuation of \autoref{tab:model-param1}: Parameters used for all 58 benchmarks in our evaluation: 50 models with errors, 6 error-free Voron-0 models, and 2 additional models we used for comparing Prusa and Cura slicers.
All dimensions, sampling gaps, and unit box sizes are given in millimeters (mm).
Running time of \tool in seconds is shown in the last column.
Times are presented as mean $\pm$ 95\% confidence interval, rounding the mean up to the next whole second and the interval up to the next-highest tenth of a second.
}
\label{tab:model-param2}
\begin{tabular}{llrlr}
\toprule
{Name} & {Dimensions} & {Sampling Gap} & {Unit Box Size} & {Time (s)} \\
\midrule
RearBedMountLeft~\cite{rearLeft} & 
$45.9\times 35.7\times 29.8$ &
0.2 &
$0.6\times 0.6\times 0.6$ 
& $594\pm 2.6$\\
RearBedMountRight~\cite{rearRight} & 
$45.9\times 56.9\times 29.8$ &
0.2 &
$0.8\times 0.8\times 0.8$ 
& $613\pm 4.2$\\
Samurai~\cite{samurai} & 
$39.4\times 63.5\times 87.3$ &
0.3 &
$0.8\times 0.8\times 0.8$ 
& $1100\pm 9.7$\\
Ship~\cite{ship} & 
$36.6\times 27.7\times 65.5$ &
0.1 &
$0.5\times 0.5\times 0.5$ 
& $1578\pm 27.5$\\
Slices~\cite{slices} & 
$148.0\times 198.4\times 40.3$ &
0.3 &
$2.0\times 2.0\times 0.5$ 
& $751\pm 2.1$\\
SpoolHolder~\cite{spoolHolder} & 
$92.0\times 60.0\times 15.0$ &
0.3 &
$1.0\times 1.0\times 0.5$ 
& $488\pm 3.6$\\
Sterm~\cite{sterm} & 
$115.9\times 147.1\times 50.7$ &
1.0 &
$2.0\times 2.0\times 2.0$ 
& $662\pm 1.2$\\
Sword~\cite{sword} & 
$198.1\times 120.6\times 35.7$ &
0.5 &
$1.0\times 1.0\times 1.0$ 
& $1077\pm 8.9$\\
T8NutBlock~\cite{t8NutBlock} & 
$39.2\times 28.0\times 25.0$ &
0.1 &
$0.5\times 0.5\times 0.5$ 
& $1287\pm 7.6$\\
Tabby~\cite{tabby} & 
$96.8\times 140.7\times 98.7$ &
0.8 &
$2.0\times 2.0\times 2.0$ 
& $1471\pm 8.8$\\
Tapa~\cite{tapa} & 
$10.7\times 61.0\times 19.5$ &
0.1 &
$0.5\times 0.5\times 0.5$ 
& $597\pm 4.2$\\
Tardis~\cite{tardis} & 
$100.0\times 129.8\times 110.0$ &
0.4 &
$1.0\times 1.0\times 1.0$ 
& $1810\pm 6.5$\\
Trash~\cite{trash} & 
$15.9\times 17.2\times 9.8$ &
0.05 &
$0.3\times 0.3\times 0.3$ 
& $264\pm 2.2$\\
Tray~\cite{tray} & 
$295.8\times 295.8\times 10.8$ &
3.0 &
$5.0\times 5.0\times 1.0$ 
& $527\pm 2.2$\\
Truss~\cite{truss} & 
$24.7\times 21.8\times 18.5$ &
0.1 &
$0.5\times 0.5\times 0.5$ 
& $263\pm 0.2$\\
Turbo~\cite{turbo} & 
$150.0\times 150.0\times 150.0$ &
0.8 &
$2.0\times 2.0\times 2.0$ 
& $1130\pm 6.2$\\
Vents~\cite{vents} & 
$140.0\times 140.0\times 22.0$ &
0.8 &
$1.2\times 1.2\times 1.2$ 
& $1306\pm 7.3$\\
VertexProblem~\cite{vertexProblem} & 
$17.2\times 4.0\times 60.0$ &
0.1 &
$0.5\times 0.5\times 1.0$ 
& $273\pm 3.0$\\
Warrior~\cite{warrior} & 
$35.7\times 34.1\times 57.2$ &
0.1 &
$0.7\times 0.7\times 0.7$ 
& $1052\pm 4.1$\\
\bottomrule
\end{tabular}
\end{table}

\clearpage
\bibliography{references}

\end{document}